\documentclass[a4paper,11pt]{article}
\pdfoutput=1 % if your are submitting a pdflatex (i.e. if you have
\usepackage{jheppub} % for details on the use of the package, please
\usepackage[utf8]{inputenc}

\usepackage{amsmath, amsthm, amsfonts, amssymb, latexsym}
\usepackage{graphicx}
\usepackage{url}

\usepackage{bm}
\usepackage{xspace}

\usepackage{enumitem}

\hypersetup{
    unicode=true,          % non-Latin characters in Acrobat's bookmarks
    pdftitle={Paths to equilibrium in non-conformal collisions},
  }

\newcommand{\mtt}[1]{\textrm{\tiny #1}}
\newcommand{\AFG}{a_\mtt{FG}}
\newcommand{\uFG}{u_\mtt{FG}}
\newcommand{\phiM}{\phi_\mtt{M}}
\newcommand{\phiH}{\phi_\mtt{H}}
\newcommand{\LIR}{L_\mtt{IR}}
\newcommand{\mIR}{{m^2_\mtt{IR}}}

\newcommand{\sac}{\, , \qquad}
\newcommand{\tcond}{t_\textrm{cond}}
\newcommand{\teos}{t_{\textrm{EoS}}}

\newcommand{\Thyd}{T_{\textrm{hyd}}}
\newcommand{\Teos}{T_{\textrm{EoS}}}
\newcommand{\Tcond}{T_{\textrm{cond}}}
\newcommand{\thyd}{t_{\textrm{hyd}}}
\newcommand{\tiso}{t_{\textrm{iso}}}
\newcommand{\tpeak}{t_{\textrm{peak}}}

\newcommand{\be}{\begin{equation}}
\newcommand{\ee}{\end{equation}}

\newcommand{\bea}{\begin{eqnarray}}
\newcommand{\eea}{\end{eqnarray}}

\newcommand{\eqn}[1]{equation~(\ref{#1})}

\newcommand{\eqq}[1]{(\ref{#1})}

\title{Paths to equilibrium in non-conformal collisions}

\author[a]{Maximilian~Attems,}
\author[b]{Jorge~Casalderrey-Solana,}
\author[a,c]{David~Mateos,}
\author[d]{Daniel~Santos-Oliv\'an,}
\author[d]{Carlos~F.~Sopuerta,}
\author[a]{Miquel~Triana}
\author[a,e]{and Miguel~Zilh\~ao}

\affiliation[a]{
Departament de F\'\i sica Qu\`antica i Astrof\'\i sica \&  Institut de Ci\`encies del Cosmos (ICC), Universitat de Barcelona, Mart\'{\i}  i Franqu\`es 1, 08028 Barcelona, Spain
}

\affiliation[b]{
Rudolf Peierls Centre for Theoretical Physics, University of Oxford, 1 Keble Road, Oxford OX1 3NP, United Kingdom
}

\affiliation[c]{
Instituci\'o Catalana de Recerca i Estudis Avan\c{c}ats (ICREA),
Passeig Llu\'is Companys 23, ES-08010, Barcelona, Spain
}

\affiliation[d]{
Institut de Ci\`encies de l'Espai (CSIC-IEEC), Campus UAB,
Carrer de Can Magrans s/n, 08193 Cerdanyola del Vall\`es, Spain
}

\affiliation[e]{CENTRA, Departamento de F\'\i sica, Instituto Superior
  T\'ecnico, Universidade de Lisboa, Avenida Rovisco Pais 1, 1049 Lisboa,
  Portugal}

\emailAdd{attems@icc.ub.edu}
\emailAdd{jorge.casalderreysolana@physics.ox.ac.uk}
\emailAdd{dmateos@icrea.cat}
\emailAdd{santos@ieec.uab.es}
\emailAdd{sopuerta@ieec.uab.es}
\emailAdd{mtriana@fqa.ub.edu}
\emailAdd{mzilhao@ffn.ub.es}

\preprint{{ ICCUB-17-009}}

\abstract{We extend our previous analysis of holographic heavy ion collisions in non-conformal theories. We provide a detailed description of our numerical code. We  study collisions at different energies in gauge theories with different degrees of non-conformality. We compare 
four relaxation times: the hydrodynamization time (when hydrodynamics becomes applicable), the EoSization time (when the average pressure approaches its equilibrium value), the isotropization time (when the longitudinal and transverse pressures approach each other) and the condensate relaxation time (when the expectation value of a scalar operator approaches its equilibrium value). We find that these processes can occur in several different orderings. In particular, the condensate can remain far from equilibrium even long after the plasma has 
hydrodynamized and EoSized. We also explore the rapidity distribution of the energy density at hydrodynamization. This is far from boost-invariant and its width decreases as the non-conformality increases. Nevertheless, the  velocity field at hydrodynamization is almost exactly boost-invariant regardless of the non-conformality. This result may be used to constrain the initialization of hydrodynamic fields in heavy ion collisions. 
}

\begin{document}
\maketitle
\flushbottom

%%%%%%%%%%%%%%%%%%%%%%%%%%%%%%%%%%%%%%%%%%%%%%%%%%%%%%%%%%%%%%%%%%%%%%%%%%%%%%%
\section{Introduction}
\label{sec:intro}
%%%%%%%%%%%%%%%%%%%%%%%%%%%%%%%%%%%%%%%%%%%%%%%%%%%%%%%%%%%%%%%%%%%%%%%%%%%%%%%

``Holographic Heavy Ion Collisions'', namely shockwave collisions in an asymptotically AdS spacetime, have provided interesting insights into the far-from-equilibrium properties of hot, strongly-coupled, non-Abelian plasmas that are potentially relevant for the quark-gluon plasma (QGP) created in heavy ion collision experiments (see e.g.~\cite{CasalderreySolana:2011us} for a review). Until recently, 
all such holographic studies (see e.g.~\cite{Chesler:2010bi,Casalderrey-Solana:2013aba,Casalderrey-Solana:2013sxa,Chesler:2015wra,Chesler:2015bba,Chesler:2015lsa,Chesler:2016ceu}) were performed in models dual to conformal field theories (CFTs). One notable lesson of this body of work is that ``hydrodynamization'', the process by which the plasma comes to be well described by hydrodynamics, can occur before ``isotropization'', the process by which all pressures become approximately equal to one another in the local rest frame. 

We have recently begun the study of holographic collisions in non-conformal theories 
\cite{Attems:2016tby,Attems:2017ezz} based on the set of models introduced in \cite{Attems:2016ugt}.\footnote{Some second-order transport coefficients \cite{Kleinert:2016nav} and the entanglement entropy \cite{Rahimi:2016bbv} have been computed for these models.} One crucial difference between the conformal and the non-conformal cases is that in the latter  the equation of state, namely the relation between the energy density and the average pressure, is not fixed by symmetry, and hence it needs not be obeyed out of equilibrium. The relaxation process therefore involves an additional channel,  namely the evolution of the energy density and the average pressure towards asymptotic values related by the equation of state. This process was dubbed ``EoSization'' in \cite{Attems:2016tby}, and once it has taken place we say that the system has ``EoSized''. The main result of  
\cite{Attems:2016tby} was that  EoSization and  hydrodynamization can  occur in any order.

The models of \cite{Attems:2016ugt} are dual to CFTs deformed by a source 
$\Lambda$ for a dimension-three operator. 
The source breaks scale invariance explicitly and triggers a non-trivial Renormalization Group (RG) flow. In this paper we will examine the relaxation process by which the expectation value (the condensate) of this scalar operator approaches its equilibrium value. We refer to the time at which this happens as the ``condensate relaxation time'', $\tcond$. It is particularly interesting to compare this relaxation time to the hydrodynamization, EoSization and isotropization  times, $\thyd$, $\teos$ and $\tiso$. The reason is that the latter three times refer to the approach to equilibrium of conserved charges (energy and momentum), whereas the former refers to the relaxation of a non-conserved quantity (the expectation value of the scalar operator). 
In all the collisions that we have examined we find that isotropization happens last, reinforcing the intuition from conformal  collisions that this process is extremely slow. For this reason, in most of the paper we will focus on the other three times and we will come back to $\tiso$ in section~\ref{sec:discussion}. In contrast, we find that the other three times  can occur in several different orderings. In particular, $\tcond$ can be much longer than $\thyd$ and $\teos$. This shows that one-point functions of non-conserved operators can remain far from equilibrium long after a plasma has hydrodynamized and EoSized. 

We also examine the physics away from mid-rapidity. For this purpose we compute the rapidity profile of the energy density at hydrodynamization. Just like in the conformal case  \cite{Casalderrey-Solana:2013aba,Chesler:2015fpa}, this profile is not boost-invariant but Gaussian. The width of this Gaussian decreases as the degree of non-conformality increases.  Although the energy profile is determined by far-from-equilibrium physics beyond hydrodynamics, this decrease seems correlated with the bulk viscosity in our models. Indeed, as the non-conformality increases the bulk viscosity grows, which reduces the longitudinal expansion and hence the width of the region where energy is deposited. 

A remarkable result of our away-from-mid-rapidity analysis is the fact that, although the energy density profile is far from boost-invariant, the velocity field is almost exactly boost-invariant even for the most non-conformal collisions. For CFTs this was first observed in \cite{Chesler:2015fpa}. Therefore our result implies that, although  the non-conformality has a large effect on the energy density profile  at hydrodynamization, it leaves the velocity field essentially unmodified.

This paper is organized as follows. In section~\ref{sec:setup} we introduce our
non-conformal models, along with its thermodynamic and transport properties. In
section~\ref{sec:numerical} we describe the numerical procedure used to evolve
the corresponding equations, and in section~\ref{sec:tests} we present tests on the
numerical code we have developed to this end. In
section~\ref{sec:collisions} we perform a detailed study of shockwave collisions
in our models. We conclude with a general discussion in 
section~\ref{sec:discussion}.

%%%%%%%%%%%%%%%%%%%%%%%%%%%%%%%%%%%%%%%%%%%%%%%%%%%%%%%%%%%%%%%%%%%%%%%%%%%%%%%
\section{Setup}
\label{sec:setup}
%%%%%%%%%%%%%%%%%%%%%%%%%%%%%%%%%%%%%%%%%%%%%%%%%%%%%%%%%%%%%%%%%%%%%%%%%%%%%%%

\subsection{The model}
\label{sec:model}
We will consider dynamics in a five-dimensional holographic model consisting of gravity coupled to a scalar field with a non-trivial potential. The action for our Einstein-scalar model is 
\begin{equation}
\label{eq:action}
S=\frac{2}{\kappa_{5}^{2}} \int d^{5} x \sqrt{-g} \left[ \frac{1}{4} \mathcal{R}  - \frac{1}{2} \left( \nabla \phi \right) ^2 - V(\phi) \right] \,.
\end{equation}
The dynamic equations resulting from it read
\begin{align}
\label{eq:EE-phi}
& R_{\mu\nu} - \frac{R}{2}g_{\mu\nu} = 8\pi T_{\mu\nu} \,, \\[2mm]
& \square \phi = \frac{\partial V}{\partial \phi} \,,
\end{align}
where
\begin{align}
\label{eq:Tmunu-phi}
8\pi T_{\mu\nu} & = 2\partial_{\mu} \phi \partial_{\nu} \phi
- g_{\mu\nu} \left(
g^{\alpha \beta} \partial_{\alpha} \phi  \partial_{\beta} \phi
+ 2V(\phi)
\right)\,,
\end{align}
and $\kappa_5$ is the five-dimensional Newton constant. The potential $V(\phi)$ encodes the details of the dual gauge theory. We choose a simple potential characterised by a single parameter, $\phiM$, which reads
\begin{equation}
\label{eq:potential}
L^2 V(\phi) = -3 -\frac{3}{2}\phi^2 - \frac{1}{3} \phi^4 + \left(\frac{1}{2\phi_{\rm M}^4} +  \frac{1}{3\phi_{\rm M}^2}\right) \phi^6- \frac{1}{12\phi_{\rm M}^4} \phi^8 \,,
\end{equation}
where $L$ is a length scale. Note that $V(\phi)$ is negative, possesses a maximum at $\phi=0$ and a minimum at $\phi=\phiM > 0$. A detailed study of this model's thermodynamics and near-equilibrium properties was presented in~\cite{Attems:2016tby}; here we will  briefly recall the most important points.

The motivation for choosing the potential \eqq{eq:potential} is that it  has three important properties. First, the resulting vacuum solution is asymptotically AdS$_5$ in the UV with radius $L$,  since $V(0)=-3/L^2$. Second, the second derivative of the potential at $\phi=0$  implies that the scalar field has mass $m^2=-3/ L^2$ therein.
%Following the standard quantization analysis,
This means that, in the UV, this field is dual to an operator in the gauge theory, $\mathcal{O}$, with dimension $\Delta_\textrm{UV}=3$. Third, the solution near $\phi = \phiM$ is again 
AdS$_5$ with a different radius
\begin{equation}
\label{eq:LIR}
\LIR= \sqrt{- \frac{3}{V\left(\phiM\right)}} = \frac{1}{1+ \frac{1}{6} \phiM^2} L \, .
\end{equation}
In this region the effective mass of the scalar field differs from its UV value and it is given by 
\begin{equation}
\label{eq:MIR}
\mIR= \frac{12}{ L ^2} \left(1+\frac{1}{9} \phiM^2 \right)= \frac{12}{\LIR^2} \frac{\left(1+\frac{1}{9} \phiM^2 \right)}{\left(1+\frac{1}{6} \phiM^2 \right)^2}\,. 
\end{equation}
As a consequence, the operator $\mathcal{O}$ at the IR fixed point has dimension 
\begin{equation}
\label{IRdim}
\Delta_\mtt{IR}=2 + 2\sqrt{ 1+ \frac{\mIR \LIR^2}{4}}=
6\, \left( 1+\frac{\phiM^2}{9}\right) \left(1+\frac{\phiM^2}{6} \right)^{-1}\,.
\end{equation}

To compute the vacuum state of these theories, one needs to first set an ansatz for the solution. In Fefferman-Graham (FG) coordinates, the solution with translation invariance and no horizon can be written in the following form, 
\begin{equation}
\label{eq:dsvac}
% ds^2 = \frac{d\uFG^2}{\uFG^2} + e^{2 \AFG(\uFG)} \left(-d x_{+} dx_{-} + d{\bf x}_\perp^2\right) \,,
ds^2 = \frac{L^2}{\uFG^2} d\uFG^2+ e^{2 \AFG(\uFG)} \eta_{\mu \nu} \, dx^\mu dx^\nu \,,
\end{equation}
with $\AFG(\uFG)$ and $\phi(\uFG)$ the non-trivial fields characterising the solution and $\uFG$ the holographic coordinate. %, and the coordinates $x_{\pm}$ correspond to $x_{\pm}=x_{L} \pm t$.
The computation of the vacuum state can be simplified when the potential is derived from a super-potential as
\begin{equation}
\label{eq:defW}
V(\phi)= -\frac{4}{3} W\left(\phi \right)^2 + \frac{1}{2} W'\left(\phi\right)^2 \,, 
\end{equation}
which for the potential selected \eqref{eq:potential} will be
\begin{equation}
\label{eq:defWs}
L\, W\left(\phi \right)=-\frac{3}{2} - \frac{\phi^2}{2} + \frac{\phi^4}{4 \phiM^2} \,.
\end{equation}
In this case, the scalar profile $\phi(\uFG)$ and the metric coefficient $\AFG(\uFG)$ can be obtained from the equations
\begin{equation}
\label{spotential}
\uFG\frac{d \, \AFG}{d\uFG}=\frac{2}{3}W, \quad
\uFG\frac{d \, \phi}{d\uFG}=-\frac{\partial W}{\partial \phi} \,, 
\end{equation}
and normalizability boundary conditions. Luckily enough, the equations have an analytic solution for the super-potential chosen,\footnote{Note that with respect to our conventions in  \cite{Attems:2016ugt} we have 
$L \, \phi_0^{\mtt{[here]}}= \phi_0^{\mtt{[there]}}$.} 
\begin{align}
\label{eq:metricsol}
e^{2 \AFG}&= \frac{\phi_0^2 L^2}{\phi^2} \,
  \left(1- \frac{\phi^2}{\phiM^2} \right)^{\frac{\phiM^2}{6}+1}e^{-\frac{\phi^2}{6}}  
  \,, \\[2mm]
\label{eq:phisol}
\phi&= \frac{\phi_0 \, \uFG}{\sqrt{1+ \frac{\phi_0^2}{\phiM^2} \uFG^2}}  \,,
\end{align}
where $\phi_0$ is an arbitrary  constant with dimensions of mass that controls the magnitude of the non-normalizable mode of the scalar field. As we will see below, 
$\phi_0$ is equal to the source of the dimension-three operator $\mathcal{O}$ in the dual gauge theory:
\be
\Lambda = \phi_0 \,.
\ee
The presence of this source breaks conformal invariance explicitly. Throughout  the paper we will use a redundant notation since we will use $\phi_0$ when we wish to emphasize the gravitational description and $\Lambda$ when we wish to emphasize the gauge theory scale. 

\subsection{\label{gtq}Gauge theory quantities}

Noticing that the small field behaviour of the superpotential~\eqref{eq:defWs} is identical to that of the GPPZ flow~\cite{Girardello:1998pd}, we can readily determine the expectation values  of the stress tensor and the scalar operator. We begin by  expanding the metric and the scalar field in powers of $\uFG$ in the $\uFG \rightarrow 0$  limit. Following \cite{Bianchi:2001kw}, we write the 5-dimensional metric for asymptotically AdS geometries in generic FG form
\begin{equation}
ds^2 = \frac{L^2}{\uFG^2} \left (d \uFG^2 + g_{\mu \nu} \, dx^\mu dx^\nu\right) \, ,
\end{equation}
and we write  the power expansions  of the metric and the scalar field as\footnote{Note that with respect to our conventions in  \cite{Attems:2016ugt} we have 
$\phi^{(2)}_{\mtt{[here]}}=\Lambda \phi^{(2)}_{\mtt{[there]}}$.} 
\begin{align}
\label{eq:pwg}
g_{\mu \nu} & = \eta_{\mu \nu} + g^{(2)}_{\mu \nu} \, \uFG^2 +  g^{(4)}_{\mu \nu} \, \uFG^4 + ... \, ,
\\[2mm] \label{eq:pwphi}
\phi & = \phi_0 \uFG + \phi^{(2)} \uFG^3 + \ldots  \,.
\end{align}
The expectation values of the field theory operators are then  given by
\begin{align}
\label{eq:Texp}
  \left< T_{\mu \nu} \right> & =
                               \frac{2 L^3}{\kappa_5^2} 
                               \left[ g^{(4)}_{\mu \nu}  + \left(\Lambda \,\phi^{(2)} -  \frac{\Lambda^4}{18} + \frac{\Lambda^4}{4\phiM^2}\right) \eta_{\mu \nu}\right] 
                               \, , \\[2mm]
  \label{eq:Oexp}
  \left< \mathcal{O} \right> & = - \frac{2 L^3}{\kappa_5^2}\,  \left(2  \phi^{(2)}  + \frac{\Lambda^3}{\phiM^2} \right)\,.
\end{align} 
As expected, equations~\eqref{eq:Texp} and \eqref{eq:Oexp} imply the Ward identity for the trace of the stress tensor  
\begin{equation}
\label{eq:TTrace0}
\left<T^{\mu}_\mu\right>= - \Lambda \left< \mathcal{O} \right> \,,
\end{equation}
and we adopt a renormalization scheme such that
$\left<T_{\mu\nu}\right>= \left< \mathcal{O} \right>=0$ in the vacuum.
Henceforth we will omit the expectation value signs and work with the rescaled
quantities
\begin{equation}
\label{eq:Tcomponents}
\Big( \mathcal{E}, J_{\mathcal{E}}, P_{x^i}, \mathcal{V} \Big)=
\frac{\kappa_5^2}{2 L^3} \, \Big(-T^t_t, T^z_t, T^{x^i}_{x^i}, \mathcal{O} \Big) \,.
\end{equation}
In these variables the Ward identity takes the form
\begin{equation}
\label{eq:Ward}
\mathcal{E} - 3 \bar P= \Lambda \mathcal{V} \,,
\end{equation}
where 
\begin{equation}
% \bar P = \frac{1}{3} \left( P_L + 2 P_T \right) 
\bar P = \frac{1}{3} \sum_i P_{x^i}
\end{equation}
is the average pressure.
Out of equilibrium the average pressure is not determined by the energy density because the scalar expectation value V fluctuates independently. 
In equilibrium, however, $\mathcal{V}$ is determined by the energy density and the Ward identity becomes the equation of state
\begin{equation}
\label{eoseos}
\bar P = P_{\rm eq} (\mathcal{E}) \,,
\end{equation}
with
\begin{equation}
  \label{eq:Peq}
P_{\rm eq} (\mathcal{E}) = \frac{1}{3} \Big[ \mathcal{E} -  \Lambda \mathcal{V}_{\rm eq} (\mathcal{E}) \Big]\,.
\end{equation}

\subsection{Thermodynamics and transport}
\label{sec:thermo}

To explore the thermal physics of our model, we search for static black brane solutions of
the action~\eqref{eq:action} following the approach of~\cite{Gubser:2008ny}.
Since for these solutions the scalar field is a
monotonic function of $\uFG$, we may use it as a coordinate when solving the dynamic equations. The value of $\phi$ at the black brane horizon, $\phiH$, univocally characterises the black brane solution. Therefore, by imposing the appropriate ``horizon'' boundary conditions at different $\phiH$ values one can compute all the equilibrium geometries. Finding the thermodynamics then amounts to finding a
family of black brane solutions parametrized by $\phiH$, and obtaining their Hawking temperatures $T$ and
entropy densities $s$. This construction is done is detail
in~\cite{Attems:2016ugt}, to where we refer the interested reader. 

\begin{figure}[tbph]
  \includegraphics[width=.48\textwidth]{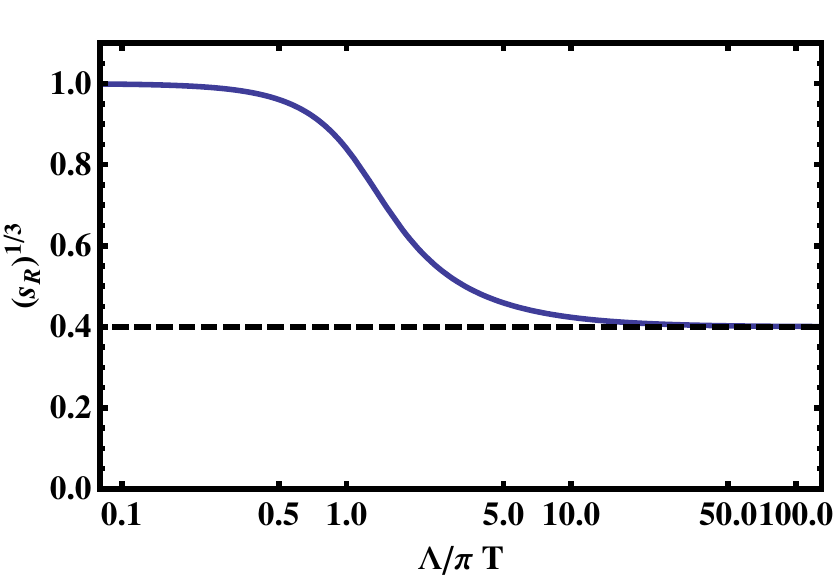}
  \hfill
  \includegraphics[width=.48\textwidth]{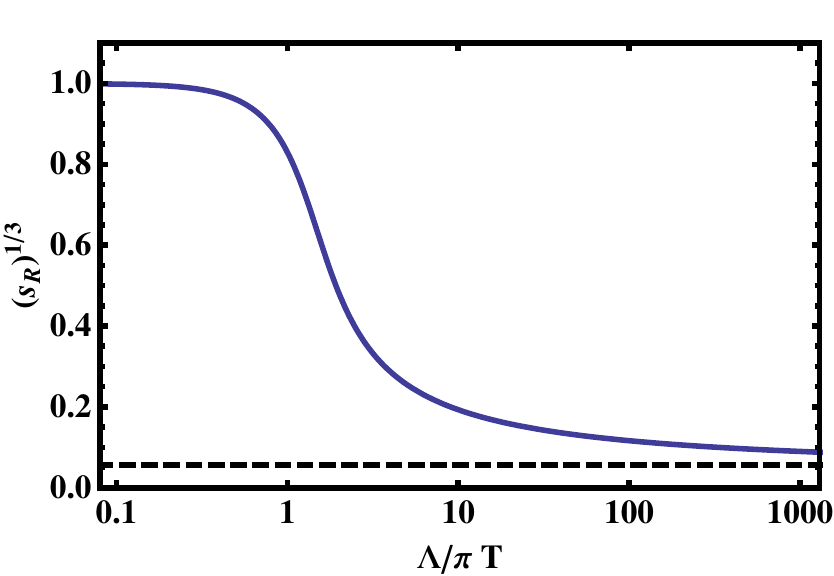}
  \caption{Ratio of entropy density to temperature for $\phiM=3$ (left) and
    $\phiM=10$ (right) as a function of the inverse temperature. The dashed line
    shows $\LIR/L$. \label{plot:sofT} }
\end{figure}

For our purposes here, it is enough to note that we find a set of values $(\phiH, T, s)$
for each model, i.e.~for each $\phiM$. With these, one can compute all
thermodynamic quantities of interest as well as the bulk viscosity $\zeta$.
In figure~\ref{plot:sofT} we plot the dimensionless quantity 
\begin{equation}
s_R = \frac{\kappa_5^2}{2 \pi^4 L^3} \frac{s}{T^3}\, , 
\end{equation}
as a function of the inverse temperature for two different values of
$\phiM$. Since the theory is conformal both at the UV and at the IR, the high
and low temperature behaviour of the entropy density must coincide with that of
a relativistic conformal theory and scale as $T^3$. In the intermediate region,
this scaling is not fulfilled and therefore we can interpret this quantity as a
measure of the non-conformality of the gauge theory. 

For a relativistic CFT, $s / T^3$ is proportional to the number of degrees of freedom
in the theory, which for an $SU(N)$ gauge theory with matter in the adjoint
representation scales as $N^2$. For example, for $\mathcal{N}=4$ SYM
\begin{equation}
\frac{s}{T^3}= \frac{\pi^2}{2}N^2,
\end{equation}
but the precise coefficient depends on the specific theory. In terms of the
parameters of the dual gravity description this quantity becomes
\begin{equation}
\frac{s}{T^3}=\frac{2\pi^4 L^3}{\kappa_5^2}.
\end{equation}
In our bottom-up setup, the above argument allows us to {\em define} the number
of degrees of freedom at the fixed points  in terms of the
effective AdS radius. In particular, the quantity $s_R$ should approach 1 at
high temperature and $(\LIR/L)^3$ at low temperature, which is confirmed by
the plots in figure~\ref{plot:sofT}.

\begin{figure}[tbhp]
\begin{center}
\includegraphics[width=.65\textwidth]{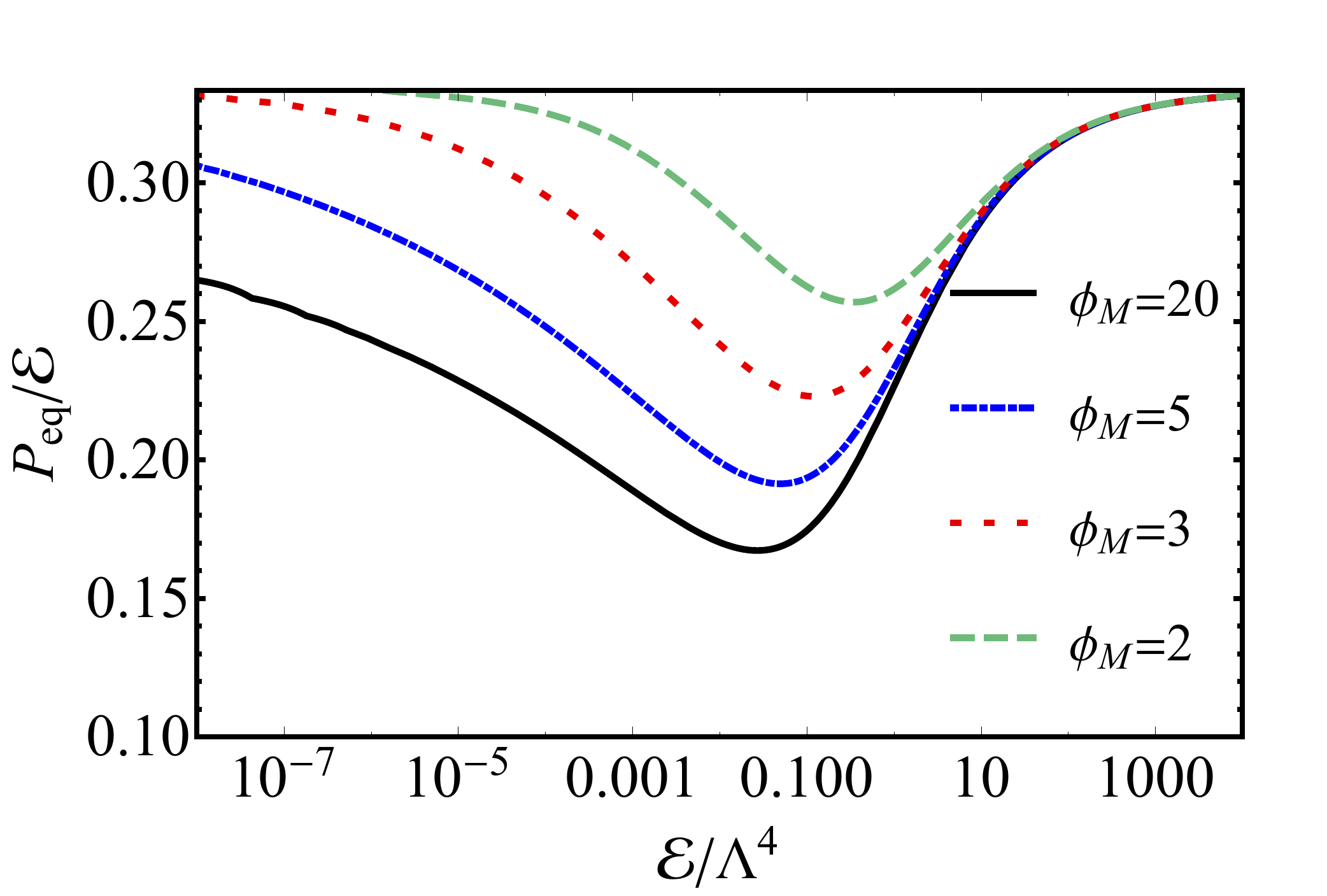}
\caption{Equilibrium pressure as a function of energy density for
$\phiM=\{2\,,3\,,5\,,20\}$. \label{fig:eos} }
\end{center}
\end{figure}

Another quantity that one can compute from $T$ and $s$ is $P_{\rm eq}(\mathcal{E})$,
introduced in~\eqref{eq:Peq}, also known as the equation of state. This quantity
gives another measure of the degree of non-conformality of the gauge theory, and
will also be necessary later on for the hydrodynamic estimations. For the
representative cases of $\phiM=2\,,3\,,5\,,20$, this quantity can be seen in
figure~\ref{fig:eos}. As expected, both at high and low energies the physics
becomes approximately conformal and $P_{\rm eq}$ asymptotes to $\mathcal{E}/3$.

\begin{figure}[htbp]
\centering
\includegraphics[width=0.65\textwidth]{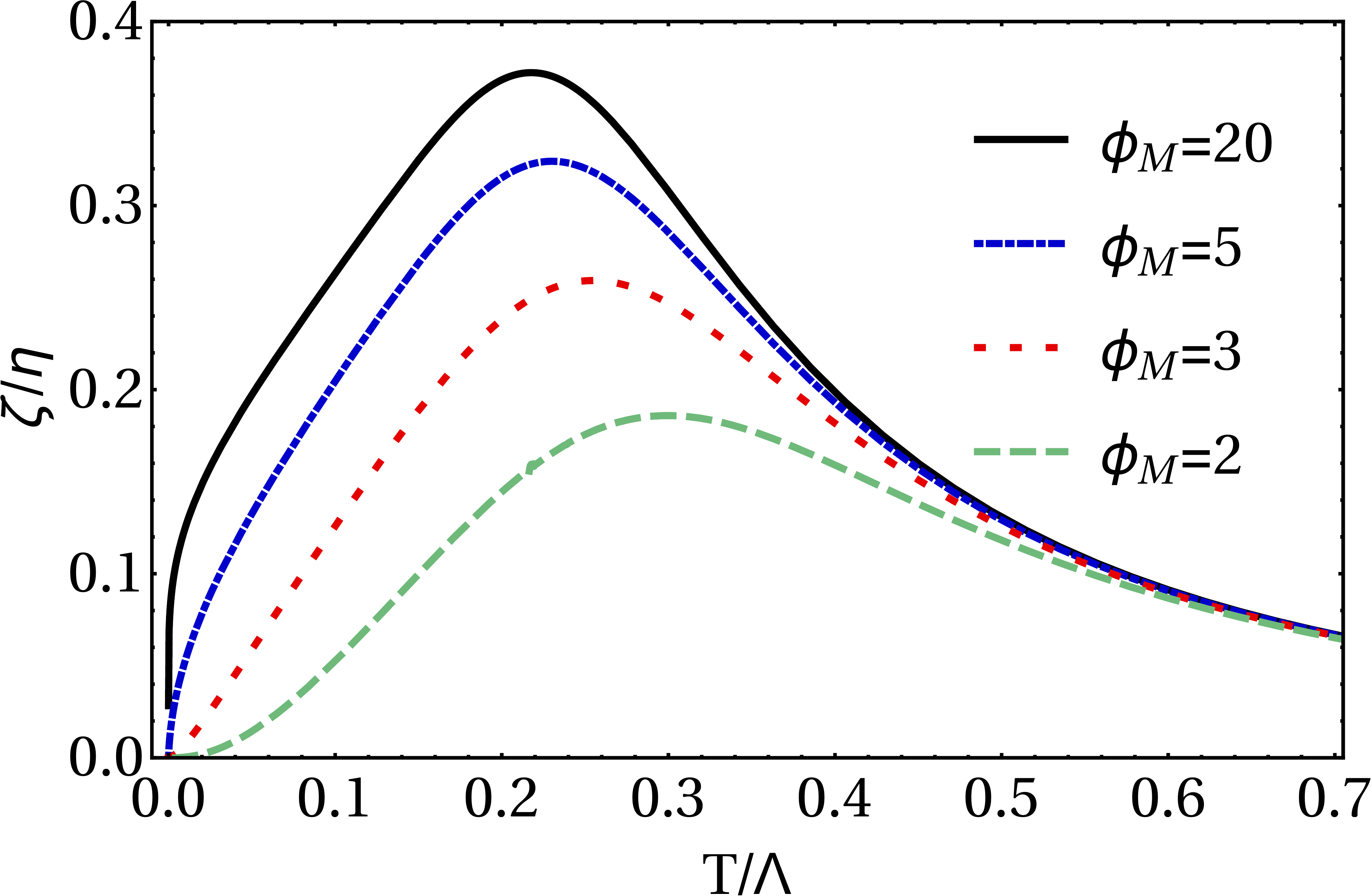}
\caption{Bulk viscosity $\zeta$ over shear viscosity $\eta$ as a function of temperature for \mbox{$\phi_{\rm M} = \{20, 5, 3, 2\}$}.
For each $\phiM$ we obtain
$ \max(\zeta/\eta) =  \{0.19,0.26,0.32,0.37\}$ at the respective temperatures
\mbox{$T/\Lambda = \{0.218, 0.220, 0.230, 0.299 \}$}.
\label{fig:zetaosT} }
\end{figure}

The transport properties of the dual gauge theory plasma also reflect the
non-conformal behaviour observed in the equation of state.
Due to the isotropy of the plasma, at leading order in gradients transport phenomena are controlled by only two coefficients: the shear viscosity $\eta$ and the bulk viscosity $\zeta$.
Because of the universality of the shear viscosity to entropy
ratio~\cite{Kovtun:2004de} in all theories with a two-derivative gravity dual,
we are ensured that this ratio in our model takes the same value as in the conformal
$\mathcal{N}=4$ theory, i.e.~$\eta/s=1/4\pi$.  On the other hand, the bulk viscosity 
(which would vanish identically in a CFT) is non-zero in our
model. Following~\cite{Eling:2011ms} we determine the bulk viscosity by studying
the dependence of the entropy on the value of the scalar field at the horizon
\begin{equation}
\frac{\zeta}{\eta}=4 \left(\frac{d \log s}{d\phiH } \right)^{-2} \, .
\end{equation}
The temperature dependence of this ratio is shown in figure~\ref{fig:zetaosT}
for different values of $\phiM$.

\subsection{Shockwave metric}
\label{sec:shockwave}
In the Fefferman-Graham frame it is possible to find a quasi-analytic solution for a single travelling shockwave on a vacuum background. The metric form will simply correspond to the vacuum metric \eqref{eq:dsvac} plus the addition of the term $f(\uFG)h(x_{\pm})dx_{\pm}^2$:
\begin{equation}
\label{metricsw}
ds^{2}= \frac{L^2}{\uFG^2}\, d\uFG^2 + f(\uFG)h(x_{\pm})dx_{\pm}^2+ e^{2 \AFG(\uFG)} \left(-d x_{+} dx_{-} + d{\bf x}_\perp^2\right) \text{,}
\end{equation}
where $x_{\pm}=z\pm t$, $z$ is the direction of propagation of the shockwave, and ${\bf x}_\perp$ are the perpendicular directions to it. The function $h(x_{\pm})$ is an arbitrary function for the waveform. 

The propagation of the shockwave at the speed of light does not alter the vacuum profiles of $\AFG$ and $\phi$, thus the only remaining function to be determined is be $f$. The equation for $f(\uFG)$ is a second-order differential equation coming from the Einstein's equations whose solution can only be obtained numerically:
\begin{equation}
\label{eqforf}
-f \left[2 \left(\uFG^2 \frac{\partial ^2\AFG}{\partial \uFG^2}+\uFG \frac{\partial \AFG}{\partial \uFG}\right)+4 \left(-\uFG \frac{\partial \AFG}{\partial \uFG}\right)^2\right]+\uFG^2 \frac{\partial ^2f}{\partial \uFG^2}+\uFG \frac{\partial f}{\partial \uFG}=0.
\end{equation} 
From the differential equation one can derive the equivalent integral expression
\begin{equation}
\label{intforf}
f(\uFG) = 4 \, e^{2\AFG(\uFG)} \,  \int_0^{\uFG} \frac{d\tilde u}{\tilde{u}} \, e^{-4\AFG(\tilde u)}.
\end{equation}
An additional difficulty for the computation of the function $f(\uFG)$ is that it grows exponentially with $\uFG$. However, inspection of \eqref{intforf} shows that this can be circumvented  by computing the redefined function 
\be
g(\uFG)=e^{2\AFG(\uFG)} f(\uFG) \,,
\ee
which takes values between 0 and 1.

Solving equation~(\ref{intforf}) order by order, we see that $f(\uFG)$ behaves as
\begin{equation}
\label{eq:fser}
 f(\uFG) = \uFG^2  + \frac{\uFG^4 \phi_0^2}{9} + O\left(\uFG^6\right) \,.
\end{equation}
With this expression, the metric~(\ref{metricsw}), and the vacuum profile of the scalar field~(\ref{eq:phisol}), one obtains from~(\ref{eq:Texp})-(\ref{eq:Oexp}) the dual gauge theory quantities of such a shockwave, namely
\begin{equation}
\label{eq:stresstensor-sw}
\mathcal{E} = P_L = \pm J_{\mathcal{E}} =  h(x_{\pm}) \,,
\qquad P_T = 0 \,,
\qquad \mathcal{V} = 0 \,,
\end{equation}
where $P_L$ is the longitudinal pressure (along the $z$~direction), and $P_T$ the transverse pressure (along the transverse directions ${\bf x}_\perp$).

%%%%%%%%%%%%%%%%%%%%%%%%%%%%%%%%%%%%%%%%%%%%%%%%%%%%%%%%%%%%%%%%%%%%%%%%%%%%%%%
\section{Numerical procedure}
\label{sec:numerical}
%%%%%%%%%%%%%%%%%%%%%%%%%%%%%%%%%%%%%%%%%%%%%%%%%%%%%%%%%%%%%%%%%%%%%%%%%%%%%%%

In this section we set $L=1$ for notational simplicity.

\subsection{Evolution equations}
\label{sec:evol}

We follow the notation of~\cite{Chesler:2010bi} and begin by writing the following 5D metric \emph{ansatz} in Eddington-Finkelstein (EF) coordinates
\begin{equation}
\label{eq:metric00}
ds^2 = -Adt^2 + \Sigma^2 \left( e^B d\bm{x}_{\perp}^2 + e^{-2B} dz^2 \right)
+ 2dt (dr + Fdz ) \,,
\end{equation}
where $A$, $B$, $\Sigma$, and $F$ are functions of the radial coordinate $r$, time $t$ and $z$. The shocks will be propagating along $z$, and $\bm{x}_{\perp}$ denotes the two perpendicular directions $\bm{x}_{\perp} = x_1, x_2$.
Note that $t$ is a null time coordinate (usually called $v$ in EF coordinates), i.e.~constant-$t$ surfaces are not spacelike but null.

Written in this form, the metric is invariant under the following transformation
\begin{equation}
\label{eq:diffeo}
\begin{aligned}
r      & \to \bar{r} = r + \xi(t,z) \,, \\
\Sigma & \to \bar{\Sigma} = \Sigma \,,\\
B      & \to \bar{B} = B \,,\\
A      & \to \bar{A} = A + 2\partial_t \xi(t,z) \,,\\
F      & \to \bar{F} = F - \partial_z \xi(t,z)\,.
\end{aligned} 
\end{equation}

Upon plugging the metric~(\ref{eq:metric00}) in~(\ref{eq:EE-phi}) the resulting system conveniently obeys a particular nested structure, consisting  of a sequence of radial ODEs at each $t=\text{const}$ null slice that can be solved in order, see e.g.~\cite{Winicour:2005ge} and references therein.

The equations of motion for our present case are given by
\begingroup
\allowdisplaybreaks
\begin{subequations}
\label{eq:fullequations}
\begin{align}
 \Sigma'' & = -\frac{1}{6} \Sigma \left(3 \left(B'\right)^2+4 \left(\phi
   '\right)^2\right) \,, \\
 \Sigma^2 F'' & = \Sigma \left(6 \tilde{\Sigma} B'+4 \tilde{\Sigma}'+3 F' \Sigma'\right)+\Sigma^2
   \left(3 \tilde{B} B'+2 \tilde{B}'+4 \tilde{\phi } \phi '\right)-4
   \tilde{\Sigma} \Sigma'  \,, \\
 12 \Sigma^3 \dot{\Sigma}' & = e^{2 B} \Big[\Sigma^2 \left(4 \tilde{B} F'-4
   \left(\tilde{\tilde{B}}+\tilde{\phi }^2\right)-7 \tilde{B}^2+2
   \tilde{F}'+\left(F'\right)^2\right) \notag \\
& \qquad \qquad +2 \Sigma \left(\tilde{\Sigma} \left(F'-8
   \tilde{B}\right)-4 \tilde{\tilde{\Sigma}}\right)+4 \tilde{\Sigma}^2\Big]-8 \Sigma^2
   \left(\Sigma^2 V(\phi )+3 \dot{\Sigma} \Sigma'\right)  \,, \\
 6 \Sigma^4 \dot{B}' & = e^{2 B} \Big[\Sigma^2 \left(-\tilde{B}
   F'+\tilde{B}^2+\tilde{\tilde{B}}-2 \tilde{F}'+4 \tilde{\phi
   }^2-\left(F'\right)^2\right) \notag \\
  & \qquad \qquad +\Sigma \left(\tilde{\Sigma} \left(\tilde{B}+4
   F'\right)+2 \tilde{\tilde{\Sigma}}\right)-4 \tilde{\Sigma}^2\Big]-9 \Sigma^3
   \left(\dot{\Sigma} B'+\dot{B} \Sigma'\right)  \,, \\
  2 \Sigma^3 \dot{\phi }'& = - 3 \Sigma^2 \left( \Sigma' \dot{\phi }
                                             + \dot{\Sigma} \phi' \right)
                           - e^{2 B} \Sigma \left(2 \tilde{B} \tilde{\phi}-\tilde{\phi } F'
                                                  +\tilde{\tilde{\phi }}\right)
                           -e^{2 B} \tilde{\Sigma} \tilde{\phi }
                           +\Sigma^3 V'(\phi )
                            \,, \\
 6 \Sigma^4 A'' & = 3 e^{2 B} \left(\Sigma^2 \left(4
   \left(\tilde{\tilde{B}}+\tilde{\phi }^2\right)+7
   \tilde{B}^2-\left(F'\right)^2\right)+8 \Sigma \left(2 \tilde{B}
   \tilde{\Sigma}+\tilde{\tilde{\Sigma}}\right)-4 \tilde{\Sigma}^2\right)
\notag \\
& +2 \Sigma^4 \left(-9
   \dot{B} B'+4 V(\phi )-12 \dot{\phi } \phi '\right)+72 \dot{\Sigma} \Sigma^2
   \Sigma'  \,, \\
 2 \Sigma^2 \dot{F}' & = -\Sigma^2 \left(2 B' \left(\tilde{A}+2
   \dot{F}\right)+2 \tilde{A}'+6 \dot{B} \tilde{B}+4 \tilde{\dot{B}}+8
   \dot{\phi } \tilde{\phi }+A' F'\right)
\notag \\
& +2 \Sigma \left(\Sigma'
   \left(\tilde{A}+2 \dot{F}\right)-6 \dot{B} \tilde{\Sigma}-4 \tilde{\dot{\Sigma}}-3
   \dot{\Sigma} F'\right)+8 \dot{\Sigma} \tilde{\Sigma}  \,, \\
 6 \Sigma^2 \ddot{\Sigma} & = e^{2 B} \left(\Sigma \left(2 \tilde{B} \left(\tilde{A}+2
   \dot{F}\right)+\tilde{\tilde{A}}+2 \tilde{\dot{F}}\right)+\tilde{\Sigma}
   \left(\tilde{A}+2 \dot{F}\right)\right) \notag \\
&{} +\Sigma^2 \left(3 \dot{\Sigma} A'-\Sigma \left(3
   \dot{B}^2+4 \dot{\phi }^2\right)\right)  \,,
\end{align}
\end{subequations}%
\endgroup
where, for any function $g$, we define 
\bea
\tilde{g} &\equiv&  \left( \partial_z  - F \partial_r  \right) g \,, \\[2mm]
g' &\equiv& \partial_r g \,, \\[1mm]
  \label{eq:dotf}
 d_{+} g &\equiv& \dot{g} \equiv \left( \partial_t + \frac{A}{2}  \partial_{r} \right) g \,.
\eea
Note that these equations are all of the general form
\begin{equation}
  \label{eq:gen-exp-final}
\left[
  \alpha_g(r,t,z) \partial_{rr} + \beta_g(r,t,z) \partial_r + \gamma_g(r,t,z)
\right] g(r,t,z) = -S_g(r,t,z) \,,
%
% \Big[
% \alpha_g \partial_{rr} + \beta_g \partial_r + \gamma_g
% \Big] g = -S_g \,,
\end{equation}
where $g = \Sigma,~F,~d_+ \Sigma,~d_+ B,~d_+ \phi,~A,~d_+ F$.
These are solved imposing reflecting boundary conditions at the AdS boundary $u
= 1/r = 0$, which take the form
\begingroup
\allowdisplaybreaks
\begin{subequations}
\label{eq:near-bdry}
\begin{align}
A(u,t,z) & = \frac{1}{u^2} + \frac{2 \xi (t,z)}{u} - 2 \partial_t\xi(t,z) +\xi (t,z)^2
           -\frac{2 \phi_ {0}^2}{3}
           + u^2 a_{4}(t,z) \notag \\
&{}-\frac{2}{3} u^3 (\phi_ {0} \partial_t\phi_ {2}(t,z) + 3 a_{4}(t,z) \xi (t,z) + \partial_zf_{2}(t,z) ) + O(u^4) \,, \\[2mm]
B(u,t,z) & = u^4 b_{4}(t,z) + O(u^5)
%           + u^5 \left(-4 b_{4}(t,z) \xi (t,z)
%           + \partial_t b_{4}(t,z)+\frac{2}{15} \partial_{z} f_{2}(t,z) \right) \\
 \label{eq:near-bdry-B} \\[2mm]
\Sigma(u,t,z) & = \frac{1}{u} +\xi (t,z)  -\frac{\phi_ {0}^2 u}{3}
           + \frac{1}{3} \phi_ {0}^2 u^2 \xi (t,z) \notag \\
         &{} + \frac{1}{54} \phi_ {0} u^3 \left(
             -18 \phi_ {0} \xi (t,z)^2-18 \phi_ {2}(t,z)+\phi_ {0}^3
           \right) + O(u^4) \,,\\[2mm]
F(u,t,z) &   = \partial_z\xi(t,z) + u^2 f_{2}(t,z) \notag \\
         &{} + u^3 \left( \frac{4}{15} \left(
               \phi_ {0} \partial_z\phi_ {2}{}(t,z) - 6 \partial_zb_{4}{}(t,z)
                                     \right)
                      -2 f_{2}(t,z) \xi (t,z)
                   \right) + O(u^4)\,,
\\[2mm]
\phi(u,t,z) &   = \phi_ {0} u - \phi_ {0} u^2 \xi (t,z)
                + u^3 \left(\phi_ {0} \xi (t,z)^2+\phi_ {2}(t,z)\right)
                % + O(u^4)
\notag \\
             &{} + u^4 \left(-\phi_ {0} \xi (t,z)^3-3 \xi (t,z) \phi_ {2}(t,z)
                          + \partial_t\phi_ {2}(t,z)\right)
                 + O(u^5) \label{eq:near-bdry-phi}  \,,\\[2mm]
d_{+} B(u,t,z) & = -2 u^3 b_{4}(t,z) + O(u^4)
                  % + u^4 \left(6 b_{4}(t,z) \xi (t,z)-\frac{3}{2} b_{4}{}^{(1,0)}(t,z)
                  % -\frac{1}{3} f_{2}{}^{(0,1)}(t,z)\right)
\,,\\[2mm]
d_{+} \Sigma(u,t,z) & = \frac{1}{2 u^2} + \frac{\xi (t,z)}{u}
                + \frac{1}{2} \xi (t,z)^2-\frac{\phi_ {0}^2}{6} \notag \\
               &{} + \frac{1}{36} u^2 \left(18 a_{4}(t,z)+18 \phi_ {0} \phi_ {2}(t,z)-5 \phi_ {0}^4\right)
                + O(u^3) \,, \\[2mm]
d_{+} \phi(u,t,z) & = -\frac{\phi_ {0}}{2}
                   + u^2 \left(\frac{\phi_ {0}^3}{3}-\frac{3}{2} \phi_ {2}(t,z)\right)
                   + O(u^3)\,, \\[2mm]
                 % &{} + u^3 \left(\frac{1}{3} (-2) \phi_ {0}^3 \xi (t,z)
                 %     + 3 \xi (t,z) \phi_ {2}(t,z)-\phi_ {2}{}^{(1,0)}(t,z)\right)\\
%
d_{+} F(u,t,z) & = \partial_{tz}\xi(t,z) - u f_{2}(t,z)
                + O(u^2)\,.
  %&{} + u^2 \left(\frac{12}{5} b_{4}{}^{(0,1)}(t,z)-\frac{2}{5} \phi_ {0} \phi_ {2}{}^{(0,1)}(t,z)+f_{2}(t,z) \xi (t,z)+f_{2}{}^{(1,0)}(t,z)\right)\\
%
\end{align}
\end{subequations}%
\endgroup
The subleading coefficient of the scalar field in EF coordinates $\phi_2$, introduced in equation~\eqq{eq:near-bdry-phi}, is related to its FG counterpart, $\phi^{(2)}$, through
\be
\phi^{(2)} = \phi_2 -\frac{1}{6} \phi_0^3 \,.
\ee
The function $\xi(t,z)$ encodes our residual gauge freedom, whereas the functions $a_4(t,z)$ and $f_2(t,z)$ are constrained to obey
\begin{subequations}
\label{eq:boundary-evol}
\begin{align}
\partial_ta_{4} & = -\frac{4}{3} \left(
                           \partial_zf_{2}
                         + \phi_{0} \partial_t\phi_{2}
                       \right) \,, \\[2mm]
\partial_tf_{2} & = \frac{1}{4} \left(
                          -\partial_za_{4}
                          -8 \partial_zb_{4}
                          +\frac{4}{3} \phi_0 \partial_z\phi_2
                        \right) \,,
\end{align}
\end{subequations}
with $b_4$ read off from $B$ through~\eqref{eq:near-bdry-B} and both $\phi_2$ and
$\partial_t \phi_2$ read off from $\phi$ through~\eqref{eq:near-bdry-phi}.

To solve the resulting system we follow the general approach
of~\cite{Chesler:2010bi,Chesler:2013lia}, with some important differences that
we will outline below.

\subsection{Expectation values from evolution variables}

With the near-boundary behaviours above, together with the Fefferman-Graham
expansions~\eqref{eq:pwg} and~\eqref{eq:pwphi}, one finds the coordinate
transformation relating the fall-off coefficients in each frame. With these, and
the expectation values~(\ref{eq:Texp}) and~(\ref{eq:Oexp}), one can write the
expressions for the gauge theory values in terms of our evolution variables
$(b_4, a_4, f_2, \phi_2)$ as
\begin{align}
\mathcal{E} & = -\left(\frac{3}{4} a_4 + \phi_0\phi_2 + \frac{9-7 \phi_{\rm M}^2}{36 \phi_{\rm M}^2} \phi_0^4\right) \,,
\label{eq:energy} \\[2mm]
P_L & =  -\frac{a_4}{4}-2 b_4+\frac{\phi_0\phi_2}{3} +\left(-\frac{5}{108} +\frac{1}{4 \phi_{\rm M}^2}\right)\phi_0^4 \,, 
\label{eq:PL}
\\[2mm]
P_T & = -\frac{a_4}{4}+ b_4+\frac{\phi_0\phi_2}{3} +\left(-\frac{5}{108} +\frac{1}{4 \phi_{\rm M}^2}\right)\phi_0^4 \,, 
\label{eq:PT}
\\[2mm]
J_\mathcal{E} & = f_{2} \,, 
\label{eq:J}
\\[2mm]
\mathcal{V} & = -2\phi_2 + \frac{\phi_0^3}{3} - \frac{\phi_0^3}{\phiM^2} \,,
\end{align}
where $P_L$ and $P_T$ are the longitudinal and transverse pressures.

\subsection{Gauge fixing}
\label{sec:gauge}

We start with the procedure to fix the residual gauge freedom~(\ref{eq:diffeo}).
A convenient choice is treating $\xi(t,z)$ as another evolved variable and choosing its evolution equation by requiring that the position of the apparent horizon  lie at some constant radial coordinate $r = r_h$.
We thus want to impose
\begin{equation}
\label{eq:horiz-condition}
  \Theta |_{r=r_h} = 0 \,, \qquad
  \partial_t \Theta |_{r=r_h} = 0 \,,
\end{equation}
at all times, where $\Theta$ is the expansion of outgoing null geodesics for the metric~(\ref{eq:metric00}). At surfaces $r=\text{const}$, $\Theta$ is given by
\begin{align}
\Theta = -\frac{1}{2} e^{2 B} F \left(3 F \partial_r\Sigma-2 \partial_{z}\Sigma\right)+e^{2 B} \Sigma \left(2 F
   \partial_zB+ \partial_zF\right)-3 \Sigma^2 d_+ \Sigma \,.
\end{align}

A simple way to impose the conditions~\eqref{eq:horiz-condition} numerically is the following
\begin{equation}
\label{eq:horiz-cond-num}
\left( \partial_t \Theta + \kappa  \Theta \right)|_{r=r_h} = 0 \,,
\end{equation}
where $\kappa$ is a positive parameter typically chosen to be 1. The advantage of imposing such a condition is that it is constructed to drive the $\Theta = 0$ surface back to $r=r_h$ whenever numerical errors accumulate. This turns out to work very well in practice.

Equation~\eqref{eq:horiz-cond-num}, when expanded, gives us an equation for $\partial_t \xi$ of the form
\begin{equation}
  \label{eq:dxidt-final}
\Big[ \alpha_\xi(t,z) \partial_{zz} +
  \beta_\xi(t,z) \partial_z + \gamma_\xi(t,z) \Big] \partial_t \xi(t,z) =
-S_{\xi}(t,z) \,,
\end{equation}
to be evaluated at $r=r_h$. This is a second-order, linear ODE in the coordinate $z$, which we solve imposing periodicity in $z$.

\subsection{Field redefinitions and evolution algorithm}
\label{sec:finite-parts}

To integrate the resulting system subject to the boundary conditions~(\ref{eq:near-bdry}), it is very convenient to introduce $u = 1/r$ as our radial coordinate and redefine the evolved variables so that the divergent pieces at $u=0$ are absent.

Motivated by~(\ref{eq:near-bdry}), we make the following definitions
\begingroup
\allowdisplaybreaks
\begin{subequations}
\label{eq:finite-parts}
\begin{align}
B(u,t,z) & \equiv u^4 B_{g_1}(u,t,z) \\
         & \equiv B_{g_2}(u,t,z) \,,\\[3mm]
\Sigma(u,t,z) & \equiv \frac{1}{u} + \xi(t,z) - u \frac{\phi_ {0}^{2}}{3}
            + u^{2} \frac{\phi_ {0}^{2}}{3} \xi(t,z) + u^3 \Sigma_{g_1}(u,t,z)
\\
              & \equiv \frac{1}{u} + \xi(t,z) + \Sigma_{g_2}(u,t,z)
\,,\\[3mm]
F(u,t,z) & \equiv \partial_z\xi(t,z) + u^2 F_{g_1}(u,t,z)\\
         & \equiv \partial_z\xi(t,z) + F_{g_2}(u,t,z) \,,\\[3mm]
A(u,t,z) & \equiv  \frac{1}{u^2} + \frac{2 \xi (t,z)}{u}
           -2 \partial_t\xi(t,z)+\xi (t,z)^2-\frac{2 \phi_ {0}^{2}}{3}
           + u^2 A_{g_1}(u,t,z) \label{eq:finite-parts-Ag1}
\\
 & \equiv  \frac{1}{u^2} + \frac{2 \xi (t,z)}{u}
           -2 \partial_t\xi(t,z)+\xi (t,z)^2-\frac{2 \phi_ {0}^{2}}{3}
           + A_{g_2}(u,t,z)  \label{eq:finite-parts-Ag2}
\,,\\[3mm]
\phi(u,t,z) & \equiv  u \phi_{0} -  u^{2} \phi_{0} \xi(t,z) + u^{3} \phi_ {0}^{3} \phi_{g_1}(u,t,z)\\
& \equiv  \phi_{0} \phi_{g_2}(u,t,z) \,,\\[3mm]
d_+ \Sigma(u,t,z) & \equiv \frac{1}{2u^2} + \frac{\xi (t,z)}{u}
              + \frac{\xi (t,z)^2}{2}
              - \frac{\phi_ {0}^{2}}{6}
+ u^2 \dot \Sigma_{g_1}(u,t,z)\\
 & \equiv \frac{1}{2u^2} + \frac{\xi (t,z)}{u}
              + \frac{\xi (t,z)^2}{2}
              - \frac{\phi_ {0}^{2}}{6}
+ \dot \Sigma_{g_2}(u,t,z)\,,\\[3mm]
d_+B(u,t,z) & \equiv u^3 \dot B_{g_1}(u,t,z)\\
            & \equiv \dot B_{g_2}(u,t,z)\,,\\[3mm]
d_+\phi(u,t,z) & \equiv  -\frac{\phi_ {0}}{2} + u^{2} \phi_ {0}^{3} \dot \phi_{g_1}(u,t,z)\\
               & \equiv  -\frac{\phi_ {0}}{2} + \dot \phi_{g_2}(u,t,z)\,,\\[3mm]
d_+ F(u,t,z) &  \equiv \partial_{tz}\xi(t,z) + u \dot F_{g_1}(u,t,z) \\
             &  \equiv \partial_{tz}\xi(t,z) + \dot F_{g_2}(u,t,z)\,.
\end{align}
\end{subequations}%
\endgroup
Our equations are then rewritten in terms of the ``$g_1$'' and ``$g_2$'' variables above. $g_1$ variables are adapted to the AdS boundary $u=0$. The corresponding resulting equations, however, are extremely long and carry terms with huge powers of the coordinate $u$. Upon trying to solve this system in the whole grid, we were finding that numerical errors would accumulate very early on in the evolution, quickly spoiling the convergence of the solution. We then decided to make use of the system $g_1$ only in the vicinity of $u \sim 0$ (grid1, spanning $u \in [0, u_0]$)---where a much simpler series expanded version of the aforementioned equations was used---and another grid (grid2, spanning $u \in [u_0, u_h]$) was introduced where the much simpler system of equations $g_2$ was used instead.

Our numerical grid thus consists of a double grid in the $u$ direction $u \in [0, u_0] \cup [u_0, u_h]$, where $u_0$ is typically chosen to be 0.1, and $u_h = 1/r_h$ is typically chosen to be 2 or 3. We integrate the $g_1$ equations with boundary conditions given by~(\ref{eq:near-bdry}) in grid1; we then read off the integrated values at $u=u_0$ and use these as boundary conditions for integrating the $g_2$ equations in grid2. Note, however, that we also need to deal with the junction point $u_0$ in our $u$-dependent hyperbolic equations $\partial_t B(u,t,z)$ and $\partial_t \phi(u,t,z)$, given by equation~(\ref{eq:dotf}). We explain this procedure in appendix~\ref{sec:match}.

We are now in possession of all the necessary equations for the evolution procedure. The evolution algorithm is then as follows:
\begin{enumerate}
\item at any given time $t_n$ (which can be the initial time after having
  performed the transformation~(\ref{eq:diffeo}) that puts the apparent horizon
  at constant $u$) we know $B(u,t_n,z)$, $\phi(u,t_n,z)$, $\xi(t_n,z)$, $a_4(t_n,z)$ and
  $f_2(t_n,z)$; \label{enum:first}
\item successively solve the elliptic equations~(\ref{eq:fullequations}) (or
  rather, the corresponding system obtained in terms of the redefined ``$g_1$''
  and ``$g_2$'' functions) in the order $\Sigma_{g_{1,2}}$, $F_{g_{1,2}}$,
  $\dot \Sigma_{g_{1,2}}$, $\dot B_{g_{1,2}}$, $\dot \phi_{g_{1,2}}$,
  $A_{g_{1,2}}$, which are a sequence of radial ODEs subjected to the boundary
  conditions~(\ref{eq:near-bdry}); \label{enum:second}
\item equation~(\ref{eq:dxidt-final}) is solved to get $\partial_t \xi(t_{n},z)$
  and afterwards $\partial_tB_{g_{1,2}}(t_n,u,z)$ and
  $\partial_t \phi_{g_{1,2}}(t_n,u,z)$ can be obtained through
  equation~(\ref{eq:dotf}) with~\eqref{eq:finite-parts-Ag1}
  and~\eqref{eq:finite-parts-Ag2} (see also appendix~\ref{sec:match});
\item obtain $\partial_t a_4(t_n,z)$ and $\partial_t f_2(t_n,z)$
  through~\eqref{eq:boundary-evol} and, together with the already obtained
  $\partial_t \xi(t_{n},z)$, $\partial_tB_{g_{1,2}}(u, t_n,z)$,
  $\partial_t\phi_{g_{1,2}}(u, t_n, z)$, advance all these quantities to time
  $t_{n+1}$ with a Runge-Kutta procedure or equivalent.
\item GOTO~\ref{enum:first}.
\end{enumerate}

\subsection{Discretization}
\label{sec:discretize}

Equations~(\ref{eq:fullequations}) are written in a form that decouples the coordinates $u$ and $z$ (the collision axis) and can therefore be solved as ODEs in the $u$~direction for each point in $z$. For this reason, both coordinates can be treated separately. The $z$ direction is discretized on a uniform grid where periodic boundary conditions are imposed, while along the $u$~direction we make use of two grids, grid1 spanning $[0, u_0]$ and grid2 spanning $[u_0, u_h]$. Both $u$ grids are \textit{Lobatto-Chebyshev} grids with $N_u+1$ points. The collocation points, given by
\begin{equation}
 X_i =  -\cos\left(\frac{\pi\,i}{N_u}\right) \qquad (i=0,1,\ldots,N_u)\,,  \end{equation}
are defined in the range $[-1:+1]$, and can be mapped to our \textit{physical} grid by
\begin{equation}
 u_i = \frac{u_R + u_L}{2} + \frac{u_R - u_L}{2} X^{}_i \qquad (i=0,1,\ldots,N_u)\,,  \end{equation}
where $u_L$ and $u_R$ are the limits of each of the grids.

As the differential equations are solved in $u$ for each $z$ point, the only important operation performed in the $z$ direction are the partial derivatives present in the equations~\eqref{eq:fullequations}. To evaluate these we use a fourth-order accurate (central) finite difference approximation. Also in this direction, we find spurious high-frequency noise common to any finite differencing schemes. In order to remove it we add numerical dissipation to damp these modes.
% We have therefore implemented $N=3$ Kreiss-Oliger dissipation~\cite{Kreiss1973} whereby,
We have therefore implemented the usual Kreiss-Oliger dissipation operator of order 6~\cite{Kreiss1973} whereby,
after each time step, all our evolved quantities $f \in \{ B_{g_{1,2}}, \phi_{g_{1,2}}, a_4, f_2, \xi \}$ are added a term of the form
\begin{equation}
\label{eq:KO-N3}
D_{\rm KO} f_i \equiv \frac{\sigma}{64}
                     \left(
                       f_{i-3} - 6 f_{i-2} + 15 f_{i-1} - 20 f_i + 15 f_{i+1} - 6 f_{i+2}
                     + f_{i+3}
                     \right) \,,
\end{equation}
where $i$ labels the grid point in the $z$ direction and $\sigma$ is a tuneable dissipation parameter which must be smaller than 1 for stability, and which we have typically fixed to be 0.2. This procedure effectively works as a low-pass filter.

In the radial direction $u$, the use of the Chebyshev-Lobatto grid allow us to use pseudo-spectral collocation methods~\cite{Boyd2001}. These methods are based in the approximation of our solutions in a basis of known functions, Chebyshev polynomials $T_n(X)$ in our case, but, in addition to the spectral basis, we have an additional \textit{physical} representation and therefore we can perform operations in one basis or the other depending on our needs.
Discretization using the pseudo-spectral method consists in the exact imposition of our
equations at the collocation points of the Lobatto-Chebyshev grid.
Thanks to the trigonometric representation of the Chebyshev polynomials, we can use the Fast Fourier Algorithm (FFT) for changing from one basis to the other. One of the uses of these method is high-accuracy interpolation of any function $f$ to values of $u$ not present in our grid.
This can be computed using the standard spectral representation of the function
\begin{equation}
f(u) = \sum_{k=0}^N \hat f_{k}\, T^{}_k(X(u)) \,, \label{eq:spectralrepresentation}
\end{equation}
where ${\hat f}^{}_{k}$ are the coefficients of the spectral basis that are computed from the values of the function in the collocation points through the FFT.
The cost of the FFT algorithm scales as $O \left( N_u \log N_u \right)$, in contrast with the matrix transformation from the physical and spectral representations, which scales as $O \left(N_u^2 \right)$.

As we mentioned previously, cf.\ equation~\eqref{eq:gen-exp-final}, the radial equations for solving the metric coefficients can be written in the form
\begin{equation*}
\left[
  \alpha_g(u,t,z) \partial_{uu} + \beta_g(u,t,z) \partial_u + \gamma_g(u,t,z)
\right] g(u,t,z) = -S_g(u,t,z) \,,
\end{equation*}
where, again, $g$ represents the metric coefficients previously mentioned. Once our coordinate is discretized, the differential operator becomes an algebraic one acting over the values of the functions in the collocation points taking the form
\begin{equation*}
\left[
  \alpha_g^i(t,z) \mathcal D_{uu}^{ij} + \beta_g^i(t,z) \mathcal D_u^{ij} + \gamma_g^i(u,t,z)
\right] g^j(t,z) = -S_g^j(t,z) \,,
\end{equation*}
where $\mathcal D_{uu}$, $\mathcal D_{u}$ represent the derivative operator for a Lobatto-Chebyshev grid in the physical representation and $i$, $j$ indices in the $u$ coordinate. We now construct the operator defined inside the brackets and then invert it to solve the function $g$. Boundary conditions are imposed by replacing full rows in this operator by the values we need to fix. In the general case, for a second order operator we replace the lines $j=0$, $j=N$ by the value of the function and its derivative at $u=0$ in the case of  grid1 and at $u=u_0$ in the case of grid2. At grid1, we obtain the boundary conditions from~(\ref{eq:near-bdry}); at grid2 these are read off from the obtained values at grid1.

Another useful feature of the spectral methods is the possibility of filtering. As we did with the dissipation in the direction $z$, we can damp high order modes but in this case directly in the spectral representation. After each time step, we apply an exponential filter to the spectral coefficients of our $u$-dependent evolved quantities $\hat f \in \{ \hat B_{g_{1,2}}, \hat \phi_{g_{1,2}} \}$. The complete scheme is
\begin{equation}
\left \{  f_i\   \right\}                          \stackrel{\rm FFT}{\longrightarrow}
\left \{  \hat f_k \right\}                               \stackrel{}{\longrightarrow}
\left \{ \hat f_k\; e^{ -\alpha (k/N_u)^{\gamma N_u} } \right\}  \stackrel{\rm FFT}{\longrightarrow}
\left \{f_i \right\}  \label{eq:u-filter}
\end{equation}
where $\alpha$ and $\gamma$ are tuneable parameters which we typically fix to $\alpha = 36.0437$, $\gamma = 8$. This effectively dampens the coefficients of the higher-order Chebyshev polynomials.

\subsection{Initial data}
\label{sec:init-data}

Our chosen formulation of Einstein's equations, known as the characteristic formulation, allows one to specify the initial data needed for an evolution through freely setting the functions $B(u,z)$, $\phi(u,z)$, $\xi(z)$, $a_4(z)$ and $f_2(z)$.
For our intended applications, we wish to have initial data resembling an ultra-relativistic projectile, such as the shockwave metric in AdS. The starting point to construct such initial data is thus the shockwave metric in FG coordinates~\eqref{metricsw}. Once the function $f(\uFG)$ therein is computed, one can proceed to transform the metric to the EF frame~(\ref{eq:metric00}) in which the numerical integration is performed. Owing to the fact that both the FG and the EF metrics have an explicit Killing vector, one can use the following ansatz for the coordinate transformation between the two frames
\begin{equation}
\label{transform1}
\begin{aligned}
 \mathbf{x}_{\perp}^{\mathrm{FG}} & =\mathbf{x}_{\perp}^{\mathrm{EF}} \,, &
  \uFG & = u + \lambda_1(u, t+z) \,, \\
  x_{+} & = t+z+\lambda_2(u, t + z) \,, &
  x_{-} & = t-z + \lambda_3(u, t + z) \,,
\end{aligned}
\end{equation}
for a left-moving shock~\cite{Chesler:2013lia}.
The differential equations for the transformation functions $\lambda_1(u,z)$, $\lambda_2(u,z)$, and $\lambda_3(u,z)$ are obtained by simply taking the slots $g_{uu}$, $g_{ut}$, and $g_{uz}$ from the equation 
\begin{equation}
\label{transform2}
g^{\mathrm{EF}}=\Lambda g^{\mathrm{FG}} \Lambda^{T} \,.
\end{equation}
Equivalently, one might use the fact that the EF coordinate $u$ is a non-affine parameter for ingoing null geodesics
\begin{equation}
\label{geodesiceqs}
\partial_{u}^{2} k^{\mu}(u)
+\Gamma^{\mu}_{\alpha \beta}\partial_{u}k^{\alpha}(u)\partial_{u}k^{\beta}(u)
=F(u)\partial_{u}k^{\mu} \,,
\end{equation}
where $k^{\mu}(u)$ is the parametrized geodesic, and $F(u)=\frac{-2}{25 u}$ is a non-affinity function set to meet the desired EF frame with $g_{tr}=1$.
The geodesic equation has the advantage of being explicitly dependent on $t+z$ and therefore its solution reduces to a set of ODEs parametrised by the boundary point $z$ for $t=0$.
We thus write our initial data for a left-moving shock as follows: 
\begin{align}
\label{eq:gprofile}
  h(z)        & = \frac{\mu^3}{\omega \sqrt{2\pi}}
                e^{ -\frac{(z-z_0)^2}{2 \omega^2}} \,, \\[2mm]
\mathcal{E}(z) & = \mathcal{E}_0 + h(z)                \,, \label{eq:eps-ID} \\[3mm]
f_2(z)      & = h(z)                             \,,  \\[2mm]
\phi(u,z)   & = \frac{\phi_0 \uFG}
                  { \sqrt{1 + \frac{\uFG^2}{3 \phi_0}
                             \left( \phi_0^3 - 6 \phi_2 \right) }  } \,,
\label{eq:phi-ID} \\
  e^{3B(u,z)}   & = \frac{ e^{2\AFG(\uFG)} }
                  {\frac{\partial_{z}\lambda_1^2}{\uFG^2}
                  -\left(\partial_{z}\lambda_2+1\right) \left(\partial_{z}\lambda_3-1\right)
                   e^{2\AFG(\uFG)}+\left(\partial_{z}\lambda_2+1\right)^2 f(\uFG) h(z)} \,,
\end{align}
where $\uFG$, $\lambda_{1,2,3}$ are functions of $u$ and $z$ obtained
from~\eqref{transform1}. Recall that the function $h$ enters the metric \eqq{metricsw} and specifies the energy density, the longitudinal pressure and the energy flux in the initial state according to \eqq{eq:stresstensor-sw}. The choice \eqq{eq:gprofile} corresponds to a Gaussian profile with width $\omega$ and height $\mu^3/\omega \sqrt{2\pi}$.  $\mathcal{E}$ is the energy density per unit volume 
of the boundary field theory and $\mu^3$ is the energy density per unit transverse area. As usual \cite{Chesler:2010bi,Casalderrey-Solana:2013aba,Casalderrey-Solana:2013sxa,Casalderrey-Solana:2016xfq,Attems:2016tby} we have added a ``regulator''  $\mathcal{E}_0$, namely a background thermal bath with energy density much smaller than all other scales of interest, in order to avoid the large gradients that develop in the deep IR. 
% (which we typically set to $\mathcal{E}_0 = 0.02 \frac{\mu^3}{\sqrt{2\pi}\omega}$).
Given $\mathcal{E}_0$, we know the solution  in the absence of shocks, which has  $B=0$. In particular, we know the subleading coefficient $\phi_2$ of the scalar operator as a function of  $\mathcal{E}_0$, and this is the value that features in equation~(\ref{eq:phi-ID}).  An important point is that the $z$-independent equilibrium value $\phi(u)$ in FG coordinates is only known numerically. Equation~(\ref{eq:phi-ID}) is a good approximation to this numerical solution. The advantage of having an analytic approximation is that, in order to locate the apparent horizon in EF coordinates in the presence of the shocks,  it is necessary to know the value of the scalar field in FG coordinates slightly beyond the position of the horizon in those coordinates. Equation~(\ref{eq:phi-ID}) provides a good approximation to this value simply by declaring that it applies beyond the horizon. We have verified that the analytic form~(\ref{eq:phi-ID}) quickly relaxes upon time evolution and therefore that this way of initializing our code has no effect whatsoever on the collision dynamics. We choose the initial value for $a_4$ by comparing~(\ref{eq:energy}) and~\eqref{eq:eps-ID}. Finally,  the function 
$\xi(z)$ is initialized by imposing that the apparent horizon lie at a constant value of the $u$ coordinate.

%%%%%%%%%%%%%%%%%%%%%%%%%%%%%%%%%%%%%%%%%%%%%%%%%%%%%%%%%%%%%%%%%%%%%%%%%%%%%%%
\section{Code tests}
\label{sec:tests}
%%%%%%%%%%%%%%%%%%%%%%%%%%%%%%%%%%%%%%%%%%%%%%%%%%%%%%%%%%%%%%%%%%%%%%%%%%%%%%%

We implement the above construction in a standalone C code, where we use the GNU Scientific Library~\cite{GSL} to solve the linear system~(\ref{eq:fullequations}), the FFTW3 library~\cite{FFTW05} for FFTs, and use a
fourth-order Adams-Bashforth method to integrate the functions $B(u,z)$,
$\phi(u,z)$, $a_4(z)$, $f_2(z)$ and $\xi(z)$ forward in time, using the
procedure outlined in section~\ref{sec:finite-parts}. The code is trivially
parallelized with OpenMP. The resulting code is quite fast, being able to evolve a configuration with 
\be
% \sac L \phi_0 = 1
\phiM=10 \sac \phi_0 \omega = 0.32 \sac 
 \frac{\mu^3}{\phi_0^4 \, \omega \sqrt{2\pi}} = 1 \sac \mathcal{E}_0 = 0.02 \, \phi_0^4\,,
\ee
 with 12 + 48 $u$-points and $\phi_0 \, \Delta z=1/20$
(400 $z$-points) from $t = 0$ to $\phi_0 \, t = 1$ in 3 minutes using two cores
Intel i7-4820K CPU @ 3.70GHz.

\subsection{Quasi-normal modes}
\label{sec:QNM}

In order to test the code and our numerical implementation we have recovered some quasi-normal frequencies reported in~\cite{Attems:2016ugt}. For these tests, we evolved a $\phi_{\rm M} = 10$ $z$-independent configuration where the energy density was set to $\mathcal{E}/\phi_0^4 = 0.379686$. $a_4$ and $\phi_2$ were initialised to their corresponding equilibrium values, whereas $B$ and $\phi$ were set to
\begin{align}
\label{eq:QNM-id}
B    &  = 0.1 u^8 \,, \\
\phi &  = \phi_0 u + \phi_2 u^3 \,.
\end{align}
Since this configuration is not in equilibrium, $b_4$ and $\phi_2$ will oscillate and relax, allowing us to compute the quasi-normal modes (QNM) of the system.

Gravitational set-ups containing a single scalar field will typically show two scalar, independent,  gauge invariant types of perturbations, each one with its own tower of modes. Hence, the system will have two independent channels to relax to equilibrium. In the model studied in this work, the two channels control independently the fluctuations of the anisotropy and the trace of the stress-energy tensor of the dual plasma respectively. Since $b_4$ only contributes to the anisotropy and $\phi_2$ only to the trace, their fluctuations will be governed by different towers of modes. Therefore, the frequencies extracted from $b_4$ should match the anisotropy tower frequencies' and the ones from $\phi_2$ should match the trace, or ``bulk'', tower \cite{Attems:2016ugt}.

\begin{figure}[htbp]
\centering
\includegraphics[width=0.7\textwidth]{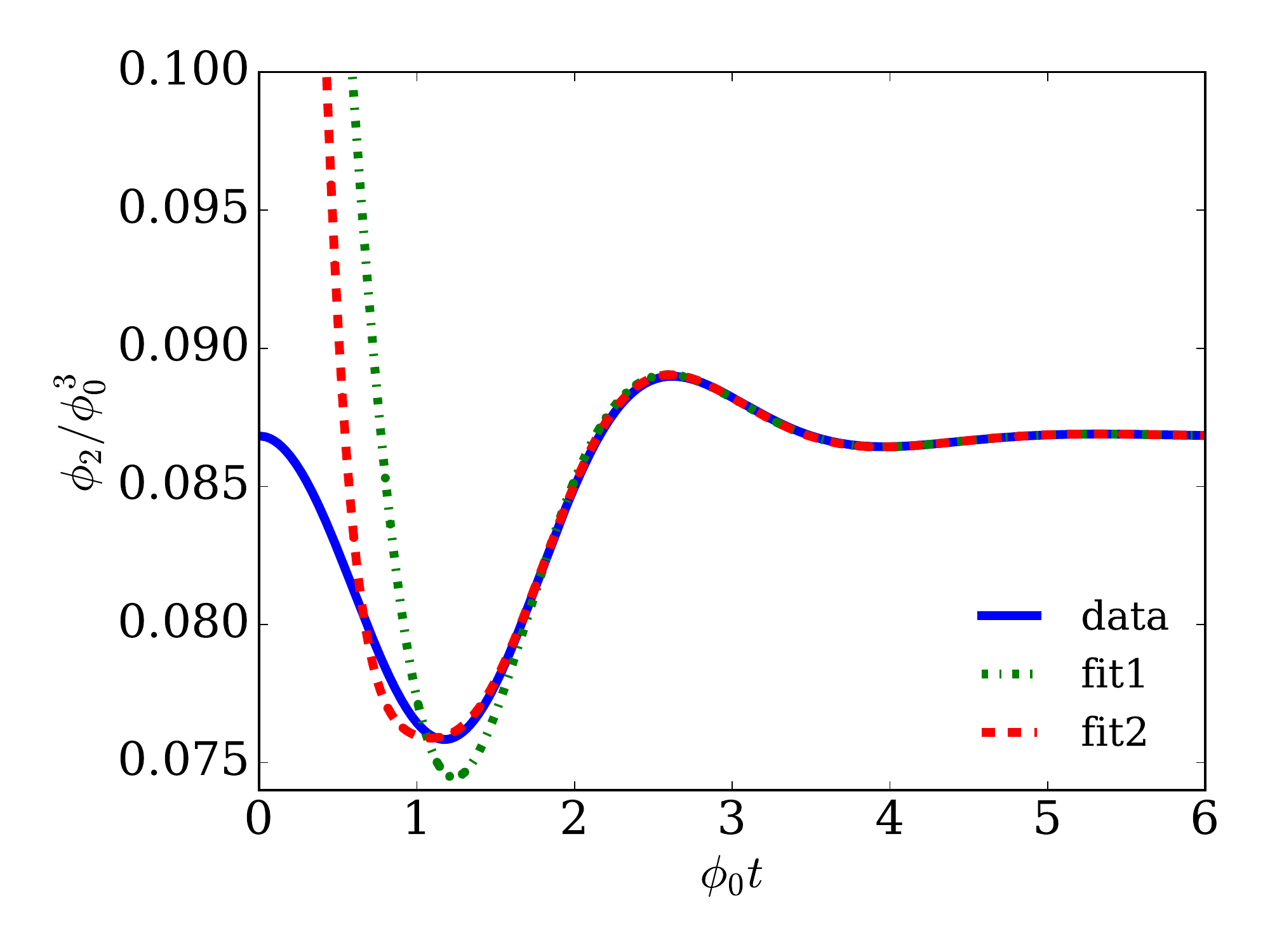} \\
\includegraphics[width=0.7\textwidth]{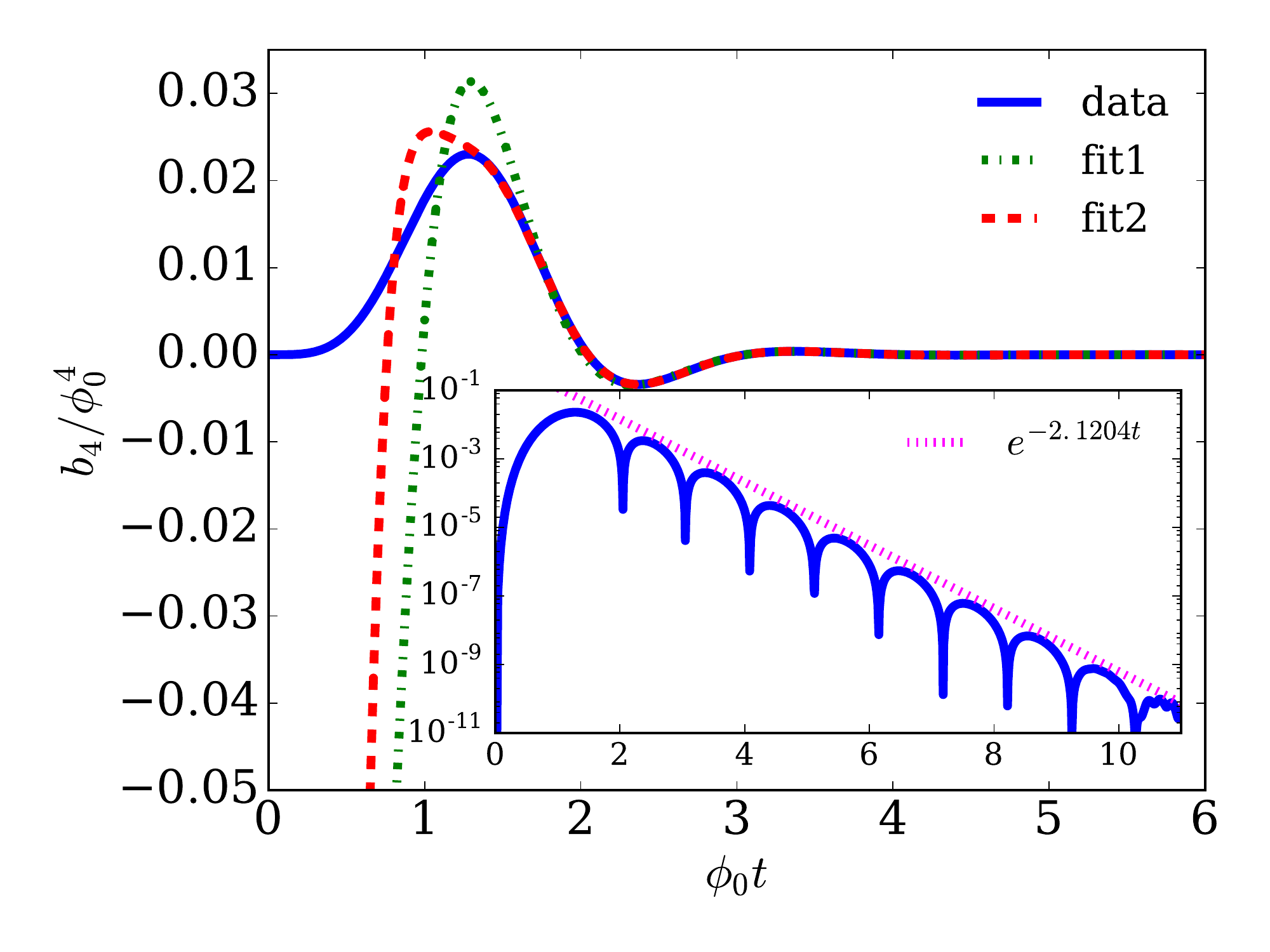}
\caption[]{$\phi_2$ and $b_4$ as functions of time for a $z$-independent configuration
  with $\phi_{\rm M} = 10$ %, $L\phi_0=1$
  and $\mathcal{E} = 0.379686 \, \phi_0^4$, with initial data as
  specified in~(\ref{eq:QNM-id}). The solid blue curve corresponds to data from the code, the dash-dotted green curve corresponds to a fit to the data using one QNM, and the dashed red curve corresponds to a fit using two QNMs as explained in the text. \label{fig:QNM_phiM10_conf6_phi2} }
\end{figure}

In figure~\ref{fig:QNM_phiM10_conf6_phi2}, we have fitted numerical data with damped sinusoidals of the form
\begin{equation}
\label{eq:damped-sin}
f(t) = C + A_1 e^{-\omega^{(1)}_i t} \cos\left( \omega^{(1)}_r t + \varphi_1 \right)
 + A_2 e^{-\omega^{(2)}_i t} \cos\left( \omega^{(2)}_r t + \varphi_2 \right) \,.
\end{equation}
In order to recover the frequencies we employed the following strategy. First, we look for the lowest frequency mode. For that, we set $A_2=0$ in equation~(\ref{eq:damped-sin}) and fit this function to our numerical data. We perform a series of fits to the data, each fit starting at a later time: we start by using the whole signal, then use only the portion $\phi_0 t \in [1, \infty[$ (say) of the signal, then only the portion $\phi_0 t \in [2, \infty[$ and so on. The frequencies $\omega^{(1)}$ thus obtained in each fit eventually converge to some value, the longest lived mode, which we are able to isolate through this process. We then fix the $C$, $A_1$, $\omega^{(1)}_r$, $\omega^{(1)}_i$, $\varphi_1$ fitting parameters obtained; the corresponding fit is labelled ``fit1'' in figure~\ref{fig:QNM_phiM10_conf6_phi2}. Having fixed these parameters we then repeat the process using equation~(\ref{eq:damped-sin}), where this time we \emph{only} allow for the $A_2$, $\omega^{(2)}_r$, $\omega^{(2)}_i$, $\varphi_2$ parameters to vary. We thus obtain the frequencies $\omega^{(2)}$; the final resulting fit is labelled ``fit2'' in figure~\ref{fig:QNM_phiM10_conf6_phi2}.

The results obtained with this procedure are displayed in figure~\ref{fig:QNM_phiM10_conf6_phi2}. For the non-conformal mode (top panel) we have obtained
\begin{align}
\label{eq:freq-phi2}
\omega^{(1)}_r & =  2.31305 \, \phi_0 \,,  & \omega^{(1)}_i & =  1.26432 \, \phi_0 \,, \\
\omega^{(2)}_r & =  4.03    \, \phi_0 \,,  & \omega^{(2)}_i & =  2.93    \, \phi_0\,,
\end{align}
which are to be compared with 
\be
\omega^{(1)} = \left(2.313106 + 1.264367 i \right) \, \phi_0 \sac
\omega^{(2)} = \left( 4.108 + 2.93141 i \right)  \, \phi_0
\ee
obtained in~\cite{Attems:2016ugt}.
For the anisotropic mode (bottom panel), we have obtained
\begin{align}
\label{eq:freq-b4}
\omega^{(1)}_r & =  3.03932 \, \phi_0 \,,  &  \omega^{(1)}_i & = 2.12048  \, \phi_0 \,, \\
\omega^{(2)}_r & =  4.9     \, \phi_0 \,,  &  \omega^{(2)}_i & =  3.6     \, \phi_0 \,,
\end{align}
which are to be compared with 
\be
\omega^{(1)} = \left(3.03944 + 2.120404 i \right)  \, \phi_0 \sac
\omega^{(2)} = \left(4.934 + 3.7393 i \right)  \, \phi_0
\ee
obtained in~\cite{Attems:2016ugt}.

We emphasise that the numbers from~\cite{Attems:2016ugt} and those of this section were obtained in a completely independent way, and the excellent agreement between them (of up to $0.004 \%$ for the lowest frequency) validates both the code presented herein as well as the method of~\cite{Attems:2016ugt}.

\subsection{Convergence analysis}
\label{sec:convergence}
%%%%%%%%%%%%%%%%%%%%%%%%%%%%%%%%%%%%%%%%%%%%%%%%%%%%%%%%%%%%%%%%%%%%%%%%%%%%%%%

Numerical simulations using finite differencing techniques typically approximate the continuum solution of the problem with an error that depends polynomially on the grid spacing $h$,
\begin{equation}
\label{eq:f-error}
f = f_h + O(h^n) \,.
\end{equation}
Different numerical implementations will give different convergence orders $n$. In our case, since we make use of fourth-order finite difference stencils, we expect to see $n=4$.
One simple way to check for consistency of a code is evolving the same configuration with coarse, medium and fine resolution, $h_c$, $h_m$ and $h_f$. One can then compute a convergence factor given by
\begin{equation}
\label{eq:Q}
Q \equiv \frac{f_{h_c} - f_{h_m}}{f_{h_m} - f_{h_f}}
      =  \frac{h_c^n - h_m^n}{h_m^n - h_f^n} \,,
\end{equation}
where $f_h$ is a chosen evolved variable obtained with numerical resolution $h$. Since in the radial direction we make use of pseudo-spectral methods, our error will be dominated by the resolution used in the $z$ direction, to which the grid spacing $h$ alludes to. For the analysis done in this section we therefore always make use of the same resolution in the radial direction.

\begin{figure}[htbp]
\centering
\includegraphics[width=0.65\textwidth]{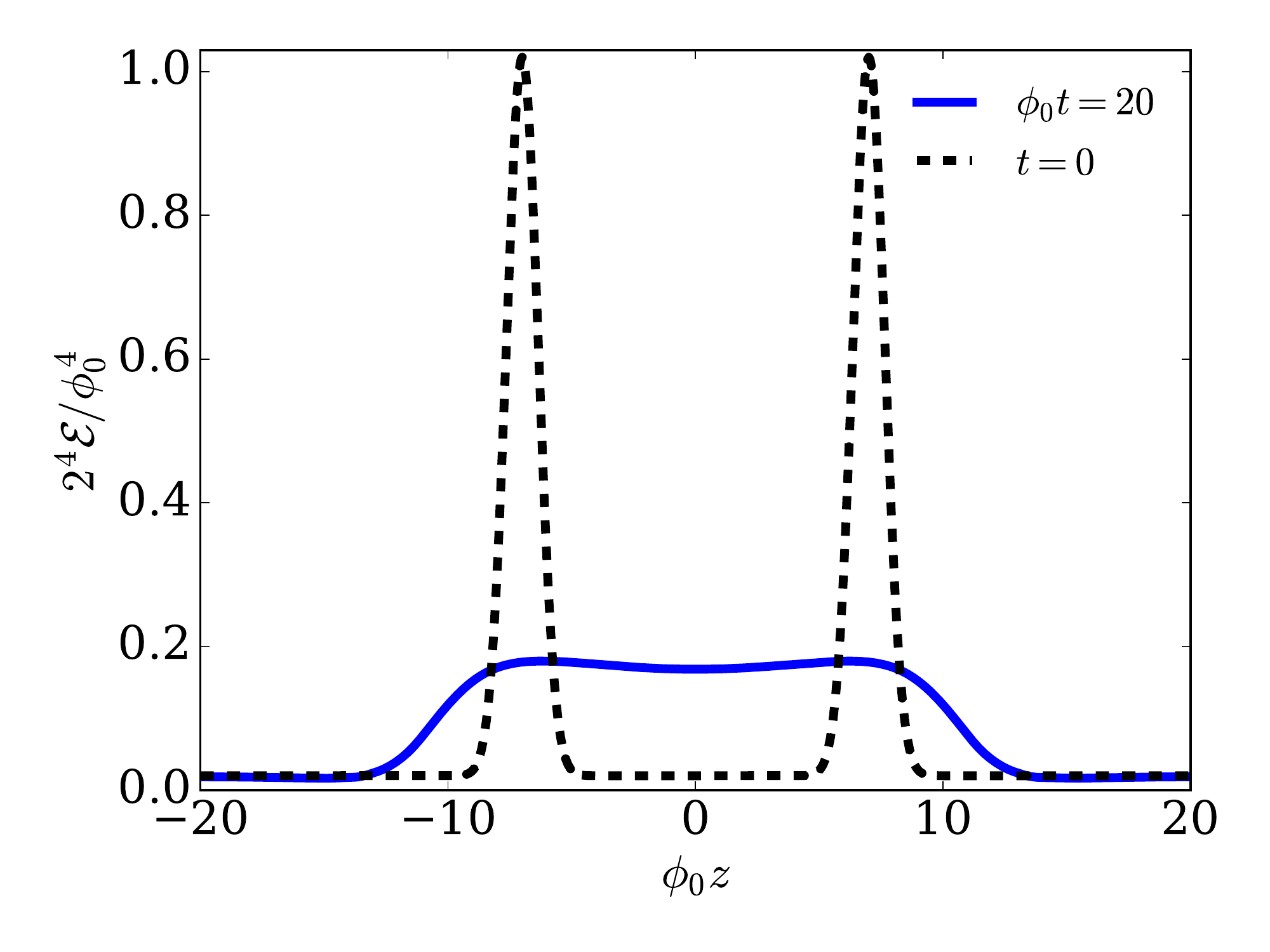} \\
\includegraphics[width=0.65\textwidth]{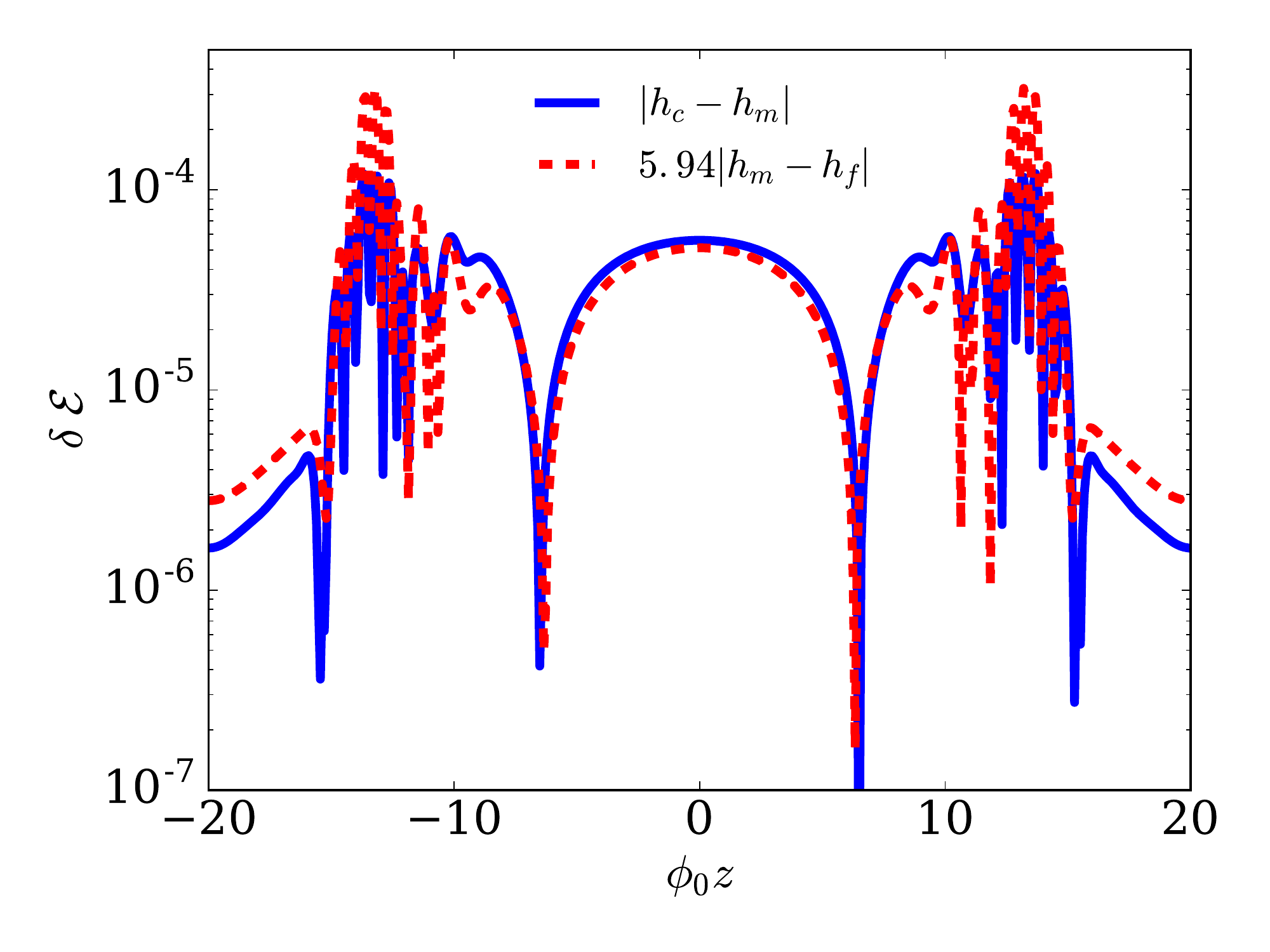}
\caption[]{Energy density at $\phi_0 t = 20$ (top panel) and correspondent
  convergence analysis (bottom panel) for a configuration with $\phiM=10$, %$L \, \phi_0 = 2$,
  $\phi_0 \, \omega = 0.64$, $\frac{2^4 \mu^3}{\phi_0^4 \sqrt{2\pi}\omega} = 1$, $\mathcal{E}_0 = \frac{0.02}{2^4} \, \phi_0^4 $.
  We plot the absolute differences between the coarse and medium resolution
  (blue solid line) and the medium and fine (red dashed line) resolution
  run. The latter has been re-scaled by the factor $Q=5.94$ expected for fourth
  order convergence. \label{fig:conf06_conv} }
\end{figure}
We show in figure~\ref{fig:conf06_conv} the convergence properties of our code obtained for a ``typical'' shockwave collision with physical parameters (cf.\ section~\ref{sec:init-data})
\begin{equation}
% \phiM=10 \sac L \phi_0 = 2 \sac \frac{\omega}{L} = 0.32 \sac 
% \frac{L^4  \mu^3}{\omega \sqrt{2\pi}} = 1 \sac L^4 \mathcal{E}_0 = 0.02 \,.
  % L \, \phi_0 = 2 \sac
  \phiM=10 \sac  \phi_0 \, \omega = 0.64 \sac
  \frac{2^4 \mu^3}{\phi_0^4 \sqrt{2\pi}\omega} = 1 \sac \mathcal{E}_0 = \frac{0.02}{2^4} \, \phi_0^4
\end{equation}
This configuration was evolved with $ \phi_0 \, h_c = 40$, $\phi_0 \, h_m = 60$ and $\phi_0 \, h_f = 80$; the expected convergence factor expected for fourth order convergence would therefore be $Q \approx 5.94$. Plotted in the figure are the results obtained for the energy density at $\phi_0 \, t = 20$, where the differences $|f_{h_m} - f_{h_f}|$ have been amplified by $Q = 5.94$. The results show fourth-order convergence. We have further verified that the values obtained for our medium resolution run are within $\sim 0.4 \%$ of the fourth-order Richardson-extrapolated ones, giving us an estimate of the error incurred in the simulation.

%%%%%%%%%%%%%%%%%%%%%%%%%%%%%%%%%%%%%%%%%%%%%%%%%%%%%%%%%%%%%%%%%%%%%%%%%%%%%%%
\section{Non-conformal collisions}
\label{sec:collisions}
%%%%%%%%%%%%%%%%%%%%%%%%%%%%%%%%%%%%%%%%%%%%%%%%%%%%%%%%%%%%%%%%%%%%%%%%%%%%%%%

\subsection{Time evolution}

Using the numerical procedure described in section~\ref{sec:numerical} we are now ready to  explore and characterise shockwave collisions in different non-conformal theories.
As in the analysis of conformal shockwave collisions in \cite{Chesler:2010bi,Casalderrey-Solana:2013aba,Casalderrey-Solana:2016xfq}, we employ Gaussian energy density profiles in the longitudinal direction, \eqn{eq:gprofile}. We choose $t=0$ as the time at which the two incoming shocks would exactly overlap in the absence of interactions. 

In a CFT scale invariance guarantees that the physics can only depend on the dimensionless product of the transverse energy scale and the width of the shock, 
$\mu \omega$. In contrast, in a non-conformal theory with an intrinsic scale $\Lambda$ the physics will also depend on the ratio $\mu /\Lambda$. We will see that, by varying this last ratio for a fixed shock profile ($\mu \omega = \mathrm{const}$), we can study the collision dynamics from low to high energies. 
%After hydrodynamization, this also translates into different initial temperatures, $\Thyd/\Lambda$, which probe different regions of the non-trivial equation of state of the model.
Indeed, our  model is specified by the value of the parameter $\phiM$, which controls the degree of non-conformality of the dual gauge theory.  For any value of $\phiM$,
when  $\mu$ and $\Lambda$ are of the same order, the formation and relaxation of the plasma happens in the most non-conformal region, while for large $\mu/\Lambda$, the early time evolution is approximately as that in a CFT. We will consider two different values $\mu \omega \simeq 0.30$ and $\mu \omega \simeq 0.12$, corresponding to what were dubbed ``1/2-shocks''  and ``1/4-shocks'' in \cite{Casalderrey-Solana:2013aba}.

Using equations \eqq{eq:Tcomponents} and 
\eqq{eq:energy}-\eqq{eq:J}
we extract the stress tensor from our numerical evolution for different values of $\mu /\Lambda$. Following the standard Landau matching procedure, we define the local energy 
density and a velocity field by determining the time-like eigenvalue of the stress tensor.
\be
T^{\mu\nu} u_\nu=-\mathcal{E}_{\rm loc} u^\mu \,, \quad u_\mu u^\mu=-1	\,.
\ee
 As a consequence of $z$-reflection symmetry, at $z=0$ the local and collision frames coincide and the local energy density is given by $\mathcal{E}$. Given the energy density, we can assign a value of the transport coefficients $\zeta(\mathcal{E})$ and 
 $\eta(\mathcal{E})$ to each spacetime point after the collisions. Since $\zeta$ vanishes in a conformal theory, we can use the assigned ratio $\zeta/\eta$ as a measure of non-conformality.

%%%%%%%%%%%%%%%%%%%%%%%%%%%%%%%%%%%%%%%%%%%%%%%%%%%%%%%%%%%%%%%%%%%%%%%%%%%%%%%
\begin{figure}[htbp]
\centering
\includegraphics[width=0.6\textwidth]{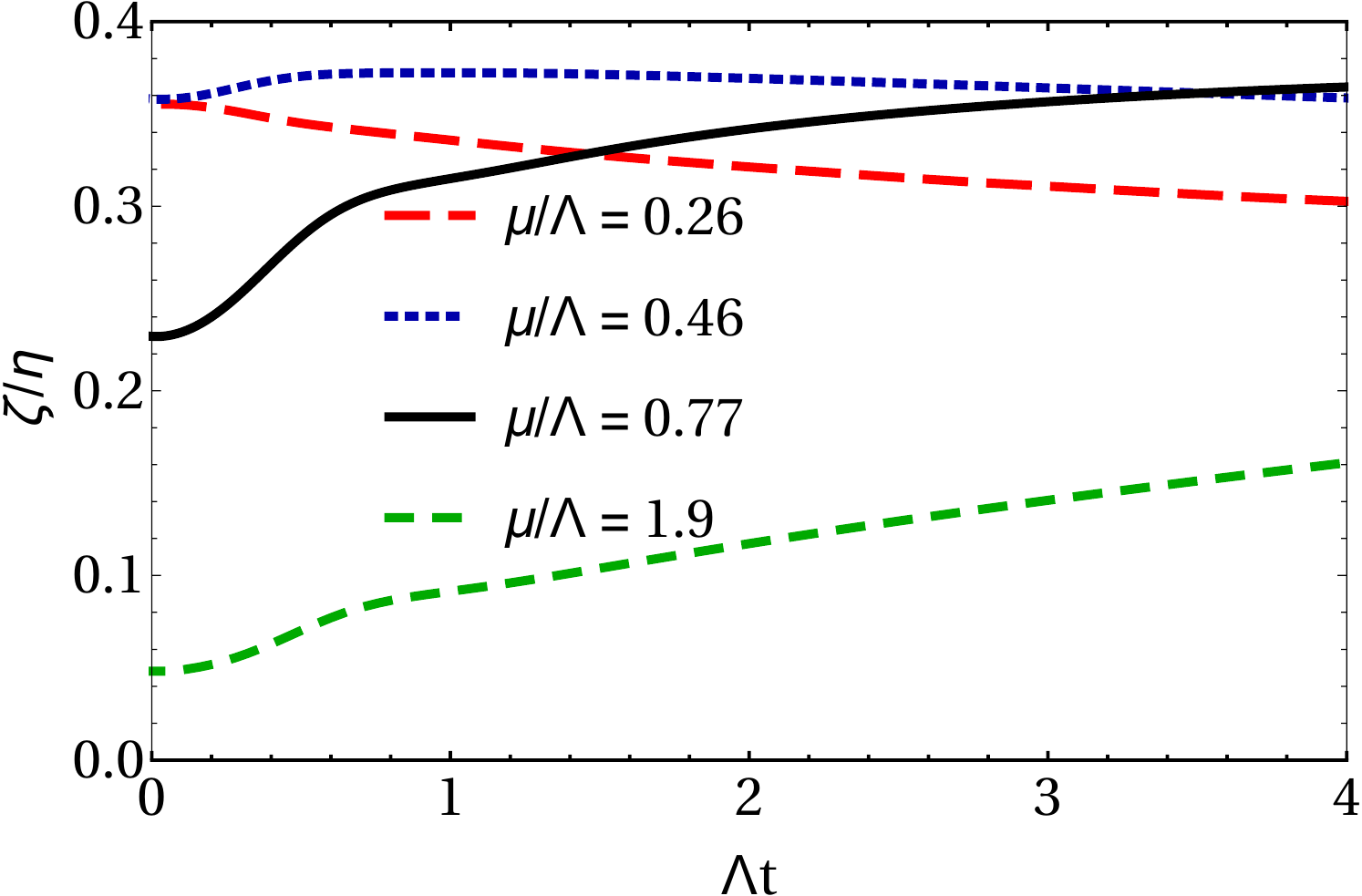}
\caption[]{Early-time evolution of the $\zeta / \eta$ ratio for runs with $\phiM = 20$
and $\mu\omega=0.30$. The times shown include times before and after  hydrodynamization.
\label{fig:tLambdabulk} }
\end{figure}
%%%%%%%%%%%%%%%%%%%%%%%%%%%%%%%%%%%%%%%%%%%%%%%%%%%%%%%%%%%%%%%%%%%%%%%%%%%%%%%

In figure~\ref{fig:tLambdabulk} we plot\footnote{For this plot and for all the
  results in this section we use several values of the energy density
  $\mathcal{E}_0$, ranging between
  $\mathcal{E}_0 = \frac{\mu^3}{\sqrt{2\pi}\omega} (0.005,0.02)$, and we
  check that the effects of this regulator are small and in the linear
  regime. Furthermore, we extrapolate all physical
  results to $\mathcal{E}_0 = 0$ checking that first and second
  order extrapolations converge to the same value.} the time dependence of this
ratio of viscosities at $z=0$ for several representative values of
$\mu/\Lambda$. For large $\mu/\Lambda$ values, the energy deposited by the
collision in this central region is also large and the system is close to 
conformality. As the system expands, the energy decreases proving regions of larger
and larger $\zeta/\eta$. For smaller values of $\mu/\Lambda$ the system stays in the
non-conformal region from an earlier time. Although after a collision the energy
at $z=0$ is continuously decreasing, we have not extended our simulation long
enough to recover conformal dynamics at late time, as we would expect to happen from the IR behaviour of our model.

The assigned values of transport coefficients also control the dynamics of the stress tensor soon after the collision. In other words, hydrodynamics becomes applicable.  At first order in the gradient expansion the  hydrodynamic stress
tensor may be expressed as 
\be
\label{eq:Thydro}
T^{\rm hyd}_{\mu\nu}=\Big[ \mathcal{E}_{\rm loc} + P_{\rm eq} \left( \mathcal{E}_{\rm loc}\right)  \Big] u_\mu u_\nu+ 
 P_{\rm eq} \left( \mathcal{E}_{\rm loc}\right) g_{\mu\nu} 
- \eta \left( \mathcal{E}_{\rm loc}\right) \sigma_{\mu \nu} - \zeta \left( \mathcal{E}_{\rm loc}\right) \Pi \Delta_{\mu\nu}\, ,
\ee
where $g$ is the Minkowski metric and $ P_{\rm eq} \left( \mathcal{E}_{\rm loc}\right)$ is the equilibrium pressure, $\sigma_{\mu \nu}$ and $\Pi$ are the shear and bulk tensors constructed from gradients of the velocity field, and $\Delta_{\mu \nu}$ is the projector on the fluid rest frame. As we will see, at sufficiently late times this expression approximates well the evolution of the full stress tensor. 

\begin{figure}[htbp]
\centering
\includegraphics[width=0.7\textwidth]{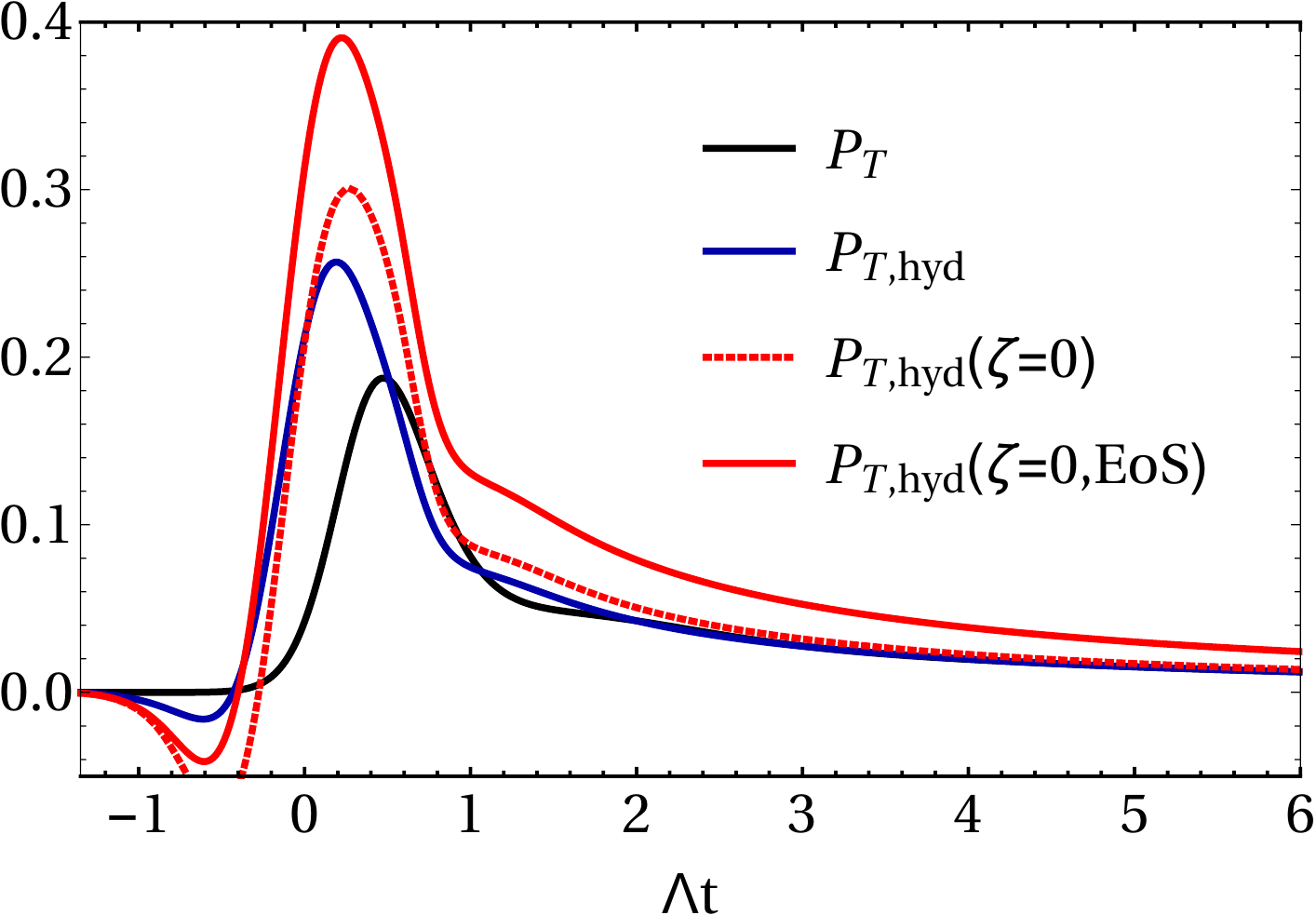}

\vspace{5mm}

\includegraphics[width=0.7\textwidth]{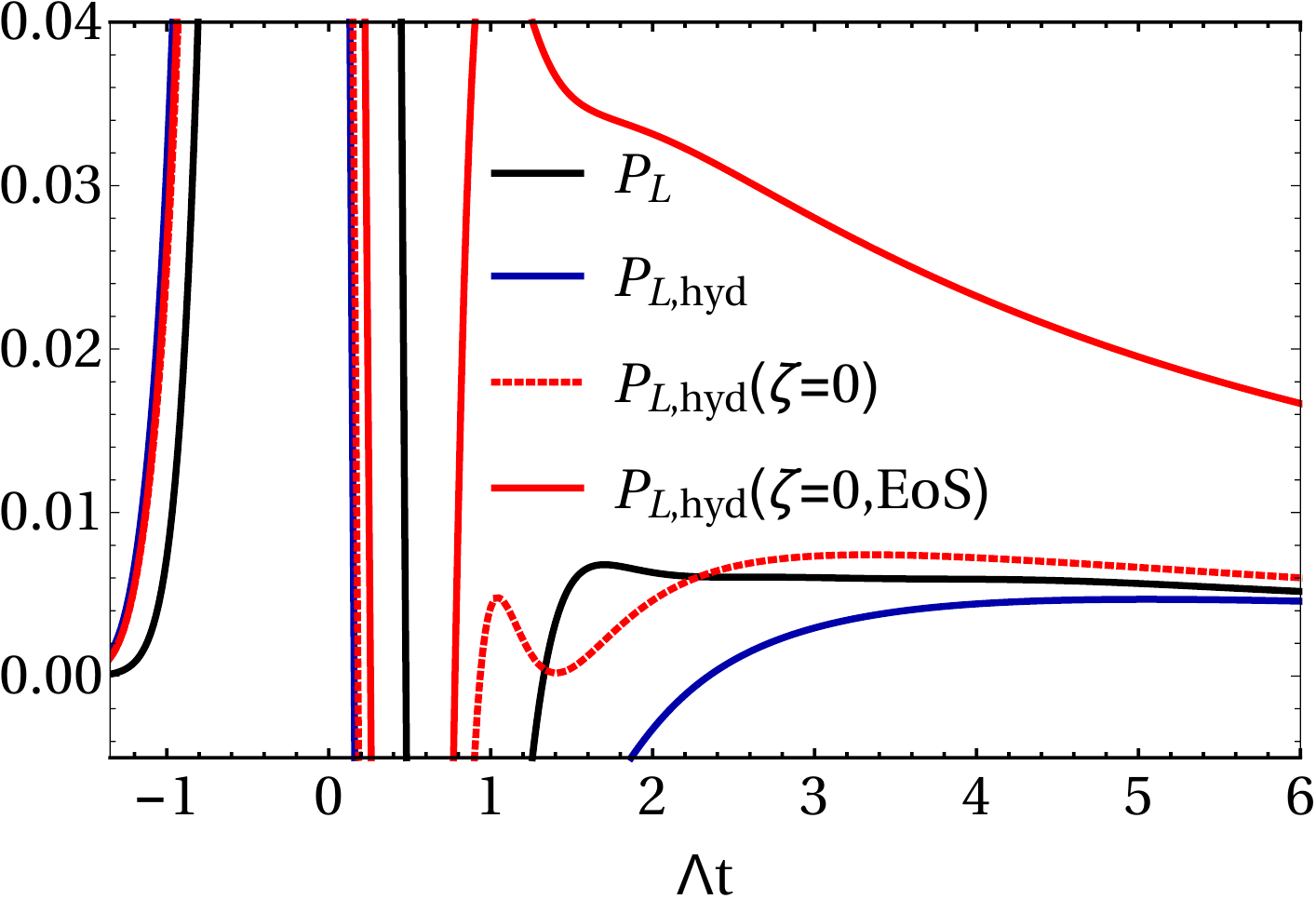}
\caption[]{Time evolution of the pressures,  in units of $\Lambda^4$, for $\phiM=20$, $\mu/\Lambda= 0.77$ and \mbox{$\mu \omega= 0.30$}. 
The evolution is compared to the hydrodynamic prediction via the constitutive relations \eqn{eq:Thydro} in different approximations: $P_{L,T}^{\rm hyd}(\zeta = 0, \mbox{EoS})$ corresponds to a conformal fluid with $P_{\rm eq}=\mathcal{E}/3$ and $\zeta=0$; $P_{\{L,T\}}^{\rm hyd}(\zeta = 0)$ includes the correct, non-conformal  equation state but still $\zeta=0$; $P_{L,T}^{\rm hyd}$ includes the correct, non-conformal equation of state and the non-vanishing $\zeta$. 
and including all non-conformal dynamics  $P_{L,T}^{\rm hyd}$.
After a time $t \Lambda=2.12$ $(4.65)$ the transverse  (longitudinal) pressure is described by non-conformal hydrodynamics with better than 10\% accuracy. 
\label{fig:landau} }
\end{figure}

%%%%%%%%%%%%%%%%%%%%%%%%%%%%%%%%%%%%%%%%%%%%%%%%%%%%%%%%%%%%%%%%%%%%%%%%%%%%%%%

To illustrate the non-conformal nature of the collision dynamics, in figure~\ref{fig:landau} we show the time evolution of the transverse (top) and the longitudinal (bottom) pressures at $z=0$ for a collision with $\mu/\Lambda= 0.77$ and $\mu \omega=0.30$ in the 
 $\phiM = 20$ model (black solid lines). We compare these evolutions with the 
 first-order hydrodynamic prediction \eqq{eq:Thydro} turning on sequentially the  two non-conformal properties in the hydrodynamic approximation, namely the non-conformal EoS and the non-zero $\zeta$. As represented by the solid red curve $P_{L,T}^{\rm hyd}(\zeta = 0, \mbox{EoS})$, we see that assuming a conformal EoS and $\zeta=0$  fails to reproduce the time evolution. The inclusion of the correct equation of state, represented by the dashed red curve $P_{\{L,T\}}^{\rm hyd}(\zeta = 0)$, brings the hydrodynamic prediction closer to the true evolution. Finally,  the inclusion of  bulk viscosity, represented by the blue solid curve 
 $P_{L,T}^{\rm hyd}$,  increases the convergence of the first order hydrodynamic prediction to the evolution of the pressures. As stated, the post collision dynamics in this regime is intrinsically non-conformal.

\subsection{Hydrodynamization and EoSization}

Inspection of figure~\ref{fig:landau} indicates that hydrodynamics provides a good description of the evolution of the stress tensor even when the difference between  the longitudinal and the transverse pressures is large, which signals the presence of large gradient corrections. This fact, first noted for conformal systems in \cite{Heller:2011ju,Chesler:2010bi}, led to the concept of ``hydrodynamization'', i.e. the process by which hydrodynamics comes to describe the dynamics of an interacting system, even if the system is far from local thermal equilibrium. In this section we systematically explore this process for different collision energies in four different non-conformal theories, parametrized by four values of the parameter $\phiM$. 

As is common in the literature, we define the hydrodynamization time as the time beyond which  both pressures  are  described by hydrodynamics within a given accuracy. However, in contrast to conformal 
dynamics, where the tracelessness of the stress tensor fixes the relation between the longitudinal and the transverse pressure, in an non-conformal theory the evolution of these two quantities is unconstrained. For this
reason, we introduce independent hydrodynamization criteria for each of the pressure components.  
As in \cite{Attems:2016tby} we define the hydrodynamization time $\thyd$  as the time beyond which the difference between the true pressures and the first-order  hydrodynamics prediction is less than $10\%$,
\begin{equation}
\label{equ:criteria_hydro}
\left| \frac{ P_{L,T} - P_{L,T}^{\textrm{hyd}} }{\bar P}\right| < 0.1 \,.
\end{equation}
Note that we have used the average pressure $\bar P$ as the characteristic scale of the stress tensor, which agrees with the criterium used in \cite{Casalderrey-Solana:2013aba} in the conformal case.

As noted in \cite{Attems:2016tby} hydrodynamization  is only sensitive to particular combinations of the shear and bulk contributions to the pressure, which we denote $P_\eta$ and $P_\zeta$. Since the shear tensor is traceless and the bulk tensor is diagonal in the local rest frame, we can write
\bea
P^{\rm hyd}_L&=& P_{\rm eq} + P_\eta + P_\zeta \,, \\
P^{\rm hyd}_T&=& P_{\rm eq} -\frac{1}{2}P_\eta + P_\zeta \,. 
\eea
Form this decomposition it is clear that the average hydrodynamic pressure is only sensitive to bulk gradients. Furthermore, as discussed in section~\ref{gtq}, the
equation of state relates the average pressure of the system to the energy density via the equilibrium value of $\mathcal{V}$, the thermal expectation value of the dimension-three operator which deforms the dual gauge theory. For this reason in \cite{Attems:2016tby} we introduced the 
 EoSization time $t_{\rm EoS}$ as the time beyond which the average pressure agrees with the equilibrium pressure with a $10\%$ accuracy,
\begin{equation}
\label{equ:criteria_eos}
\left|\frac{\bar P - P_{\rm eq}}{\bar P}\right| < 0.1 \,.
\end{equation}

%%%%%%%%%%%%%%%%%%%%%%%%%%%%%%%%%%%%%%%%%%%%%%%%%%%%%%%%%%%%%%%%%%%%%%%%%%%%%%%
\begin{figure}[htbp]
\begin{tabular}{cc}
\includegraphics[width=0.48\textwidth]{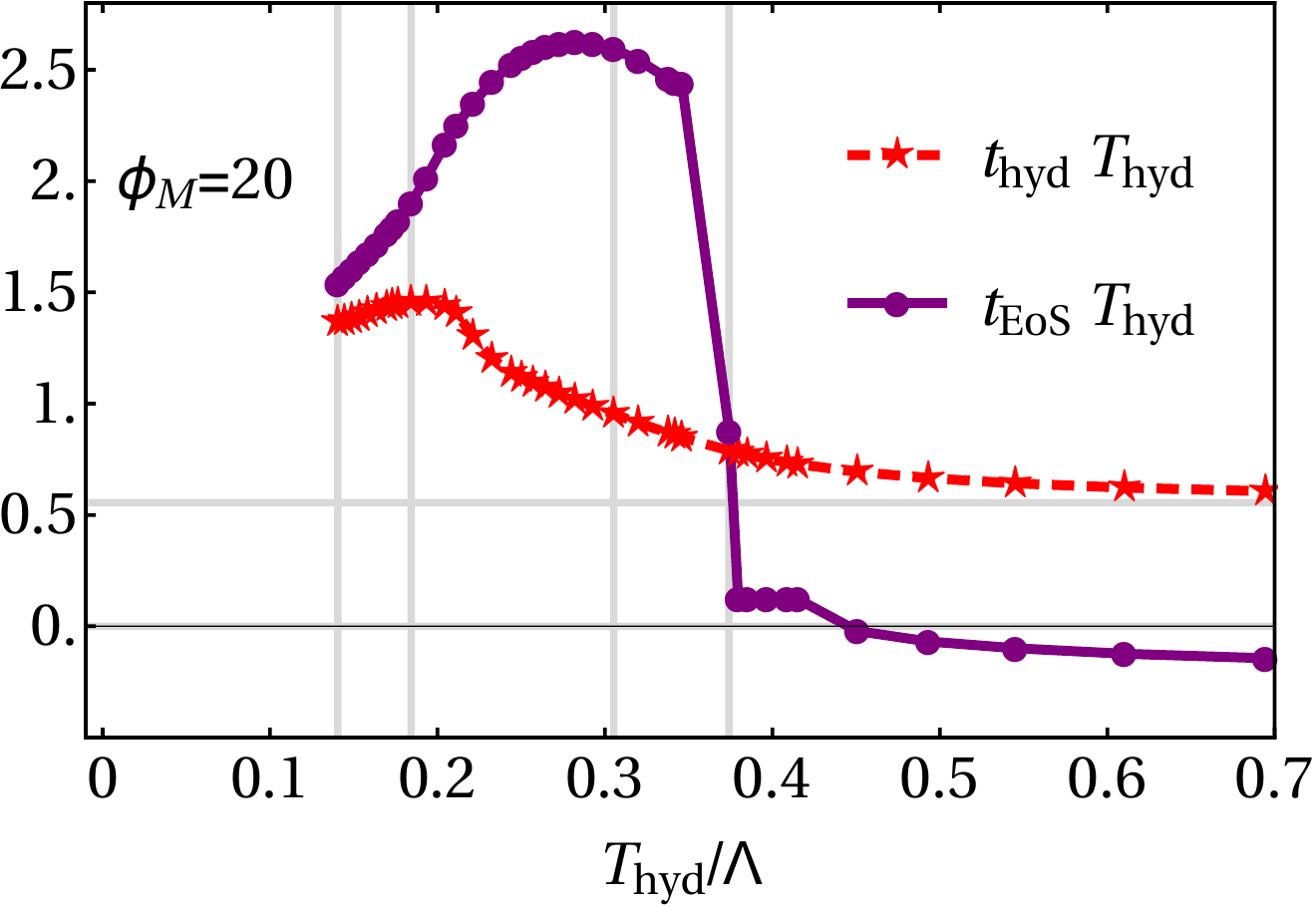}
& \includegraphics[width=0.48\textwidth]{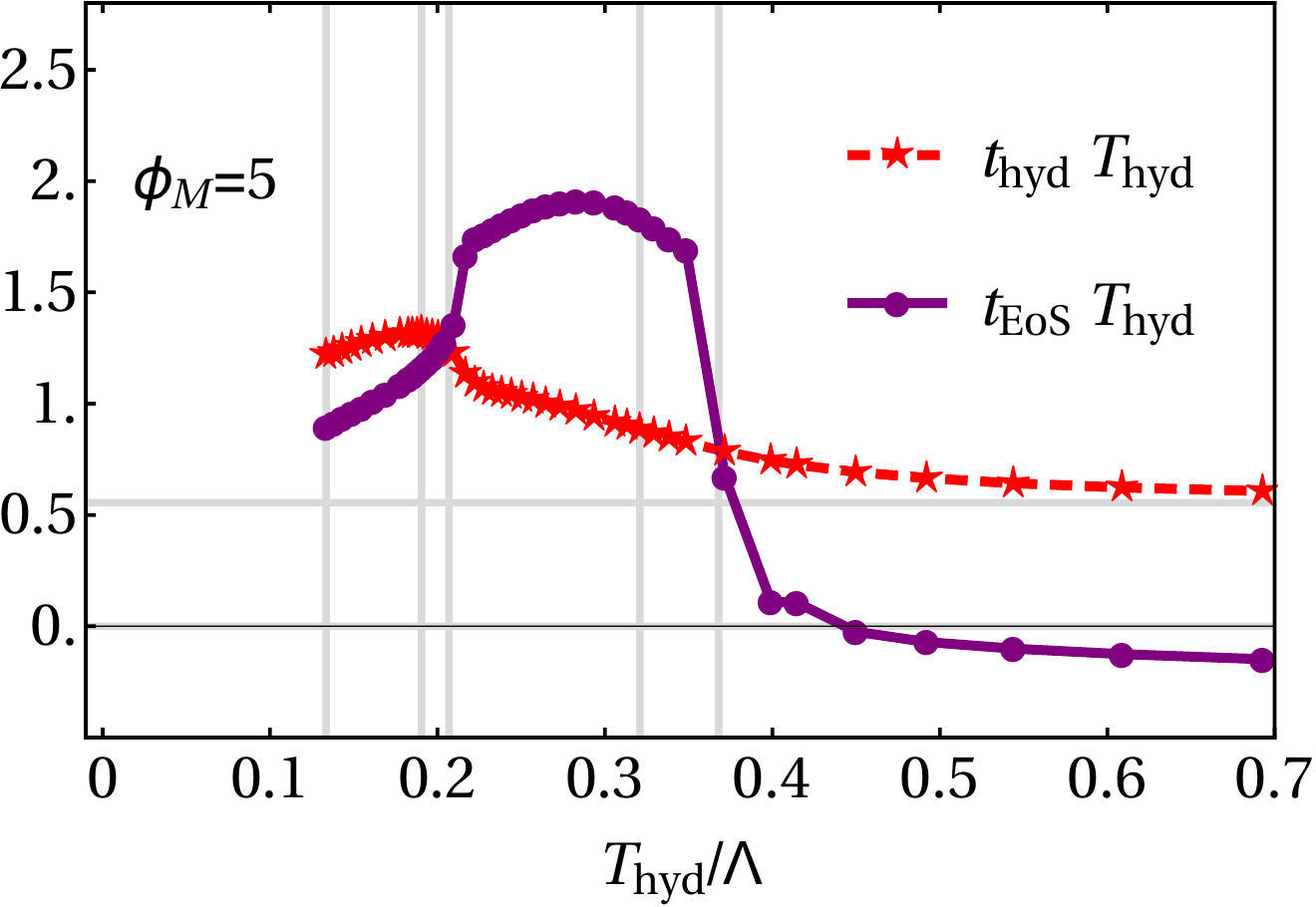} \\
\includegraphics[width=0.48\textwidth]{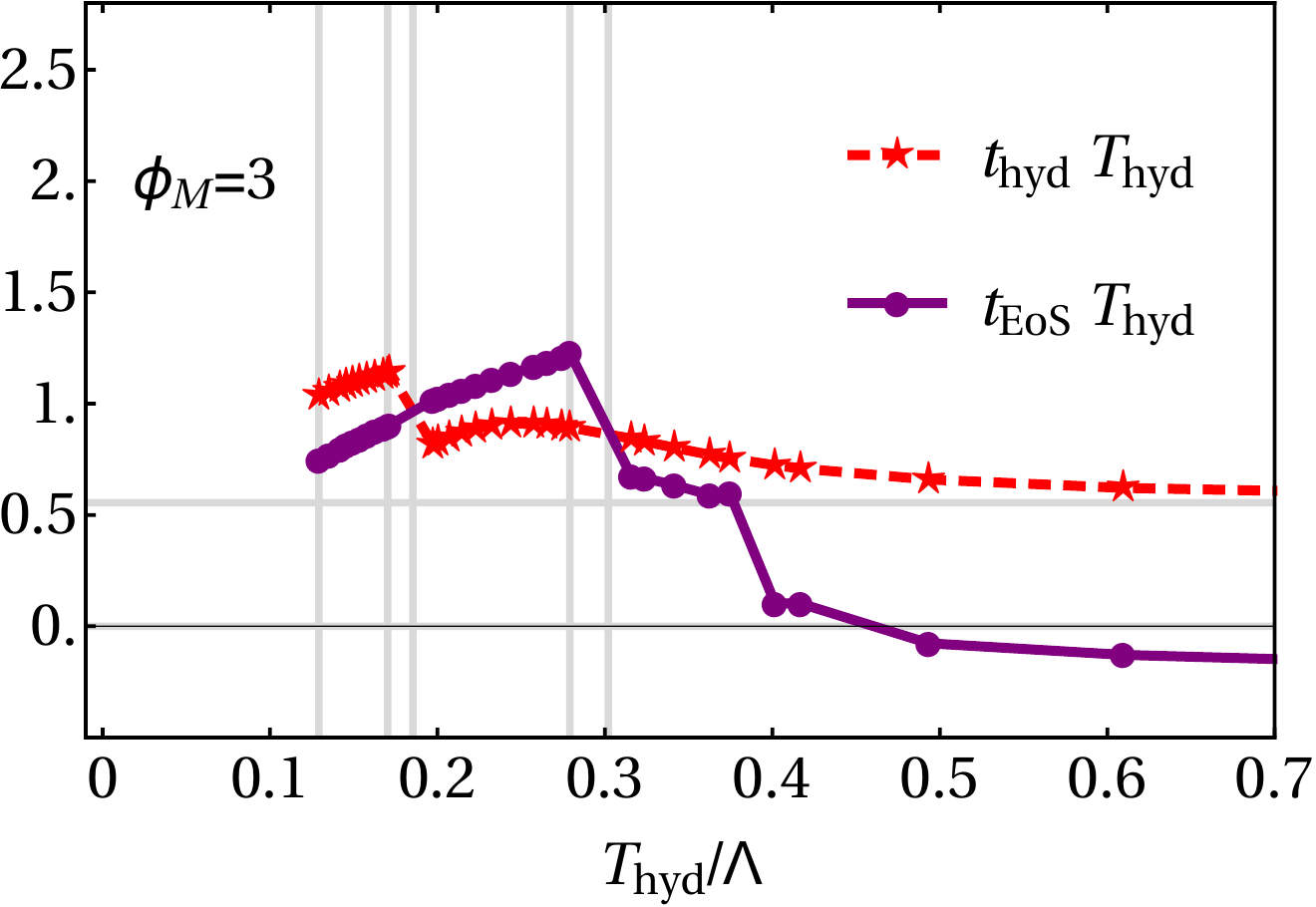}
& \includegraphics[width=0.48\textwidth]{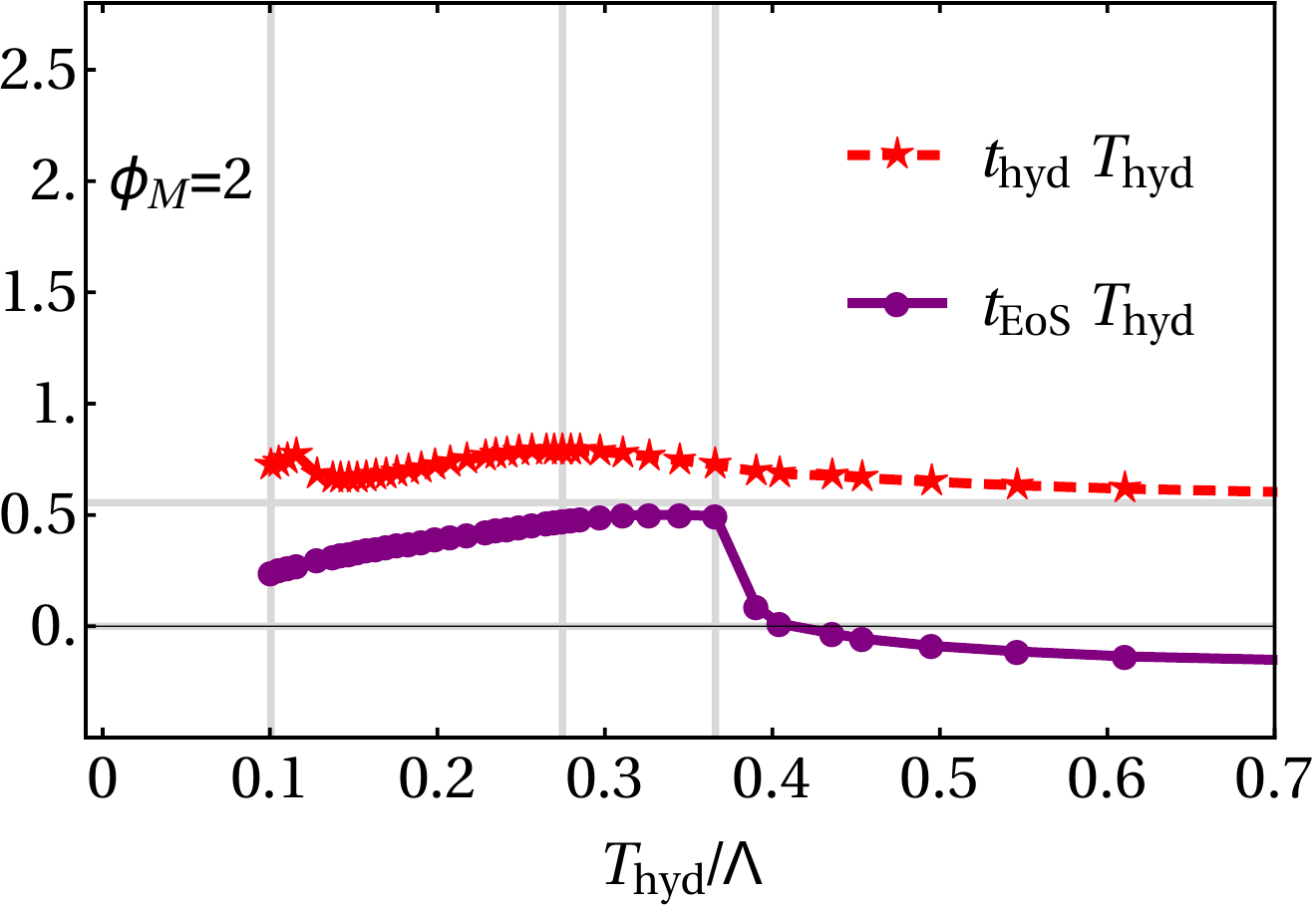}
% & \includegraphics[width=0.45\textwidth]{figs/times_T_phiM1}
\end{tabular}
\caption[]{ Hydrodynamization and EoSization times as a function of the hydrodynamization temperature for collisions of shocks with $\mu \omega=0.32$ for $\phiM = \{20, 5 , 3, 2\}$. The horizontal grey line lies at $\thyd \Thyd = 0.56$ and  corresponds to the conformal limit of the $1/2$ shocks. From left to right, the first three vertical grey lines indicate the hydrodynamization temperatures for the collisions with the minimal value of  $\Thyd/\Lambda$, the maximum value of $\thyd \Thyd$, and the maximum value of the ratio $t_{\textrm{EoS}} / \thyd$. The rightmost vertical grey line indicates the high-temperature crossing at which $t_{\textrm{EoS}}=\thyd$. The temperatures and the values of the $\zeta / \eta$ ratio at these vertical lines for each value of $\phi_M$ is as follows. 
For $\phi_M=20$ 
we have \mbox{$\Thyd / \Lambda = \{ 0.141, 0.184, 0.346, 0.374 \}$} 
and 
\mbox{$\zeta / \eta = \{0.31,0.36,0.30,0.22\}$}. 
For $\phiM = 5$ 
we have  
$\Thyd / \Lambda = \{0.129, 0.193, 0.202, 0.322, 0.366\}$ 
and 
$\zeta / \eta = \{0.23,0.31,0.32,0.26,0.22\}$. 
For $\phiM =3$ we have  
\mbox{$\Thyd / \Lambda = \{0.129, 0.170, 0.185, 0.279, 0.302\}$} 
and 
\mbox{$\zeta / \eta = \{0.16,0.21,0.23,0.26,0.25\}$}. 
And  for $\phiM =2$ 
we have  $\Thyd / \Lambda = \{0.101, 0.275, 0.366 \}$ 
and
$\zeta / \eta = \{0.05,0.18,0.17\}$. 
%
% The vertical grey lines are hydrodynamization temperatures of $\Thyd = \{0.100, 0.291, 0.364 \}$ with $\zeta / s = \{ 0.00021, 0.0036, 0.0050 \}$.
%
\label{fig:tTbulkall} }
\end{figure}
%%%%%%%%%%%%%%%%%%%%%%%%%%%%%%%%%%%%%%%%%%%%%%%%%%%%%%%%%%%%%%%%%%%%%%%%%%%%%%%

In figure~\ref{fig:tTbulkall} we plot the hydrodynamization time, $\thyd$,  (red dashed line with stars for each run) and the EoSization time, $t_{\rm EoS}$, (purple full line with dots for each run) for the different non-conformal theories. There are two observable effects. First, the hydrodynamization time increases with the non-conformality. Second,  hydrodynamization can happen before EoSization.
The conformal value of the hydrodynamization time is indicated in each panel of figure~\ref{fig:tTbulkall} with a horizontal line at $\thyd \Thyd = 0.56$. 
For a slightly non-conformal theory the increase of $\thyd$ with respect to the conformal value is minimal, as illustrated in the $\phiM = 2$ temperature scan. In this case the maximal increase is just a factor of $1.43$ larger than the conformal value.
For $\phiM = 3$ ($\phiM = 5$) the increase of the hydrodynamization time is a factor  of $2.05$ ($2.38$), and it takes place for a collision for which the ratio $\zeta/ \eta$ at the  
time of hydrodynamization is $0.21$ ($0.31$).
The expected increase of the hydrodynamization time is maximal in the $\phiM = 20$ temperature scan. In this case the maximum occurs for a collision with  $\Thyd / \Lambda = 0.184$ and the increase is a factor of $2.6$ with respect to the conformal  result. This maximal hydrodynamization time is reached with a bulk viscosity over entropy density ratio of $\zeta / \eta \approx 0.36$. We have verified that the  $\phiM = 20$ results are almost identical to those corresponding to $\phiM = 30$ or $\phiM = 100$. This is consistent with the fact that thermodynamic and transport properties such as the bulk viscosity and the speed of sound squared saturate with big positive values of $\phiM$.

We see that for sufficiently large $\mu/\Lambda$ the EoSization time becomes negative, meaning that the  average and the equilibrium pressures  differ by less than 10\% even before the shocks collide. The reason is simply that in these cases the energy density in the Gaussian tails in front of the shocks, which start to overlap at negative times, becomes much higher than $\Lambda$. At these energy densities the physics becomes approximately conformal and  the equation of state becomes approximately valid as a consequence of this symmetry. 

The equilibrium pressure and the average pressure are not within $10\%$ of one  another for a wide range of runs with $\phiM = \{20, 5 ,3, 2\}$.
For runs for which the EoSization criterion is fulfilled at all post-collision times the extracted EoSization time is either null or negative. Those specific runs show negligible non-conformal effects for the created plasma. The reason for the sharp rise of the EoSization times at low temperatures is due to a cut-off effect of the fixed $10\%$ criterion. Runs with a slightly higher temperature do easily reach $>5\%$ non-conformal effects, but do not yet trigger a later EoSization time extraction. The shockwave literature~\cite{Chesler:2013lia,Casalderrey-Solana:2013aba} typically uses a hydrodynamization criterion between $15\%$ and $20\%$, whereas here we have  settled for $10\%$. We found that changing this number implies no qualitative changes to our conclusions.

Reference~\cite{Attems:2016tby} showed that, in the model with  $\phiM =10$,  hydrodynamization precedes EoSization for collisions in a certain range of hydrodynamization temperatures. Figure~\ref{fig:tTbulkall}  shows that this also happens in the models with $\phiM = \{20, 5 ,3\}$. For $\phiM = 20$ this ordering is maintained up to the highest hydrodynamization temperature, $\Thyd / \Lambda \approx 0.37$. Since the models with  $\phiM = 20$ and $\phiM = 5$ have approximately the same thermodynamic properties~\cite{Attems:2016ugt} at $T/\Lambda \approx 0.4$, at the crossing of the EoSization and the hydrodynamization times the bulk viscosity-to-entropy ratio is also approximately the same,  $\zeta / \eta \approx 0.22$.  For $\phiM = 3$ one notices that at the high-temperature crossing between the two  times the ratio is  $\zeta / \eta \approx 0.25$. Models with $\phiM =2$ and $\phiM=1$ (not shown explicitly) show no crossing. For $\phiM = 2$ the maximal ratio is $\zeta / \eta \approx 0.18$. For $\phiM = 5$, the low-temperature crossing has $\zeta / \eta \approx 0.32$ and for $\phiM = 3$ one gets $\zeta / \eta \approx 0.23$, but the lower crossing is not yet reached with $\phiM = 20$ at the minimal temperature with $\zeta / \eta \approx 0.31$. These differences are explained by the accumulating effects of the bulk viscosity along the entire evolution of the collision. We therefore confirm our prior conservative estimate \cite{Attems:2016tby} of $\zeta / s \gtrsim 0.025$ in order to have  hydrodynamization before EoSization.

Furthermore, the maximal value for the ratio $t_{\rm EoS} / \thyd \approx \{0.684,1.10,2.07,2.87\}$ with $\phiM = \{2,3,5,20\}$ %and 1/2 collision
 is reached at $\Thyd / \Lambda \approx \{0.275, 0.279, 0.322, 0.346 \}$. The resulting temperature of the maximal values, when comparing the different non-conformal theories, increases. This again shows evidence for the accumulating effect of the bulk viscosity. It is important to stress that the non-conformal equation of state has to be taken into account for more than twice the hydrodynamization time with a bulk viscosity over entropy ratio of $\zeta / s \geq 0.025$.

%%%%%%%%%%%%%%%%%%%%%%%%%%%%%%%%%%%%%%%%%%%%%%%%%%%%%%%%%%%%%%%%%%%%%%%%%%%%%%%
\subsection{Dynamics of the scalar condensate}
\label{scalar}
\begin{figure}[htbp]
\begin{center}
\includegraphics[width=0.8\textwidth]{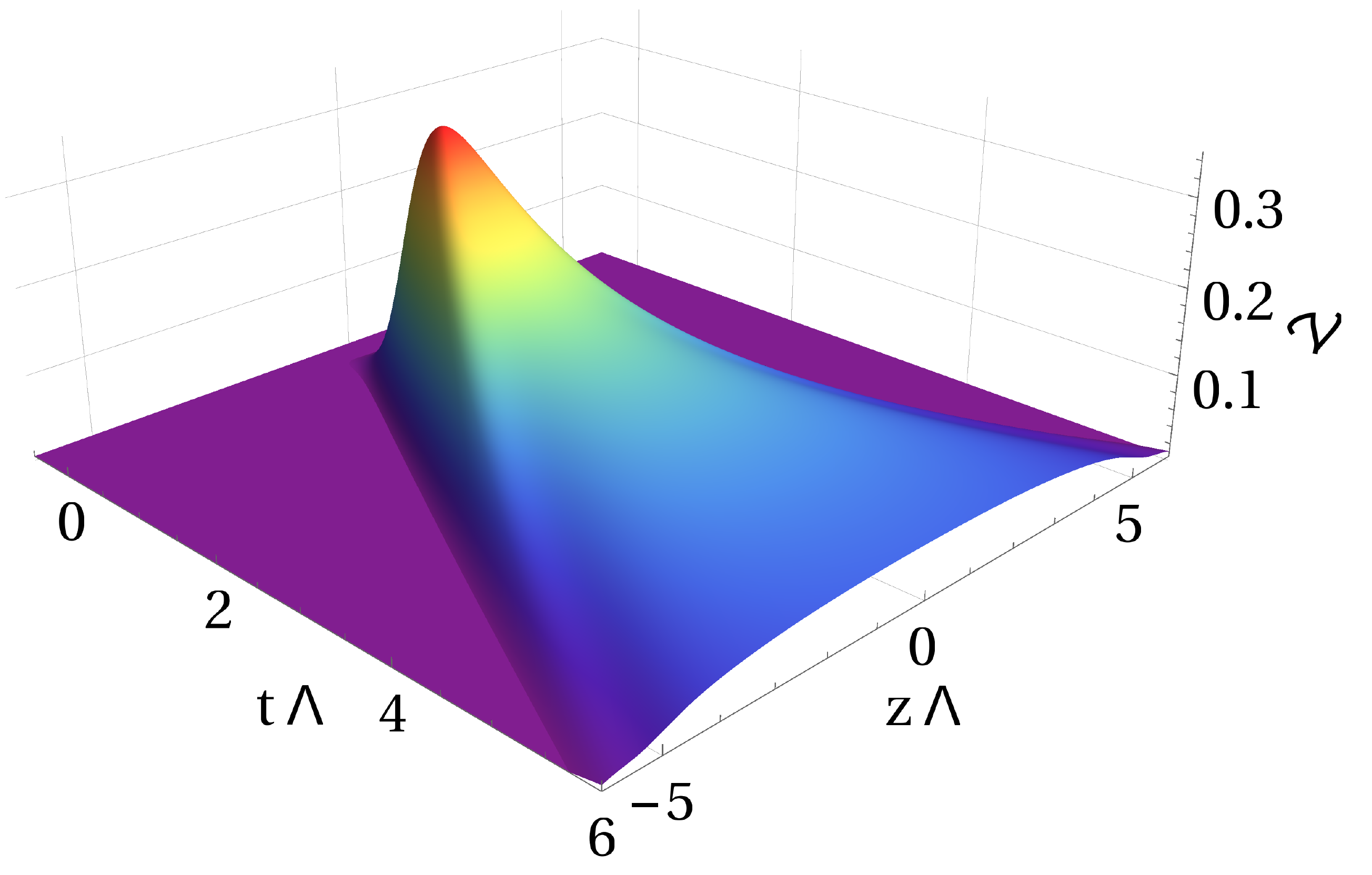}  
\caption[]{Spacetime evolution of the scalar condensate, in units of $\Lambda^3$, in a collision in the $\phiM= 20$ model with  $\mu \omega= 0.30$ and $\mu/\Lambda=0.93$. 
	 }
\label{3Dvev}  
\end{center}
\end{figure}

As we have seen, the dynamics of  the  longitudinal, transverse and  average pressures provides  information about different process in the evolving plasma. 
To clarify further the process of EoSization we focus here on the evolution of the scalar expectation value $\mathcal{V}$, since   
inspection of \eqn{eq:Ward} and \eqn{eq:Peq} shows
that EoSization is in part controlled by how $\mathcal{V}$ approaches its equilibrium value. In figure \ref{3Dvev} we show the spacetime evolution of the condensate for a characteristic collision.

In analogy with the hydrodynamization and EoSization times,  
we define the condensate relaxation time $\tcond$ as
the time beyond which the normalized difference of the true expectation value  of the scalar operator, $\mathcal{V}$, and its equilibrium value $\mathcal{V}_{\rm eq}$, is less than 10\%: 
\begin{equation}
\label{equ:criteria_O}
\left|	\frac{ {\mathcal V} - {\mathcal V}_{\rm eq}  }{ {\mathcal V} }\right| < 0.1 \,.
\end{equation}
This time is a measure of how fast this one-point function reaches its equilibrium value.  We explore  $\tcond$ for the different collision configurations studied in the previous section.

\begin{figure}[tbp]
\begin{tabular}{cc}
{\small 1. EoS $\to$ Hyd $\to$ Cond} & {\small 2. Hyd $\to$ EoS $\to$ Cond} \\
\includegraphics[width=0.48\textwidth,height=0.33\textwidth]
{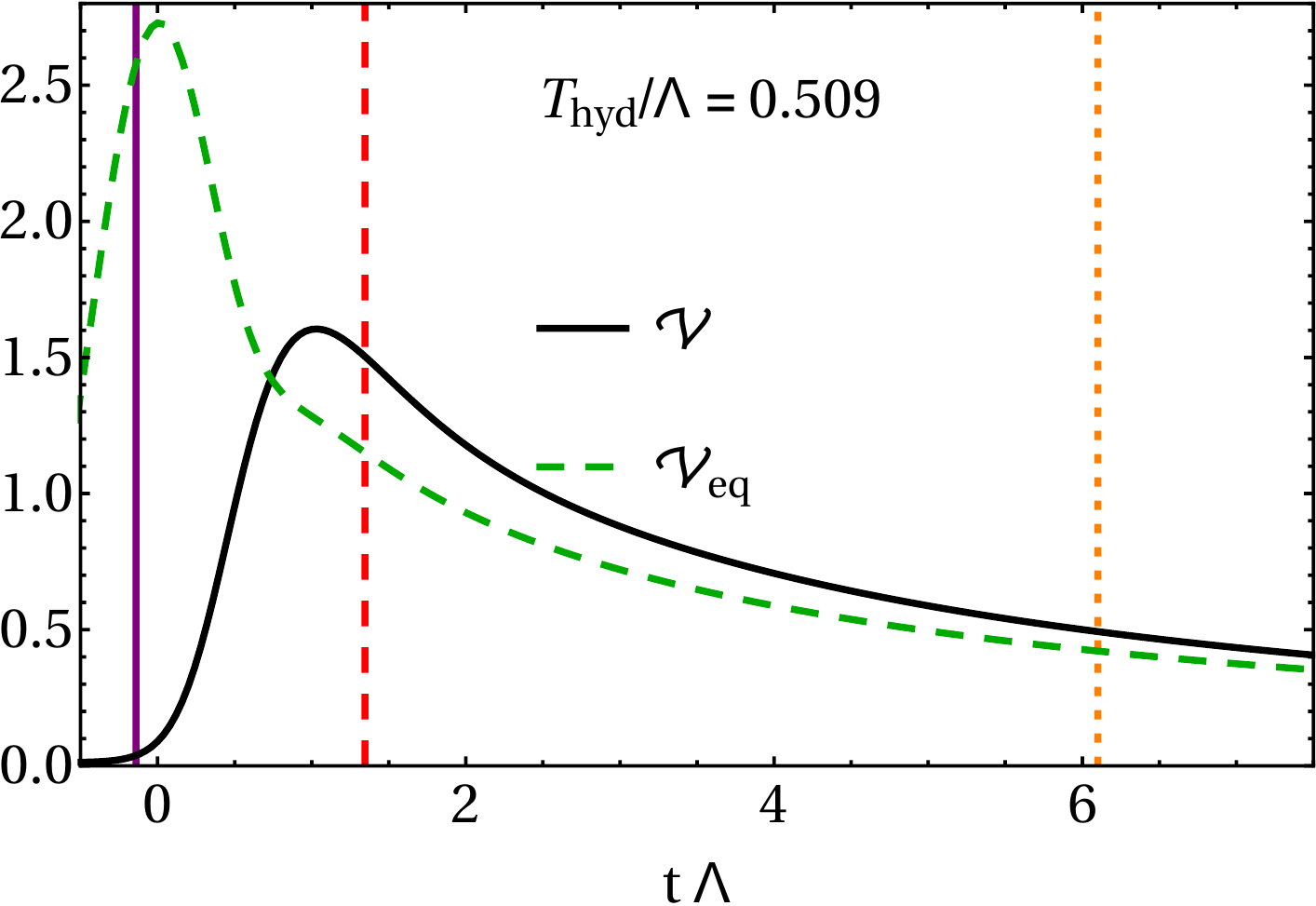}
& 
\includegraphics[width=0.48\textwidth,height=0.33\textwidth]
{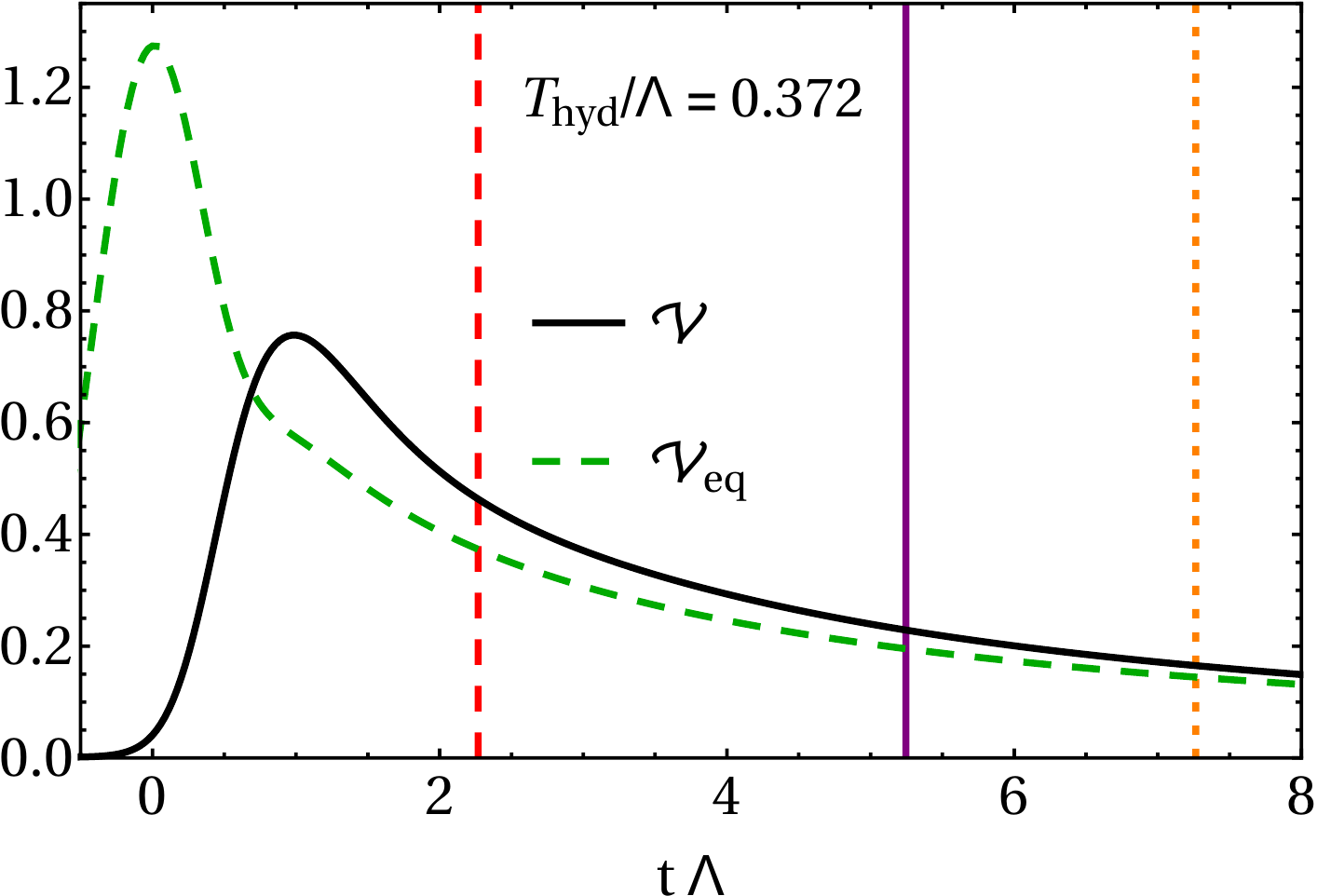} 
%\put(-370,145){1. EoS $\to$ Hyd $\to$ Cond}
%\put(-170,145){2. Hyd $\to$ Eos $\to$ Cond}
\\
%\vspace{5mm}
\\
{\small 3. Hyd $\to$ Cond $\to$ EoS} & {\small 4. Cond $\to$ Hyd $\to$ EoS} \\
\includegraphics[width=0.48\textwidth,height=0.33\textwidth]
{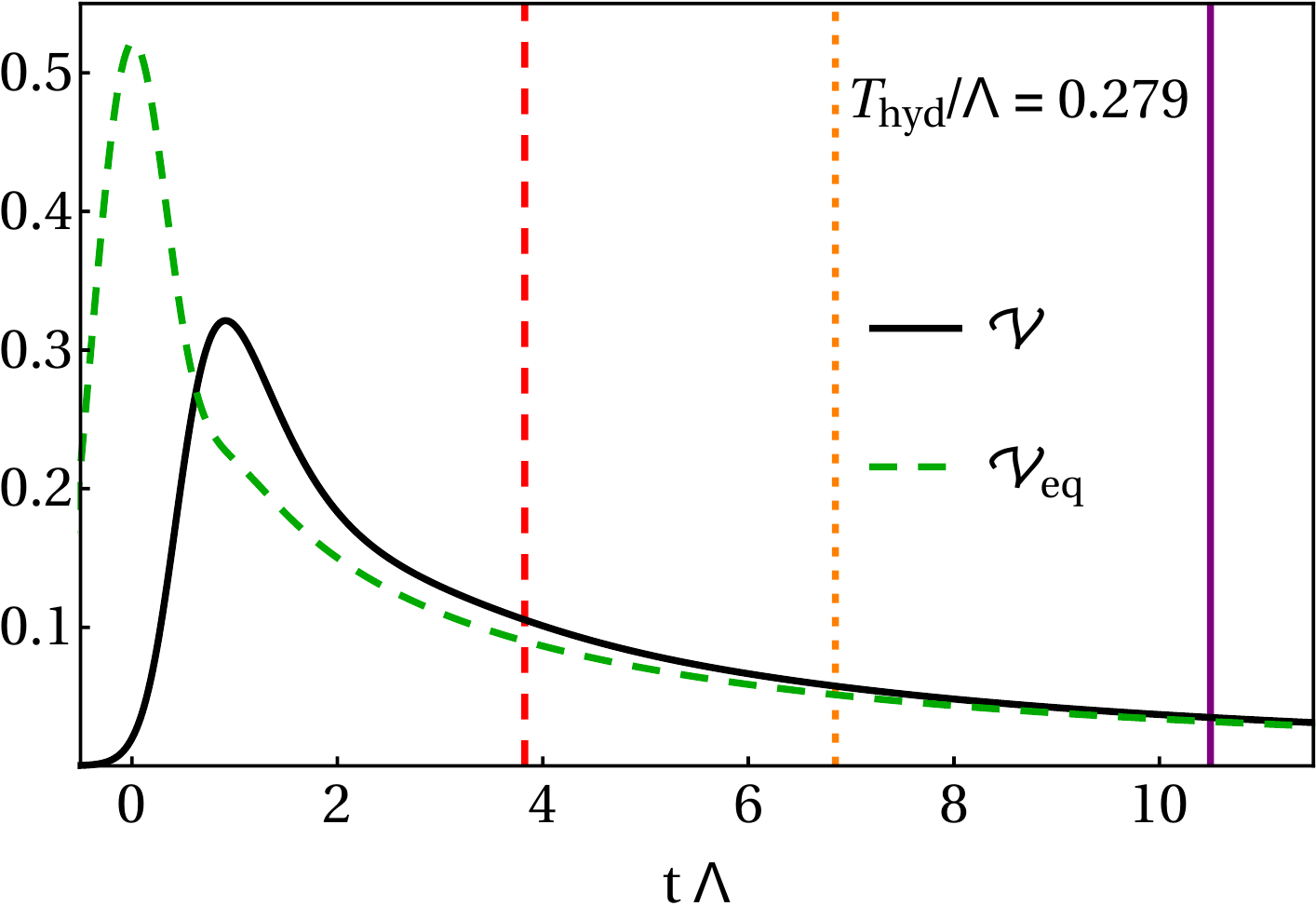}
& 
\includegraphics[width=0.48\textwidth,height=0.33\textwidth]
{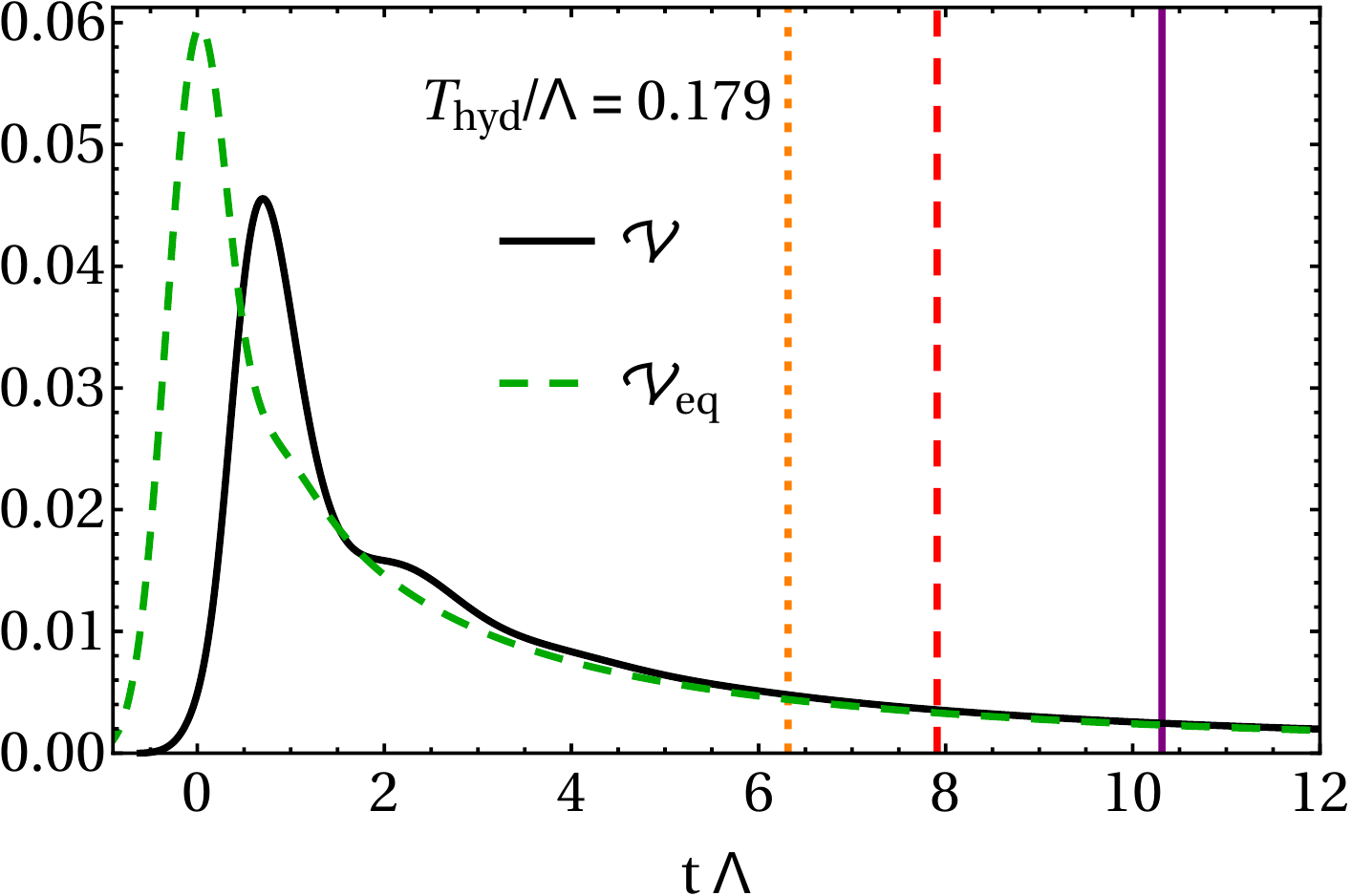}
\end{tabular}
	\caption[]{Comparison between the time evolution of the true scalar condensate $\mathcal{V}$ at $z=0$ and 	the equilibrium value $\mathcal{V}_\mtt{eq}(\mathcal{E})$ that would correspond to the instantaneous energy density,  in units of $\Lambda^3$, for collisions in the model with $\phiM = 20$ with $\mu \omega= 0.30$ and $\mu/\Lambda=\{ 1.85,1.33,0.93,0.46\}$ from top to bottom and from left to right. The vertical lines indicate the hydrodynamization time (red dashed), the EoSization time (purple solid) and the condensate relaxation  time (orange dotted). These times, in units of $1/\Lambda$,  take the following values in each panel. 
\mbox{(1) $t_{\textrm{EoS}}  = -0.134  < 0 < t_{\textrm{hyd}}  =  1.34 < \tcond  = 6.10 $.}  
\mbox{(2) $\thyd  = 2.27 < t_{\rm EoS}  = 5.25  < \tcond  = 7.26$.}
\mbox{(3) $\thyd  = 3.82 < \tcond  = 6.85 < t_{\rm EoS}  = 10.5$.}
\mbox{(4) $\tcond  = 6.31 < \thyd  = 7.91 < t_{\rm EoS}  = 10.3$.}
\label{fig:vev}}
\end{figure}

The scalar condensate is fully out of equilibrium for most of the studied shockwave collisions. In most cases,  the condensate $\mathcal{V}$ takes a much longer time to equilibrate than the system takes  to hydrodynamize. With three different relaxation times---hydrodynamization, EoSization and condensate relaxation---one can in theory find six possible orderings between these times. 
%However, since when the condensates reaches its equilibrium value the equation of state is satisfied, the condensate relaxation time must be always larger or equal than the 
However, we have found  no configuration in which hydrodynamization comes last.  Since there seems to be no obstacle of principle for this, the reason is presumably that our collisions do not generate a sufficiently large anisotropy; we will come back to this point in section \ref{sec:discussion}. 
% as the scalar relaxation only crosses below hydrodynamization times in the very non-conformal regime of $\phiM=20$, which sets the EoSization time to be last.
Thus, our models give rise to  the following four orderings:
\begin{enumerate}
 \item
 EoSization $\to$ Hydrodynamization $\to$ Condensate relaxation,
 \item 
 Hydrodynamization $\to$ EoSization  $\to$  Condensate relaxation,
 \item 
 Hydrodynamization  $\to$ Condensate relaxation $\to$  EoSization, 
\item
 Condensate relaxation $\to$  Hydrodynamization $\to$  EoSization. 
 \end{enumerate}
Each of these cases is illustrated by one of the plots in figure~\ref{fig:vev}. 
The fact that hydrodynamization and EoSization can happen in any order was the main result of \cite{Attems:2016tby}. Here we see that the situation is richer once condensate relaxation is included.  We note in figure~\ref{fig:vev} that the energy of the collision, $\mu/\Lambda$, or equivalently the hydrodynamization temperature, $\Thyd/\Lambda$, decrease monotonically from case 1 to case 4. The reason is that at $T\gg\Lambda$ the condensate grows as 
$\Lambda \mathcal{V} \sim \Lambda^2 T^2$ \cite{Attems:2016ugt}, whereas the stress tensor grows as $T^4$. As a consequence  the relative magnitude of the $\mathcal{V}$-induced correction in the average pressure through the Ward identity \eqq{eq:Ward} 
decreases and the dynamics of the condensate decouples from the dynamics of the stress tensor. Indeed, using equations \eqq{eq:Ward}, \eqq{eoseos} and \eqq{eq:Peq} we see that 
\be
\label{dividing}
3\left(\bar P - \bar P_{\rm eq}\right) = 
\Lambda \left( \mathcal{V} - \mathcal{V}_{\rm eq} \right) \,.
\ee
For EoSization to take place the left-hand side must be small in units of $\bar P$. Dividing this equation by $\bar P$ and using the scalings above we find that at  high temperature  
\be
\frac{\bar P - \bar P_{\rm eq}}{\bar P} = 
\frac{\Lambda}{3} \frac{\left( \mathcal{V} - \mathcal{V}_{\rm eq} \right)}{\bar P} \sim
\frac{\Lambda^2}{T^2} \ll 1 \,.
\ee
In this regime the  scalar condensate can still  very far from its equilibrium value according to our criterion \eqq{equ:criteria_O}, since all terms in \eqq{equ:criteria_O} scale as $\Lambda T^2$.  In conclusion, at high temperature 
the equation of state approaches the conformal equation of state and the Ward identity is no impediment for the system to EoSize (and hydrodynamize) while the scalar condensate is still far from its equilibrium value. This possibility is realized in the first two plots of figure~\ref{fig:vev}. Note that the hydrodynamization temperature in these cases is certainly not asymptotically high, but it is higher than the temperature at which the non-conformal effects are maximal, which for $\phiM=20$ is $T \sim 0.2 \Lambda$, as indicated in the caption of figure \ref{fig:zetaosT}. This seems to suffice for the asymptotic argument above to apply. 

In contrast, at $T\sim 0.2 \Lambda$ the product $\Lambda \mathcal{V}$ can be numerically  larger than  $3 \bar{P}$.  For this reason it is possible for the right-hand side of \eqq{dividing} to be smaller than 0.1 in units of $\Lambda \mathcal{V}$ while the left-hand side is larger than 0.1 in units of $\bar P$. This is why at temperatures at which non-conformal effects are sufficiently large scalar relaxation can precede EoSization (and also hydrodynamization, since the latter can precede EoSization). This is illustrated by the last two plots in figure~\ref{fig:vev}, for which the hydrodynamization temperature is close to the value at which non-conformal effects are maximal. 

%%%%%%%%%%%%%%%%%%%%%%%%%%%%%%%%%%%%%%%%%%%%%%%%%%%%%%%%%%%%%%%%%%%%%%%%%%%%%%%
\begin{figure}[htbp]
\begin{tabular}{cc}
\includegraphics[width=0.48\textwidth,height=0.34\textwidth]
{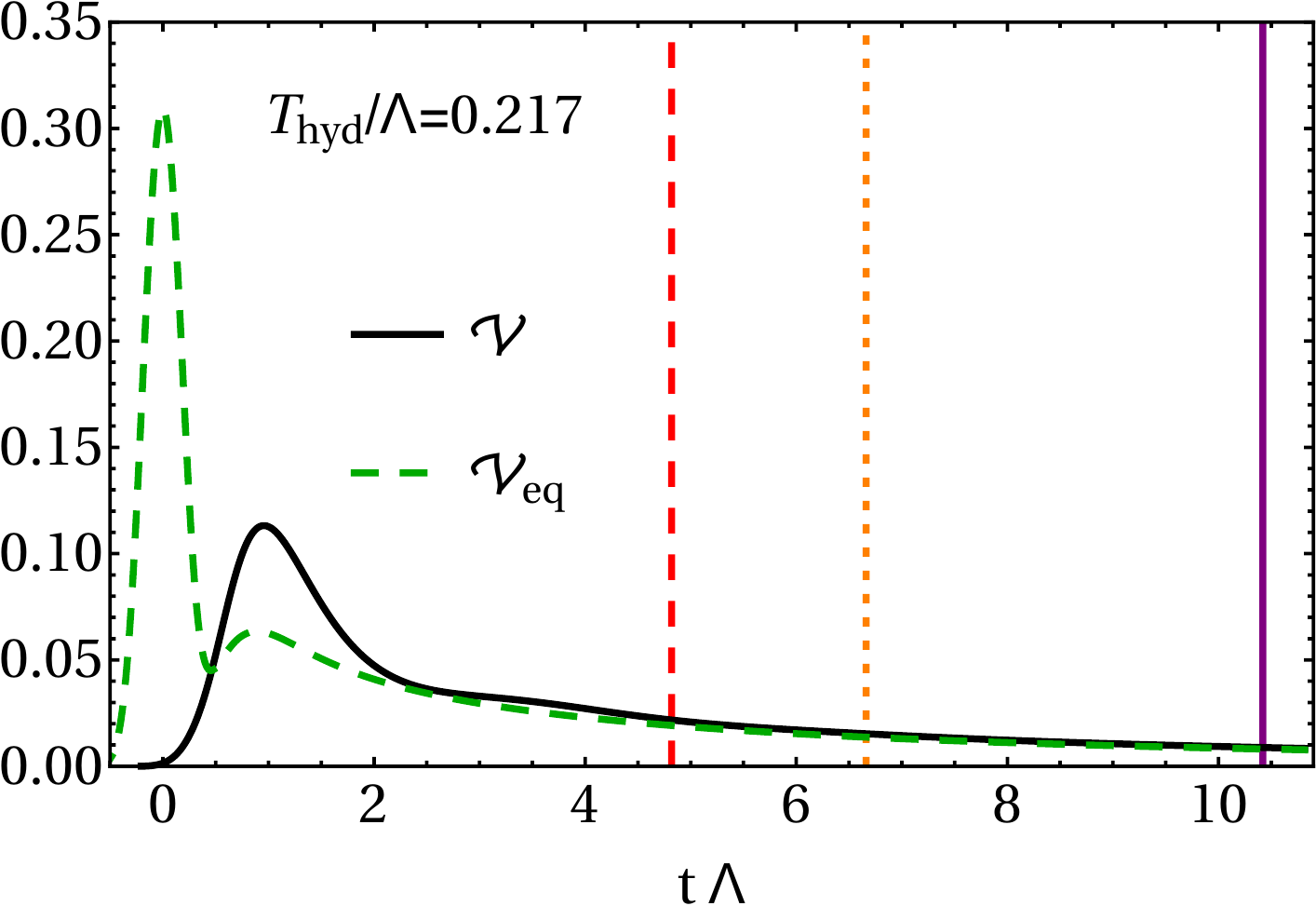}
& 
\includegraphics[width=0.48\textwidth,height=0.34\textwidth]
{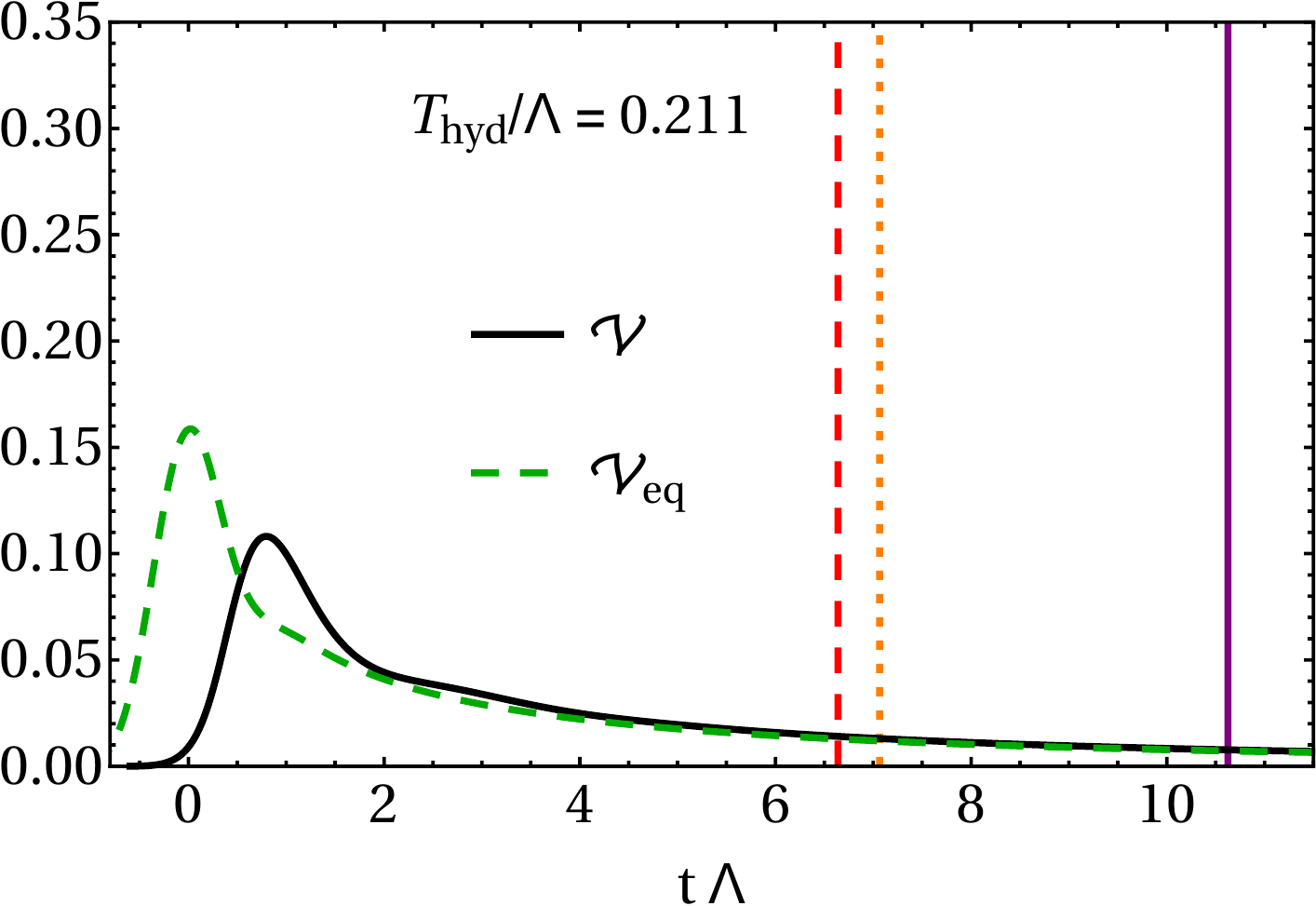}
\end{tabular}
	\caption[]{Comparison between the time evolution of the true scalar condensate $\mathcal{V}$ at $z=0$ and 	the equilibrium value $\mathcal{V}_\mtt{eq}(\mathcal{E})$ that would correspond to the instantaneous energy density,  in units of $\Lambda^3$, for collisions in the $\phiM=20$ model of shocks with the same transverse energy  
	\mbox{$\mu /\Lambda=0.62$} but different widths  $\mu \omega = 0.12$ (left) and  $\mu \omega = 0.30$ (right). 
The vertical  lines indicate the hydrodynamization time (red dashed), the EoSization time (purple solid) and the condensate relaxation time (orange dotted).  These times, in units of $1/\Lambda$,  take the following values in each panel. 
Left: $t_{\rm hyd}  = 4.82 < \tcond  = 6.66 < t_{\rm EoS}  = 10.42$.
Right: $t_{\textrm{hyd}}  = 6.64  < t_{\mathcal{V}}  =  7.07 < t_{\rm EoS}  = 10.6 $.
	\label{fig:vevquarterhalfs} }
\end{figure}
%%%%%%%%%%%%%%%%%%%%%%%%%%%%%%%%%%%%%%%%%%%%%%%%%%%%%%%%%%%%%%%%%%%%%%%%%%%%%%%

In figure~\ref{fig:vevquarterhalfs} 
we explore the relaxation dynamics for two collisions with the same incident transverse energy, $\mu / \Lambda=0.62$, but with different widths, $\mu \omega = 0.12$ (left) and $\mu \omega = 0.30$ (right).
In both cases, at late times $\mathcal{V}$ approaches its equilibrium value (green dashed) from above. 
As in figure \ref{fig:vev}, we see that  the equilibrium value $\mathcal{V}(\mathcal{E})$ begins to rise  before $t=0$ and reaches its maximum shortly after $t=0$. This is simply  because this value tracks the energy density, which begins to rise before $t=0$ because of the forward tails of the Gaussian shocks. Instead, the true condensate would be exactly undisturbed by a single shock, and therefore it begins to respond only once a significant amount of collision dynamics has taken place. For this reason, the true condensate begins to rise almost exactly at $t=0$. 
Shortly after the collision the spike in the 
equilibrium value reflects the initial large energy density of the passing shocks, which is larger in the narrower shocks since the transverse energy density is fixed. In contrast, the peak in the true condensate is very similar in both collisions. As in the conformal case \cite{Casalderrey-Solana:2013aba}, the final hydrodynamization temperature  is mostly  determined by the transverse energy scale, and therefore $\Thyd$ is almost identical for the two collisions. This is remarkable, since $\zeta/\eta$ at that  $\Thyd$ is almost maximal, indicating large non-conformal effects.   
%The fact that  $\thyd$ is also larger for the wider shocks is also consistent with the conformal case.  
We observe that a similar statement  holds true for the EoSization time, which is essentially the same in both cases, and less accurately but still approximately so for the relaxation times of the scalar condensate. It may be possible to understand these  effects as finite-resolution effects, as discussed in \cite{Casalderrey-Solana:2013sxa}.

%%%%%%%%%%%%%%%%%%%%%%%%%%%%%%%%%%%%%%%%%%%%%%%%%%%%%%%%%%%%%%%%%%%%%%%%%%%%%%%
\begin{figure}[htbp]
\begin{tabular}{cc}
\includegraphics[width=0.48\textwidth]{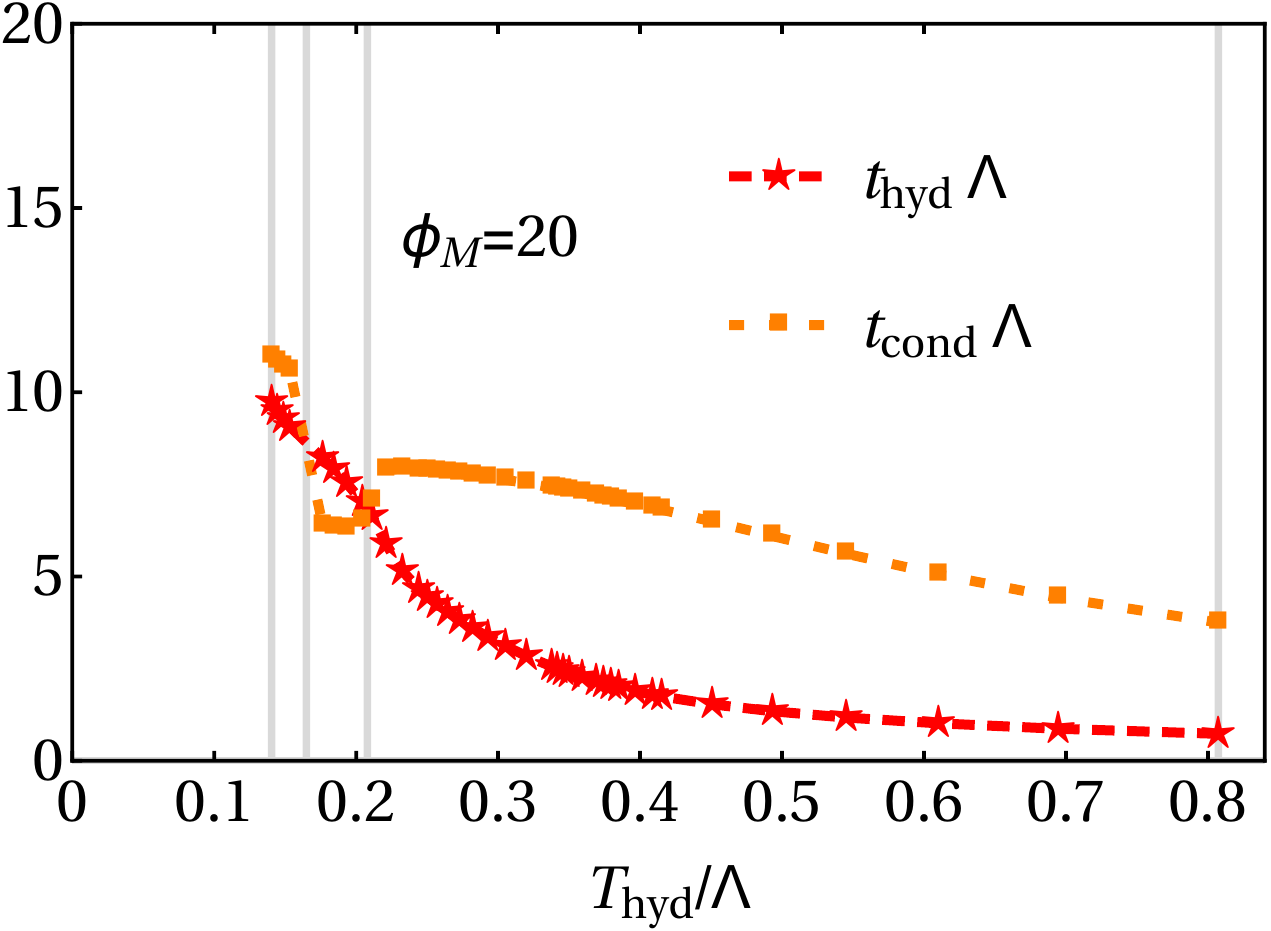}
& \includegraphics[width=0.48\textwidth]{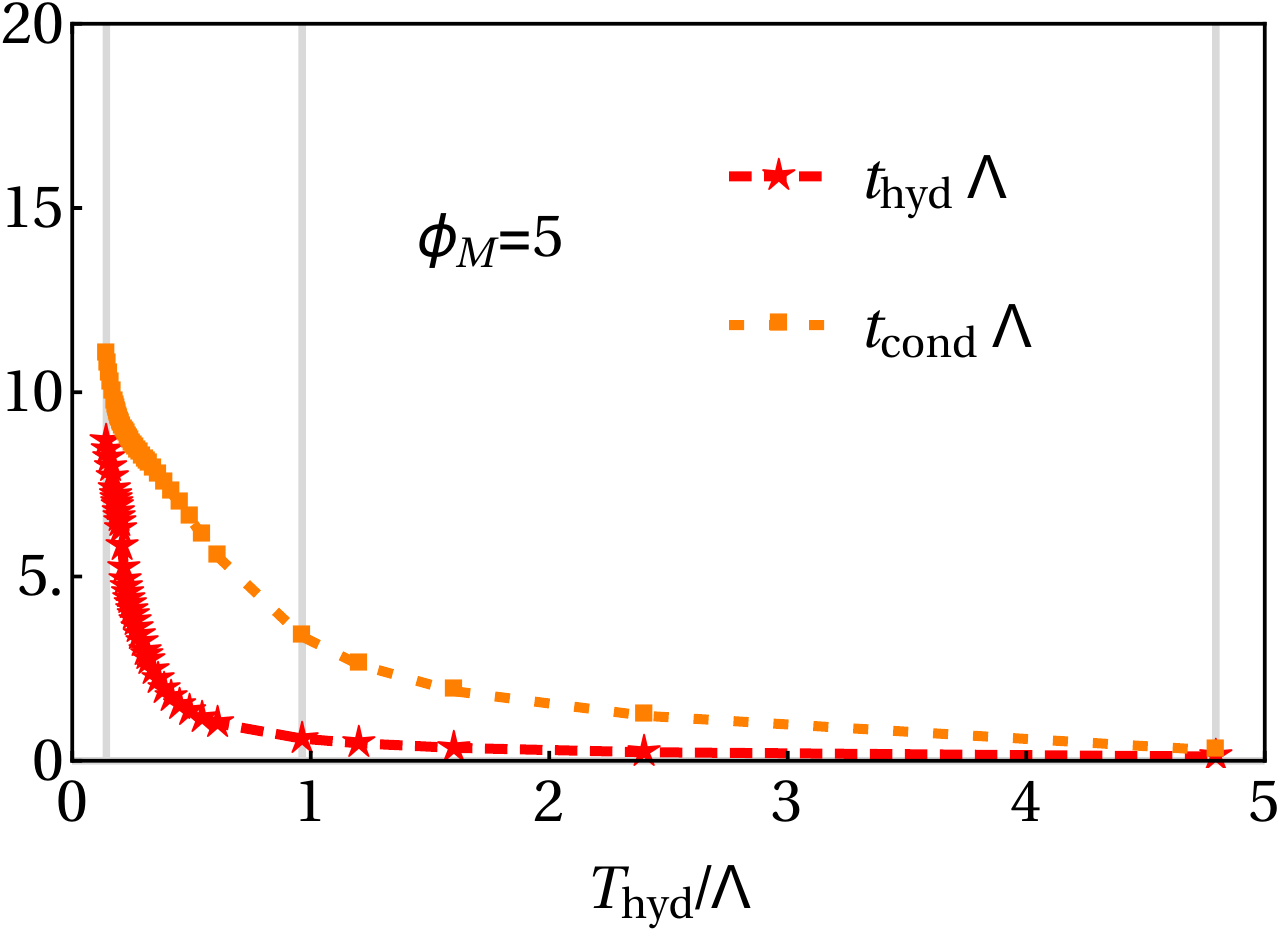} \\
\includegraphics[width=0.48\textwidth]{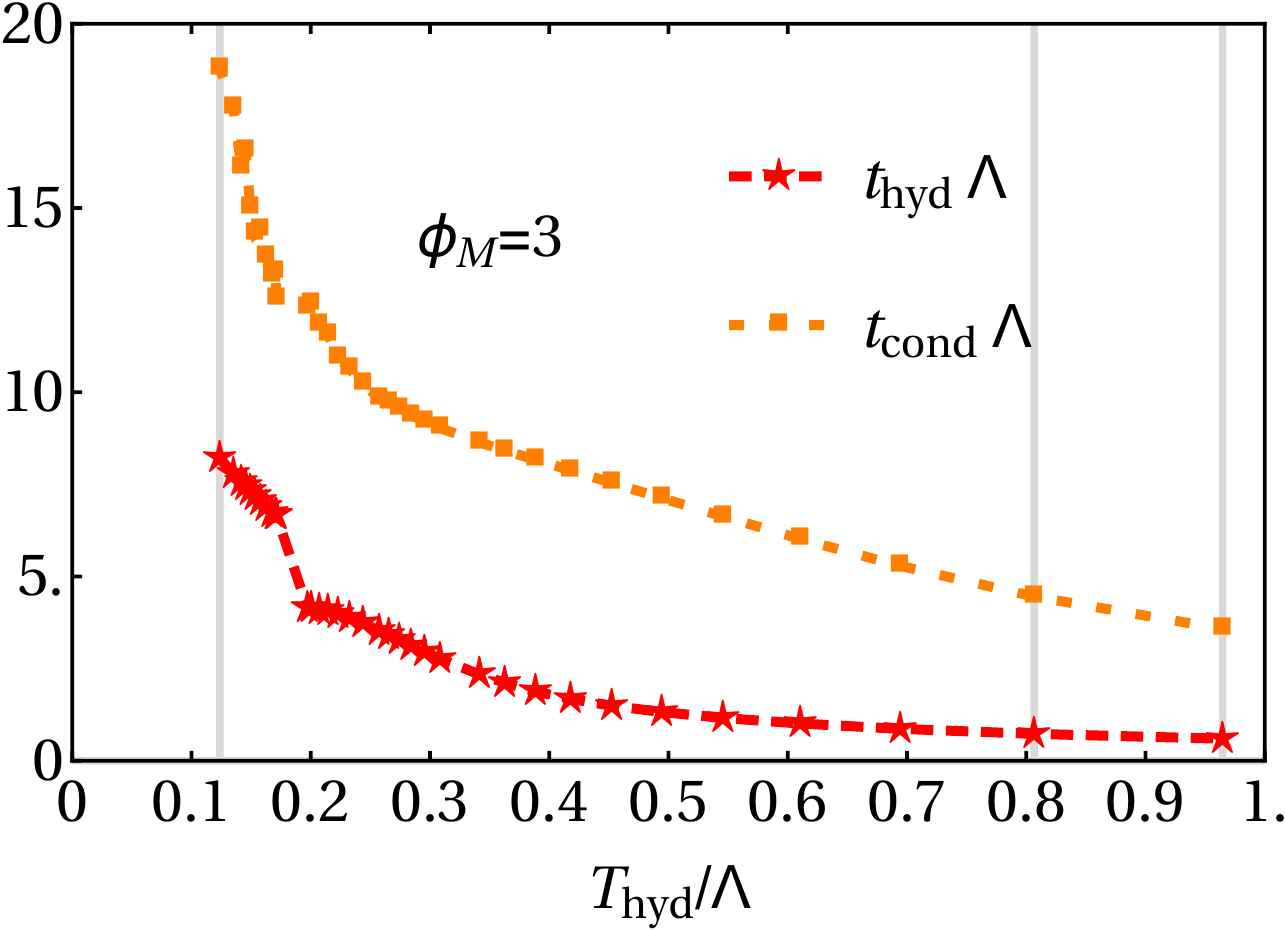}
& \includegraphics[width=0.48\textwidth]{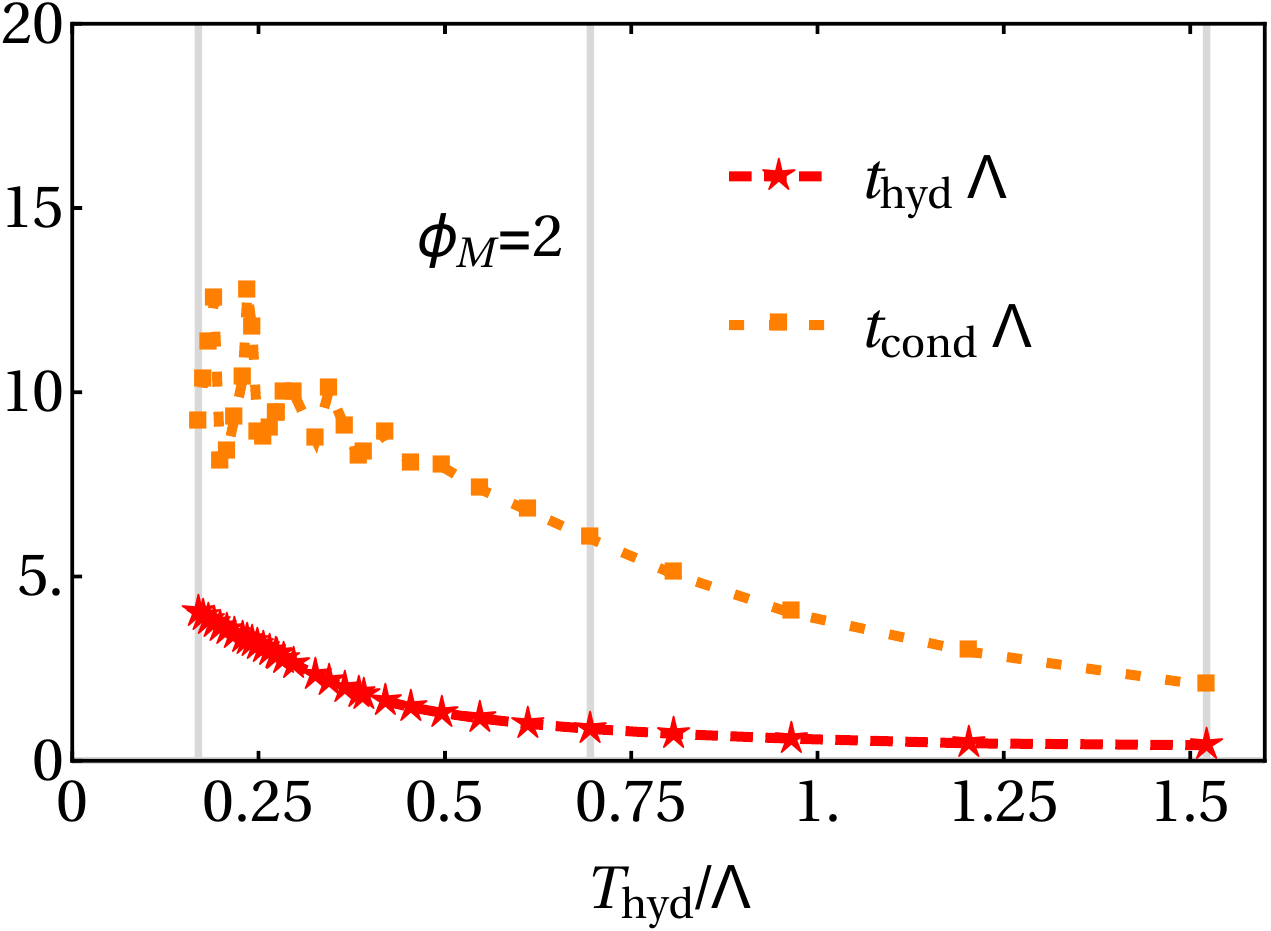}
\end{tabular}
	\caption{Condensate relaxation times and hydrodynamization times for collisions with $\mu \omega = 0.30$ in  models with $\phiM = \{20,5,3,2\}$. The leftmost (rightmost) grey vertical line indicates the lowest (highest) temperature that we probed. The other grey vertical lines indicate either points at which  
$\tcond = \thyd$ or points at which the ratio  $\tcond / \thyd$ is maximal. The positions of these lines in each panel is as follows. 
Top left: $\Thyd / \Lambda = \{ 0.141, 0.165, 0.208, 0.807 \}$. The highest temperature in this case is the one at which the ratio $\tcond / \thyd$ is maximal. Top right: $\Thyd / \Lambda = \{ 0.143, 0.964, 4.80\}$.
Bottom left: $\Thyd / \Lambda = \{ 0.124, 0.807, 0.965 \}$.
Bottom right:  $\Thyd / \Lambda = \{0.169,0.695,1.52\}$. 
\label{fig:tOL} }
\end{figure}

%%%%%%%%%%%%%%%%%%%%%%%%%%%%%%%%%%%%%%%%%%%%%%%%%%%%%%%%%%%%%%%%%%%%%%%%%%%%%%%
\begin{figure}[htbp]
\begin{tabular}{cc}
\includegraphics[width=0.48\textwidth]{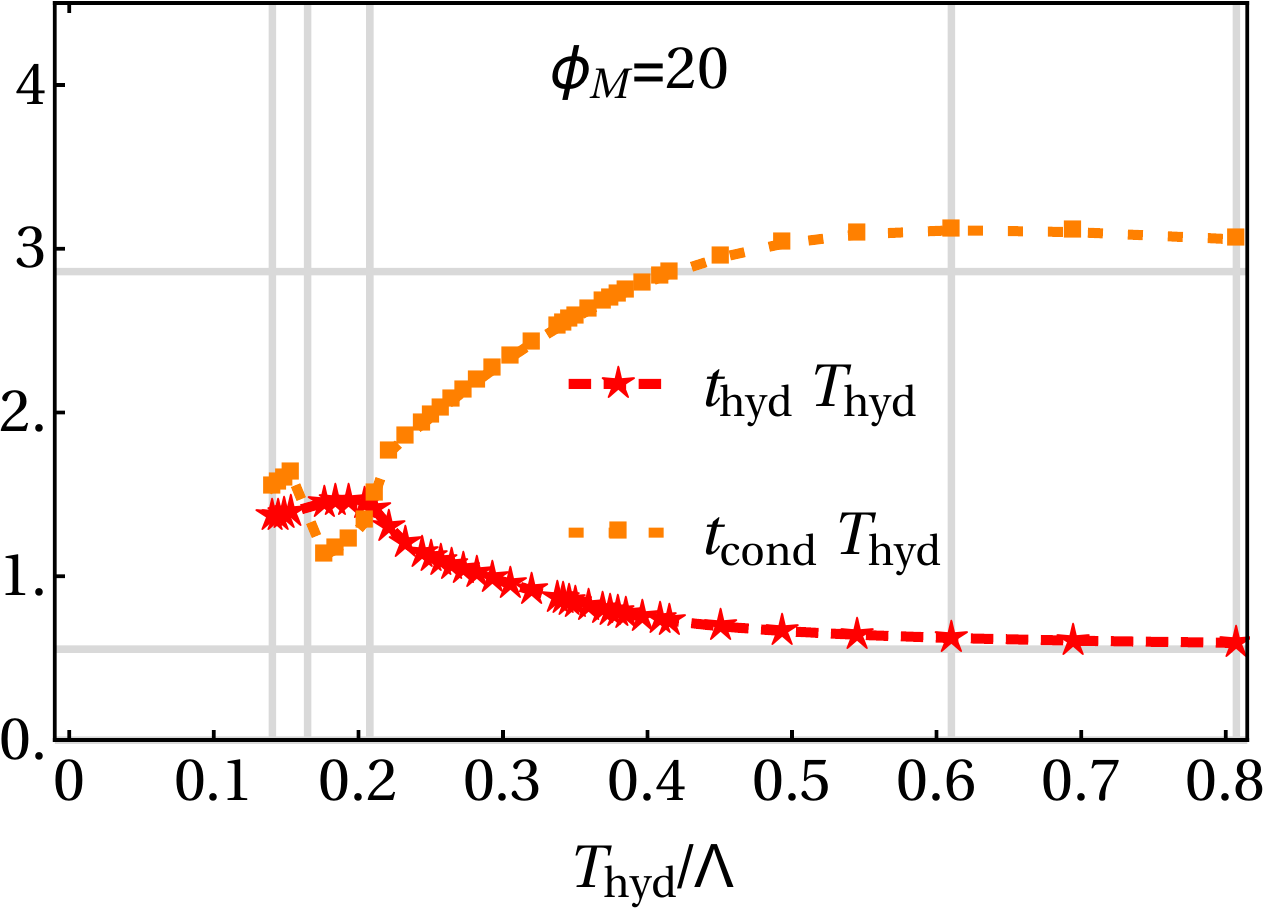}
& \includegraphics[width=0.48\textwidth]{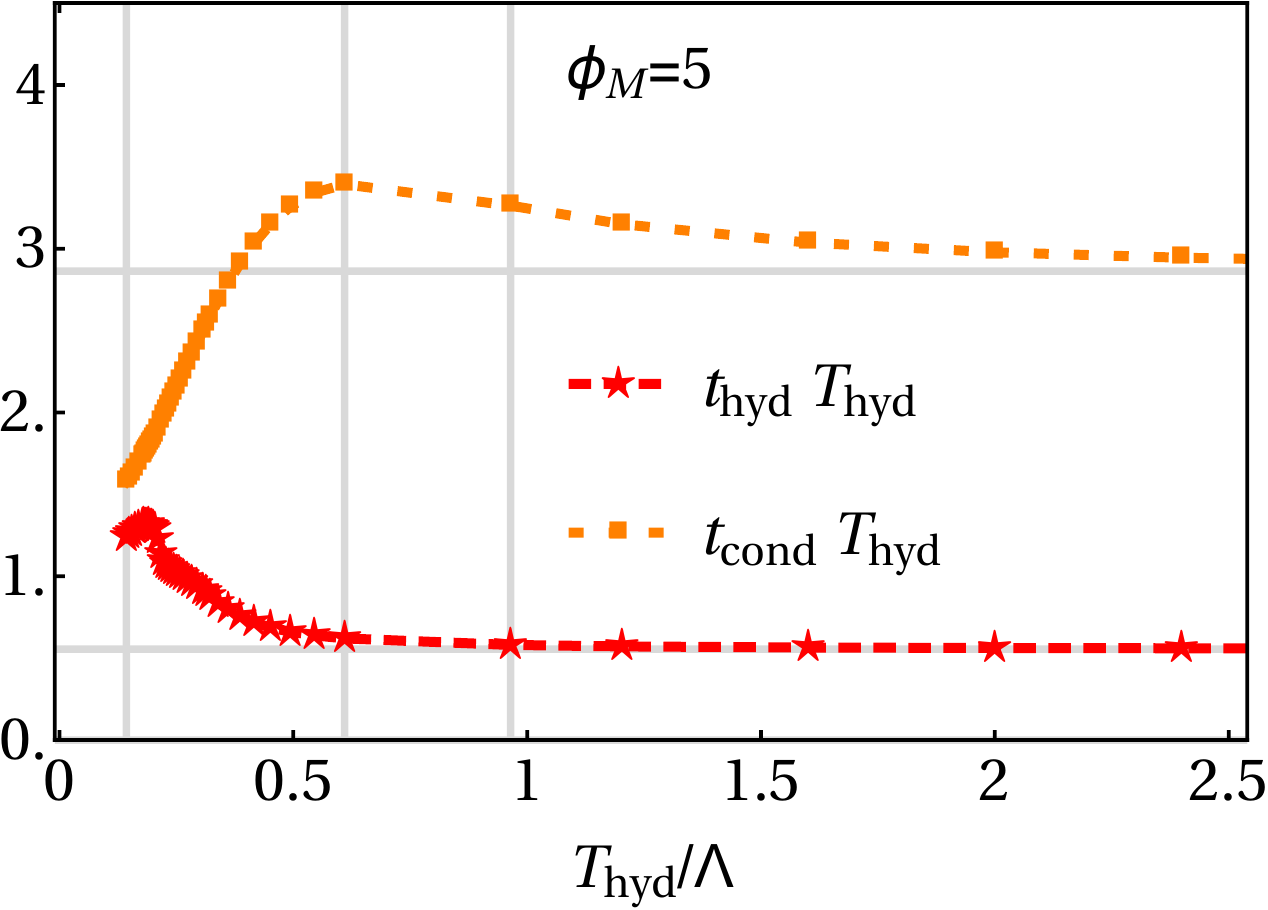} \\
\includegraphics[width=0.48\textwidth]{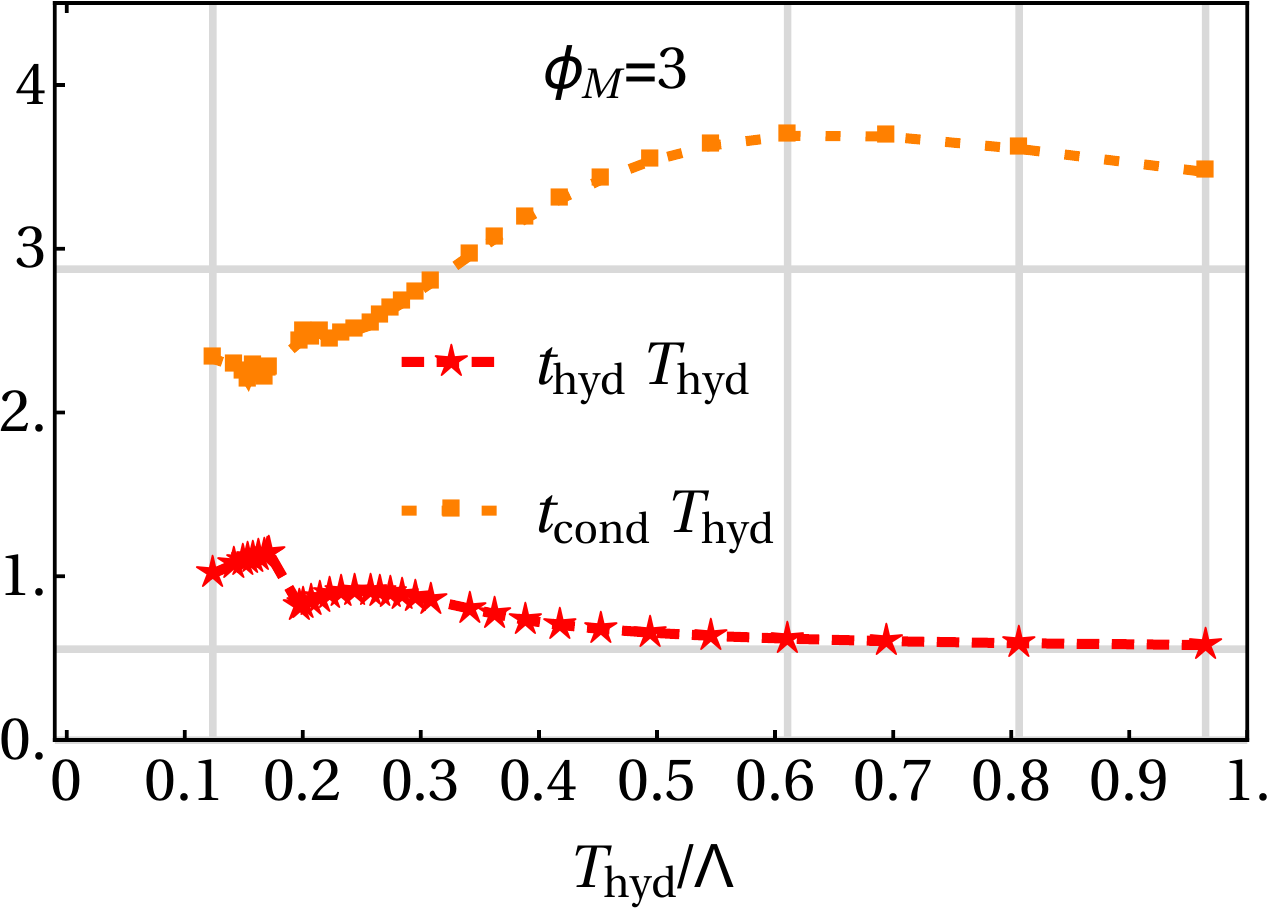}
& \includegraphics[width=0.48\textwidth]{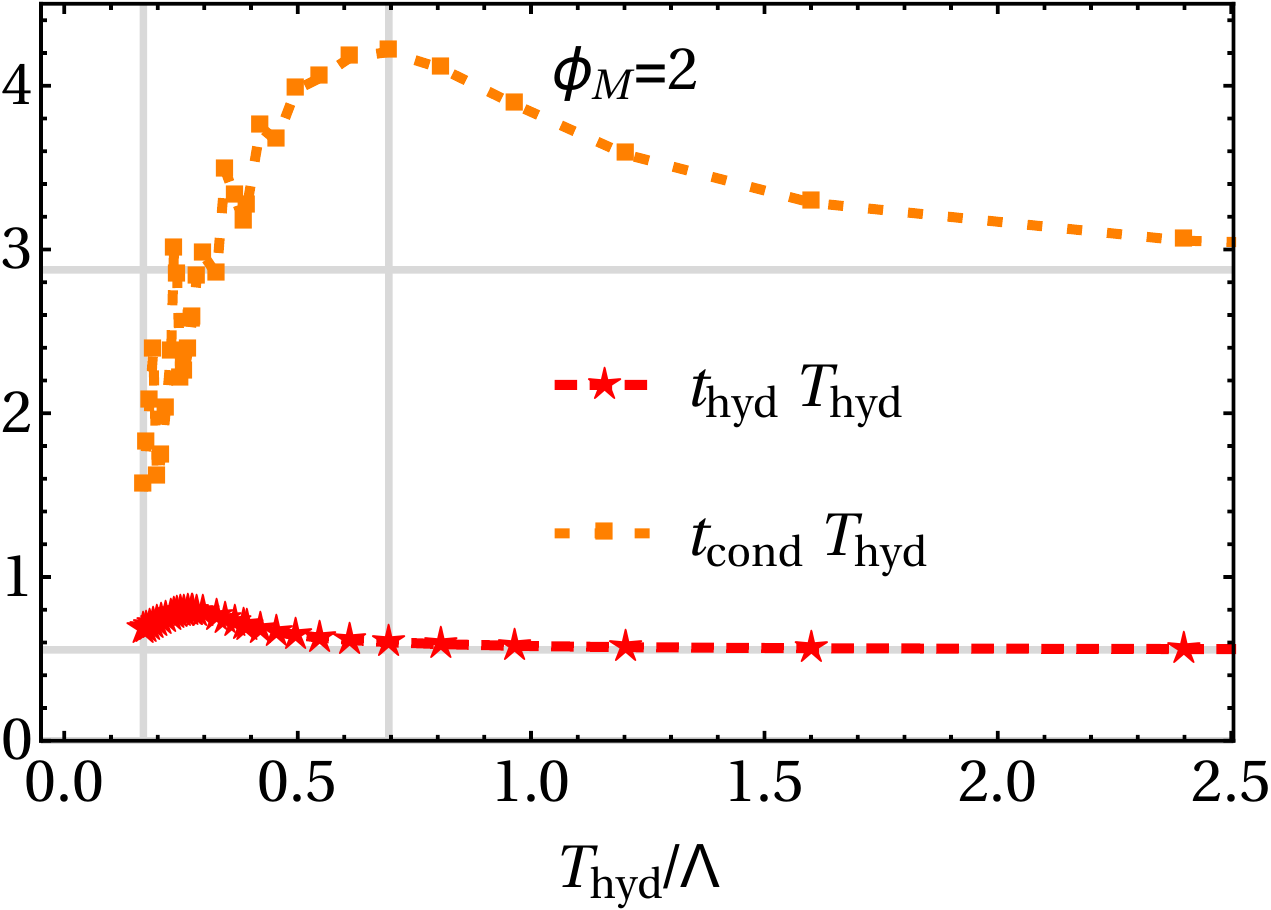}
\end{tabular}
	\caption{Condensate relaxation times and hydrodynamization times for collisions with $\mu \omega = 0.30$ in models with $\phiM = \{20,5,3,2\}$. The bottom horizontal grey line lies at $\thyd \Thyd = 0.56$ and  corresponds to the conformal limit of $\thyd$ for $1/2$ shocks. The top horizontal line lies at $\thyd \Thyd = 2.9$ and corresponds to the conformal limit of $\tcond$ for $1/2$ shocks. The leftmost (rightmost) grey vertical line indicates the lowest (highest) temperature that we probed. The other grey vertical lines indicate either points at which  
$\tcond = \thyd$, points at which the ratio $\tcond / \thyd$ is maximal, or points at which $\tcond \Thyd$ is maximal. The positions of these lines in each panel is as follows. 
Top left: $\Thyd / \Lambda = \{ 0.141, 0.165, 0.208, 0.610, 0.807 \}$. The highest temperature in this case is the one at which the ratio $\tcond / \thyd$ is maximal.
Top right: $\Thyd / \Lambda = \{ 0.143, 0.610, 0.964\}$.
Bottom left: $\Thyd / \Lambda = \{ 0.124, 0.611, 0.807, 0.965 \}$.
Bottom right:  $\Thyd / \Lambda = \{ 0.169, 0.695 \}$. The maximal values of $\tcond \Thyd$ and  $\tcond / \thyd$ take place  at the same temperature.
\label{fig:tO} }
\end{figure}

\begin{figure}[tbp]
\begin{center}
\includegraphics[width=0.6\textwidth]{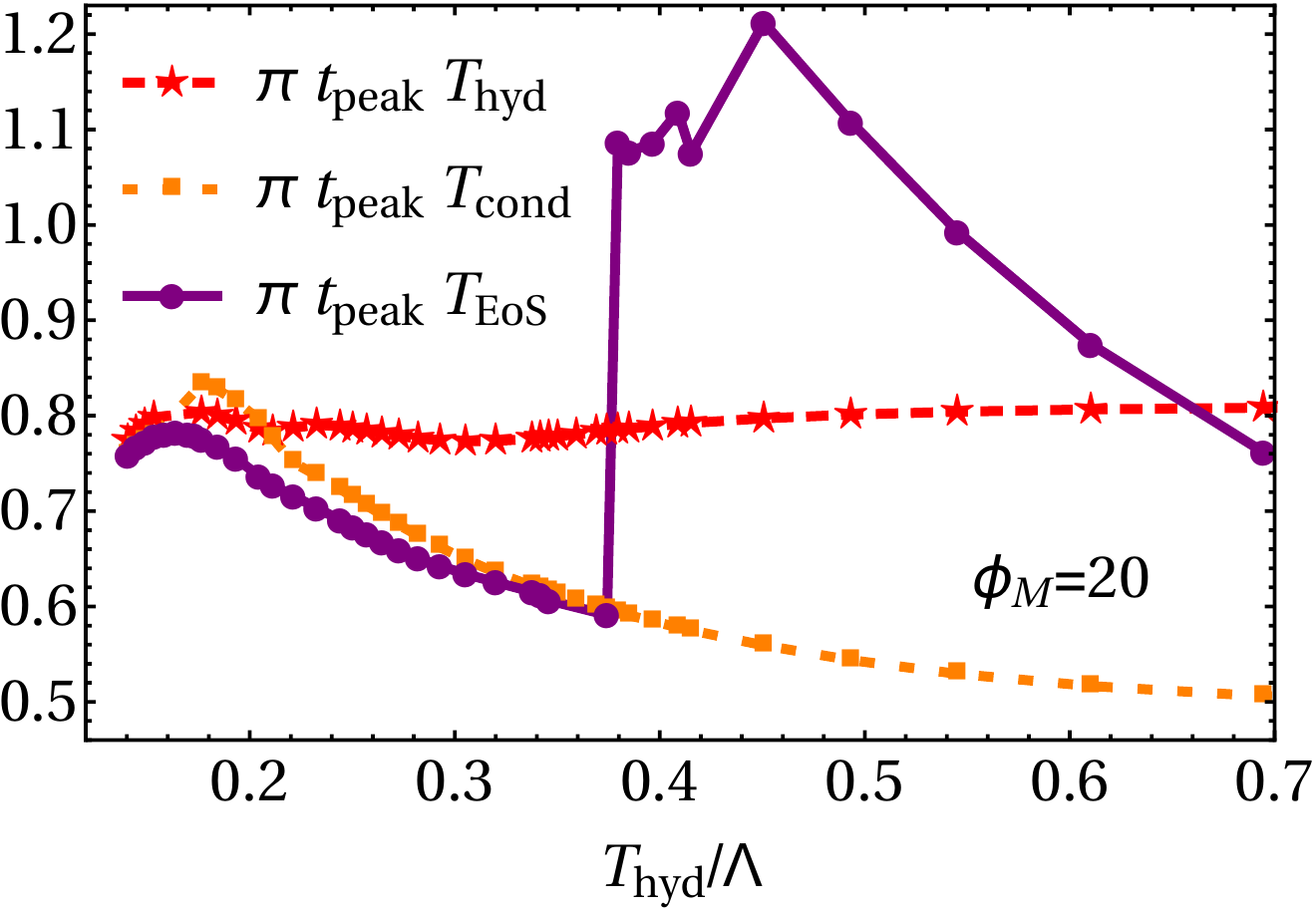} 
\caption[]{Comparison of the delay in the peak of the scalar condensate, $\tpeak$, and the effective temperatures at the times of hydrodynamization, scalar relaxation  and EoSization.}
\label{delay}  
\end{center}
\end{figure}

On general grounds, in a CFT one would expect the time at which the true condensate reaches its peak value, $\tpeak$, to be given by 
\be
\tpeak \sim \frac{c}{\pi \Thyd} \,,
\ee
with $c$ an order-one constant. The intuitive reason on the gravity side is that it takes a time of order $1/\pi \Thyd$ for the effects of the dynamics near the horizon that forms deep in the bulk when the shocks collide to reach the boundary. This delay is also observed in e.g.~the true drag force on a quark compared to the force that it would experience in an equilibrium plasma with the same instantaneous energy density \cite{Chesler:2013urd,Lekaveckas:2013lha}. In a non-conformal theory one would expect $c$ to be constant for high-energy collisions in which non-conformal effects are small but to deviate from a constant for collisions in which non-conformal effects are significant. These expectations are confirmed in our model, as illustrated by figure \ref{delay}, where we plot $\pi \tpeak   \Thyd$ and, for comparison, also $\pi \tpeak   \Tcond$ and $\pi \tpeak   \Teos$.
We see that the latter two vary significantly as a function of the collision energy (represented here by its proxy, $\Thyd$) and do not become constant at high energies. Also, in these cases one must bear in mind that $\Tcond$ and $\Teos$ are only well defined when hydrodynamization precedes scalar relaxation and EoSization, respectively. In contrast, we see that 
$\pi \tpeak   \Thyd$ does approach a constant of order $c \simeq 0.8$ at high energies, and that it deviates slightly from it at low energies.

In figure~\ref{fig:tOL} we compare the hydrodynamization time and the condensate relaxation time, in units of $\Lambda^{-1}$, as a function of the hydrodynamization temperature.
Both times attain their maximum values  at the lowest temperatures we were able to probe, where non-conformal effects are large.
Comparing different theories, we see that the maximal $\thyd \Lambda$ happens for 
$\phiM = 3$. For $\phiM = 20$ we observe a  crossing of the scalar relaxation time and the hydrodynamization time, as illustrated above in figure~\ref{fig:vev}, 
meaning that  condensate relaxation can precede  hydrodynamization or vice~versa. In contrast, models with $\phiM = \{5 , 3, 2\}$ show scalar relaxation times that are always significantly longer than the corresponding hydrodynamization times.
 In particular, for small non-conformality (small $\phiM$) and small temperatures the condensate may still be out of equilibrium at hydrodynamization.
Also for small non-conformality ($\phiM=2$) the oscillations in the scalar condensate cause jumps in the relaxation times extracted with the constant criterion \eqq{equ:criteria_O}. From this scan we can extract two
characteristic numbers: 
%the maximal value for the ratio $\tcond / \thyd \approx 5.6$ occurs for a $\phiM = 5$ and $\mu\omega=0.30$ collision, which is reached at $\Thyd / \Lambda \approx 0.96$; 
the maximal value of $\tcond \Lambda \approx 18.8$ is reached at low temperatures with $\phiM = 3$, whereas 
the maximal value for the ratio $\tcond / \thyd \approx 6.09$ occurs for $\phiM = 3$ and  is reached at $\Thyd / \Lambda \approx 0.81$.

One conclusion of figure~\ref{fig:tOL} is that both the condensate relaxation time and the hydrodynamization time, when measured in units of the intrinsic scale in the theory, decrease as the energy of the collision, or equivalently the hydrodynamization temperature. In fact, these values approach zero at asymptotically high energies, as is clear from the top-right panel in figure~\ref{fig:tOL}, where we have extended the range of the horizontal axis to high values in order to illustrate this effect.  In figure~\ref{fig:tO} we show these times measured in units of the hydrodynamization temperature itself. These plots clearly show how at high temperatures the systems behaves effectively as a conformally invariant system.  Indeed, if $T\gg \Lambda$ the temperature becomes the only relevant scale and 
both $\tcond \Thyd$  and $\thyd \Thyd$  approach constant values. Furthermore, these asymptotic values are the same in all four models, which reflects the fact their UV properties are identical. Nevertheless, the temperature at which this asymptotic sets in depends on the model. As discussed around equation \eqq{dividing}, at  high temperatures the dynamics of the condensate decouples from the dynamics of the stress tensor. The fact that in this  asymptotic regime  $\tcond \Thyd$ is 5.18 times larger than $\thyd \Thyd$ explicitly shows that a hydrodynamized plasma can be far from equilibrium, since between $\thyd$ and $5\thyd$ hydrodynamics provides a good description of the stress tensor but the expectation value of the scalar operator is still far from its equilibrium value.

\subsection{Rapidity profile}

Up to now we have focused on the mid-rapidity region, $z=0$. We will now study the energy deposition along the collision axis.
To make contact with hydrodynamic simulations of ultra-relativistic heavy ion collisions, we explore the local energy density in the fluid rest frame, 
$\mathcal{E}_{\rm loc}$, at a fixed proper time $\tau=\thyd$, 
with $\thyd$ the hydrodynamization time at $z=0$, as a function of the spacetime rapidity $y$, with 
\be
\tau =\sqrt{t^2-z^2}\, ,  \quad \quad  y=\frac{1}{2}\ln \frac{t+z}{t-z}\,.
\ee

\label{sec:rapidity}
%%%%%%%%%%%%%%%%%%%%%%%%%%%%%%%%%%%%%%%%%%%%%%%%%%%%%%%%%%%%%%%%%%%%%%%%%%%%%%%
\begin{figure}[htbp]
\begin{center}
\includegraphics[width=0.7\textwidth]{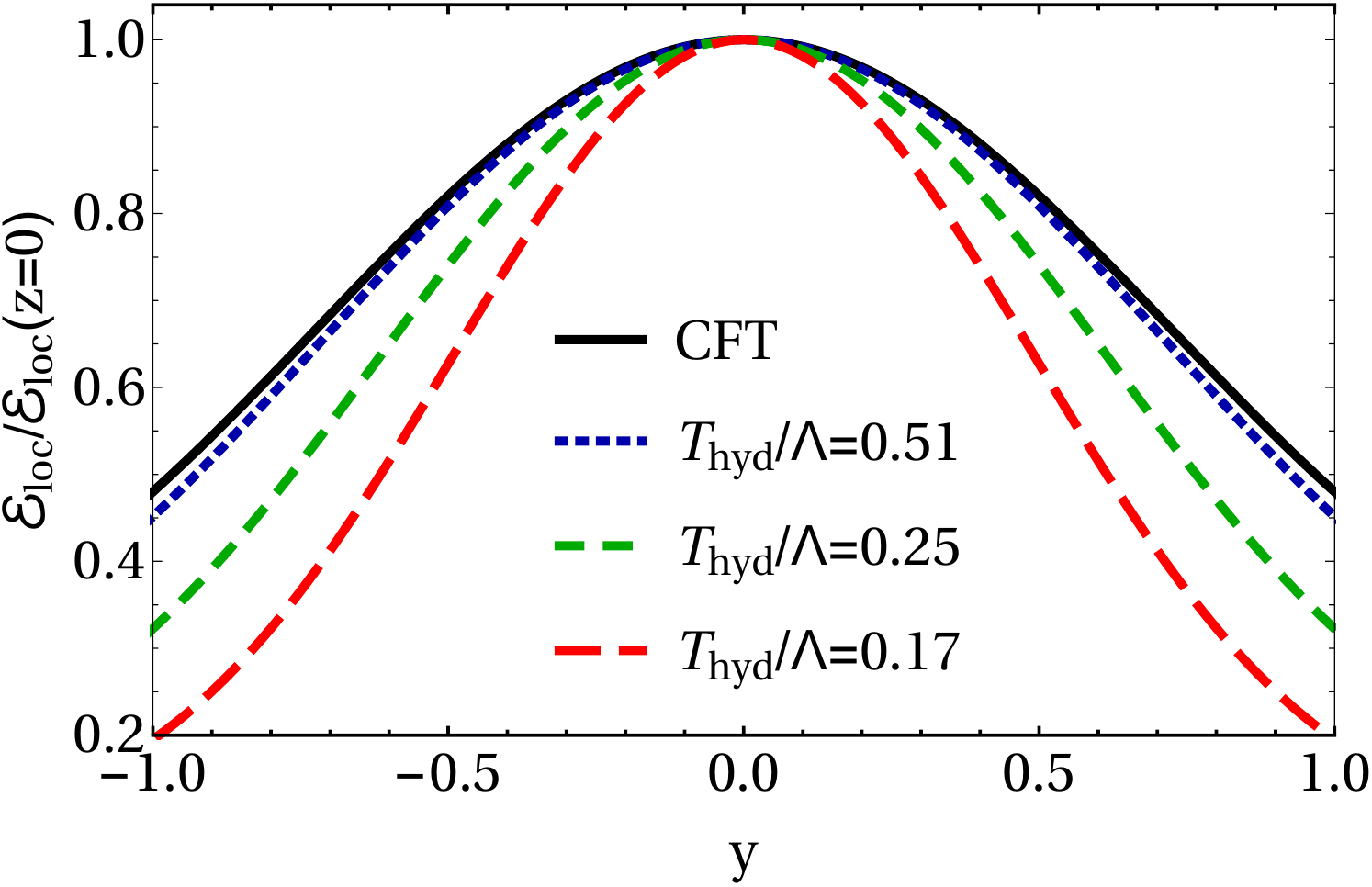}  
\caption[]{Rapidity distribution of ${\mathcal{E}}_{\rm loc}$ at fixed proper time $\tau=\thyd$, with $\thyd$ the hydrodynamization time at $z=0$  in the $\phiM = 20$ model for collisions with $\mu \omega = 0.30$ and $\mu/ \Lambda = \{ 0.29, 0.77, 1.9\}$. 
	 }
\label{fig:rapidity}  
\end{center}
\end{figure}
%%%%%%%%%%%%%%%%%%%%%%%%%%%%%%%%%%%%%%%%%%%%%%%%%%%%%%%%%%%%%%%%%%%%%%%%%%%%%%%
\begin{figure}[htbp]
\begin{center}
\includegraphics[width=0.85\textwidth]{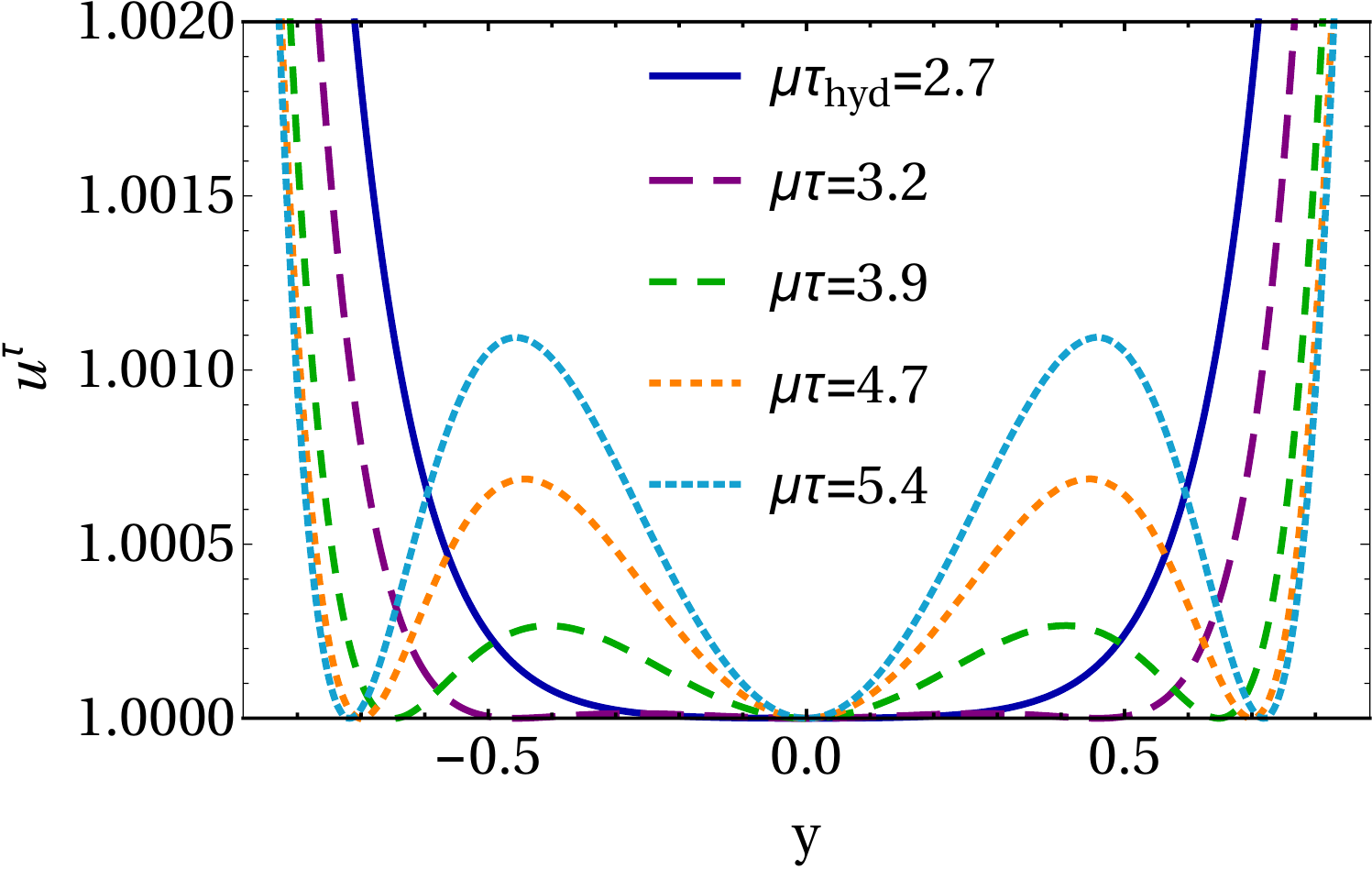} 
\caption[]{The component of the four-velocity field along the proper time direction at fix proper times 
 for the most non-conformal configuration of figure~\ref{fig:rapidity}, $\phiM = 20$ and $\mu/ \Lambda = 0.29$. }
\label{fig:rapidity_v}  
\end{center}
\end{figure}
%%%%%%%%%%%%%%%%%%%%%%%%%%%%%%%%%%%%%%%%%%%%%%%%%%%%%%%%%%%%%%%%%%%%%%%%%%%%%%%

In  figure~\ref{fig:rapidity} we show the rapidity distribution of the local energy density (normalised to the central energy density) for different collision energies in the 
$\phiM=20$ model. For comparison, we also show the same distribution for collisions in  $\mathcal{N}=4$ SYM \cite{Casalderrey-Solana:2013aba,Chesler:2015fpa}.  As in that conformal  case, here the deposited energy density exhibits a strong rapidity dependence which is well approximated by a Gaussian within a 1-2\% accuracy. The width of the Gaussian, however, depends on the transverse energy scale 
$\mu/ \Lambda$. For smaller values of the collision energy, 
the hydrodynamization temperature is also smaller and the non-conformal behaviour is more pronounced. As the collision energy increases, the rapidity width of the energy deposition grows, approaching the conformal 
distribution asymptotically at large collision energies. Although this energy density profile is controlled by  non-hydrodynamized dynamics, the observed dependence with  $\Thyd$ is consistent with the expectations from 
bulk viscosity. Similarly to the reduction of transverse expansion observed in hydrodynamic simulations of ultra-relativistic plasmas \cite{Ryu:2015vwa}, the bulk viscosity reduces the longitudinal pressure, reducing the transport of energies at large rapidities.  It is  interesting that the increase in the width of the energy rapidity profile is in qualitative agreement with the rapidity distribution of matter in heavy ion collisions as a function of $\sqrt{s}$.

Despite the fact that the system  is manifestly non-boost invariant, specially in the most non-conformal region, to a very good approximation the initial velocity field at hydrodynamization is. 
In figure~\ref{fig:rapidity_v} we show the component of the velocity field along the proper time direction,
\begin{align}
u^\tau = \cosh \left( y \right) u^t - \sinh \left( y \right) u^z\,,
\end{align}
as a function of rapidity for several proper times after $\tau=\thyd$. The fact that 
$u^\tau$ is so close to 1 in the two units of rapidity that we have plotted shows that 
the  four-velocity field is well  aligned with the proper time direction, with small deviation at the sub-percent level. 
This result was first observed in shockwave collisions in conformal gauge theories~\cite{Chesler:2015fpa} for a variety of initial Gaussian shock widths. What we are observing here is  that this result survives the introduction of large non-conformal effects, even though those same effects do cause a narrowing of the energy density rapidity distribution.
 At later times, the fact that the rapidity deposition of energy is not boost-invariant will change the velocity field, increasing the rapidity component of the velocity; nevertheless, this change is completely  predicted by hydrodynamics. Our simulations imply than even for non-conformal dynamics, in order to completely predict the stress tensor dynamics in different configurations only the rapidity distribution of energy density needs to  be specified at hydrodynamization, since the initial velocity field is given, to a very good approximation, by $u^\tau=1$. 
 This observation can be translated into consequences for hydrodynamic modellers of heavy ion collisions: even for configurations with significant rapidity dependence the initialization of the velocity field after the collision in a boost-invariant manner is well supported by our simulations.

%%%%%%%%%%%%%%%%%%%%%%%%%%%%%%%%%%%%%%%%%%%%%%%%%%%%%%%%%%%%%%%%%%%%%%%%%%%%%%%
\section{Discussion}
\label{sec:discussion}
%%%%%%%%%%%%%%%%%%%%%%%%%%%%%%%%%%%%%%%%%%%%%%%%%%%%%%%%%%%%%%%%%%%%%%%%%%%%%%%

Following the procedure described in section~\ref{sec:numerical} we have simulated 565  shockwave collisions in the gravity-plus-scalar models of \cite{Attems:2016ugt}. Via holography, we have used the results to perform a thorough analysis of the out-of-equilibrium dynamics of the dual set of non-conformal gauge theories with different degrees of non-conformality. 

One of the most astonishing results of this analysis is the tremendous success of hydrodynamics to describe the out-of-equilibrium evolution. 
This fact has been extensively studied in many settings for conformal theories in the past, where it was found that, at strong coupling, hydrodynamization typically precedes isotropization. We have verified that this is also the case in our non-conformal plasmas. In fact, in all the collisions that we have examined we have found that isotropization is always the last process to take place of the four that we have considered. To illustrate this quantitatively, we note that the ratio $P_T/P_L$ at the latest of the three equilibration times shown in each of the four panels of figure~\ref{fig:vev} is $2.0, 1.9, 1.9$ and $1.7$, respectively. In other words, at the latest equilibration time shown in the panels the transverse pressure is still at least 70\% larger than the longitudinal one, indicating that the plasma is still significantly anisotropic. 

It is remarkable  that hydrodynamics works so well even with a non-trivial equation of state. In particular, in our most non-conformal models the number of degrees of freedom changes by several orders of magnitude between the high- and the low-temperature phases---three in the $\phiM=10$ case shown in figure \ref{plot:sofT}(right). Yet, the dynamics of the system is well described very soon after the collision by a hydrodynamic expansion around this non-trivial equation of state. The break-down of the  different components of the hydrodynamic  estimator displayed in figure~\ref{fig:landau} clearly illustrates this point.  The success is such that in our extensive exploration of the parameter space of non-conformal collisions   we have never encountered a case in which the hydrodynamization time exceeds the value in the conformal case by factor larger than 2.6. This is in agreement with the expectations based on the near-equilibrium analysis in terms of quasi-normal modes \cite{Buchel:2015saa,Attems:2016ugt}.

The success of hydrodynamics is even more surprising in cases in which hydrodynamization precedes all other equilibration processes. In these cases, which correspond to the panels 
2 and 3 of figure~\ref{fig:vev}, hydrodynamics provides an accurate description of the evolution of the plasma despite the fact that ``everything else is far from equilibrium'', meaning that  the average pressure and the condensate are still far from their equilibrium values and the plasma is still highly anisotropic.  

Focusing on the particular ordering of hydrodynamization and EoSization, 
our results  confirm that the former precedes the latter as long as the system is sufficiently non-conformal. What is perhaps surprising is that, as measured by the bulk viscosity-to-entropy ratio, a ``sufficient'' degree of non-conformality requires only a fairly moderate value 
$\zeta/s \gtrsim 0.025$, as estimated in \cite{Attems:2016tby}. 
This  indicates that similar phenomena may also occur in real-world heavy ion collisions, where both calculations \cite{Paech:2006st,Arnold:2006fz,Karsch:2007jc,Denicol:2014vaa}  and data-driven parametrization \cite{Ryu:2015vwa, Bernhard:2016tnd}  yield larger values than this estimate in a significant part of the time evolution of the resulting plasma. It would be interesting to extend existing  
phenomenological studies \cite{Torrieri:2008ip,Monnai:2009ad,Song:2009rh,Rajagopal:2009yw,Dusling:2011fd,McDonald:2016vlt} of the effect of bulk viscosity in heavy ion collisions to investigate the possibility that hydrodynamization may precede EoSization.

Although the Ward identity \eqq{eq:Ward} implies that  EoSization and  condensate relaxation  are related, we have seen that nevertheless these two processes  can occur in any ordering.  The reason for this is easy to understand in  two limits, one in which the temperature is much higher than the intrinsic scale in the theory and another in which it is comparable to this scale. In the first case the different scalings with the temperature of $\bar P\sim T^4$ and of $\Lambda \mathcal{V} \sim \Lambda^2 T^2$ imply that at high temperature the contribution of the condensate to the Ward identity is subleading, and the dynamics of the stress tensor decouples from that of the condensate. Thus, in this limit the system can EoSize and hydrodynamize while the condensate remains far from equilibrium.  This is clearly illustrated by figures \ref{fig:tTbulkall} and \ref{fig:tO}, in which we see that in the high-temperature limit 
\be
\Thyd \teos \to 0 \sac \Thyd \thyd \to 0.56 \sac \Thyd \tcond \to 2.9 \,.
\ee
Our simple example suggests that other one- or higher-point functions of non-conserved operators may take a long time to relax even in an almost-conformal, hydrodynamized and EoSized plasma---for example, a similar  delay in  the relaxation of fluctuations in a non-conformal plasma undergoing a process of isotropization has been analysed in  \cite{Chesler:2012zk}. 
  This may have important implications for processes depending on non-hydrodynamic properties of the plasma created in heavy-ion collisions, such as emission rates and the reaction of the plasma to probes, which are typically assumed to be quantified in terms of equilibrium plasma properties.  It would be interesting to explore the deviations from equilibrium of these  phenomenologically relevant quantities with holography. 

In the second case, when the temperature is close to the value at which the non-conformal effects are maximal, the value of the pressures and of the condensate are all parametrically the same. However, numerically we find that in some situations 
$\Lambda \mathcal{V} > 3 \bar P$ at $t=\tcond$. This means that at this time the condensate is within 10\% of its equilibrium value but its contribution though the Ward identity still causes a larger-than-10\% deviation between the average pressure and its equilibrium value.  

In section \ref{scalar} we determined the possible orderings once the three times 
$\thyd, \teos$ and $\tcond$ are simultaneously considered. We found that in our model only the four orderings illustrated in figure \ref{fig:vev} seem to be realized. Out of the six orderings that are logically possible, the two missing ones are 
\begin{enumerate}[resume]
 \item
 EoSization  $\to$ Condensate relaxation $\to$ Hydrodynamization,
 \item 
Condensate relaxation $\to$  EoSization $\to$ Hydrodynamization, 
 \end{enumerate}
namely the two orderings in which hydrodynamization happens last. Presumably the reason is simply that our collisions do not produce a plasma that is sufficiently anisotropic. Indeed, EoSization, and indirectly condensate relaxation through the Ward identity, is controlled by the bulk gradient corrections to the equilibrium pressure. Therefore it is conceivable that, in a dynamical situation in which shear corrections are much larger than bulk corrections, the average pressure and the condensate may relax to their equilibrium value at a time at which the difference between the pressures is still not well predicted by hydrodynamics. 

\begin{figure}[tbp]
\begin{tabular}{cc}
\includegraphics[width=0.48\textwidth]
{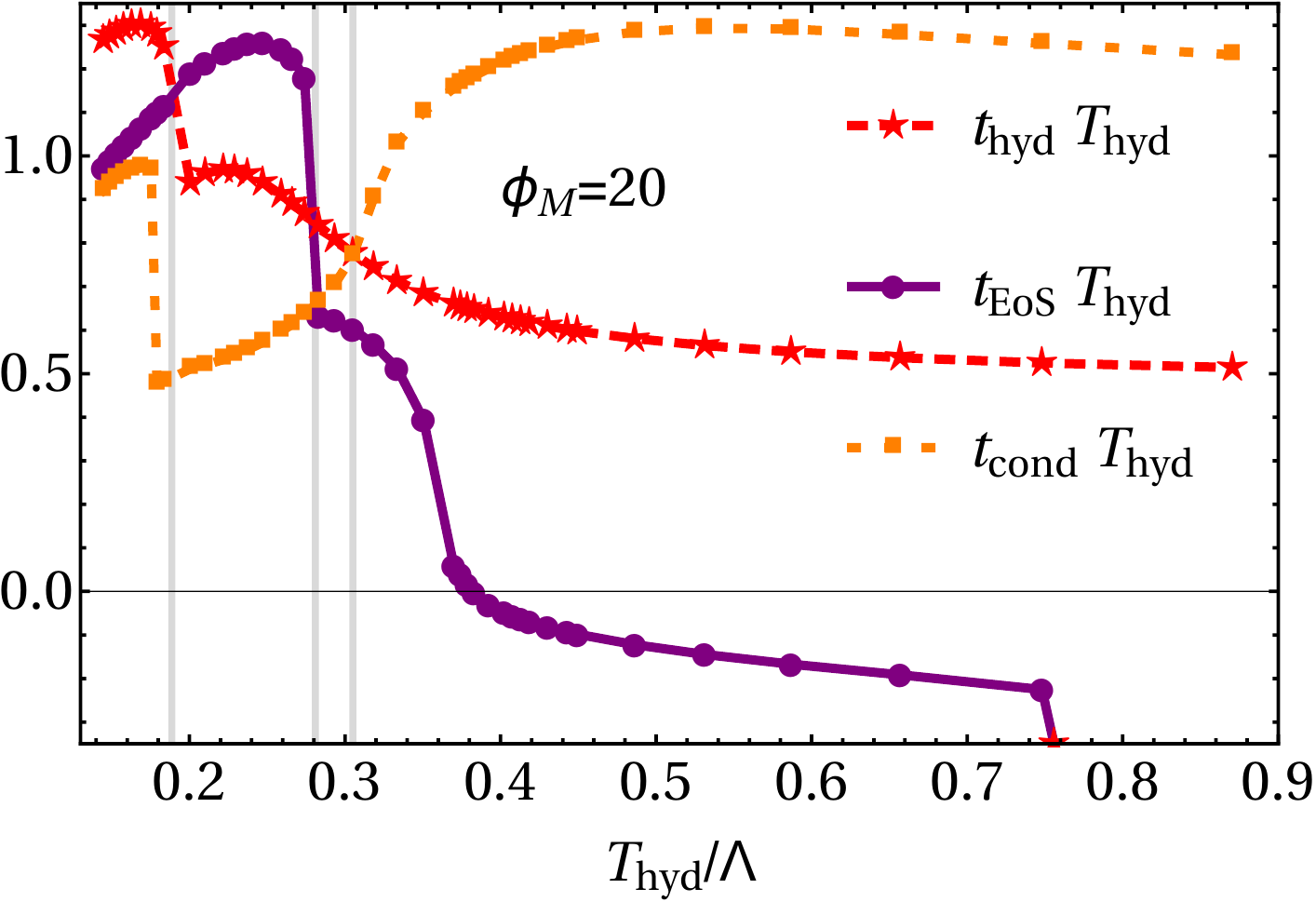}
& \includegraphics[width=0.48\textwidth]
{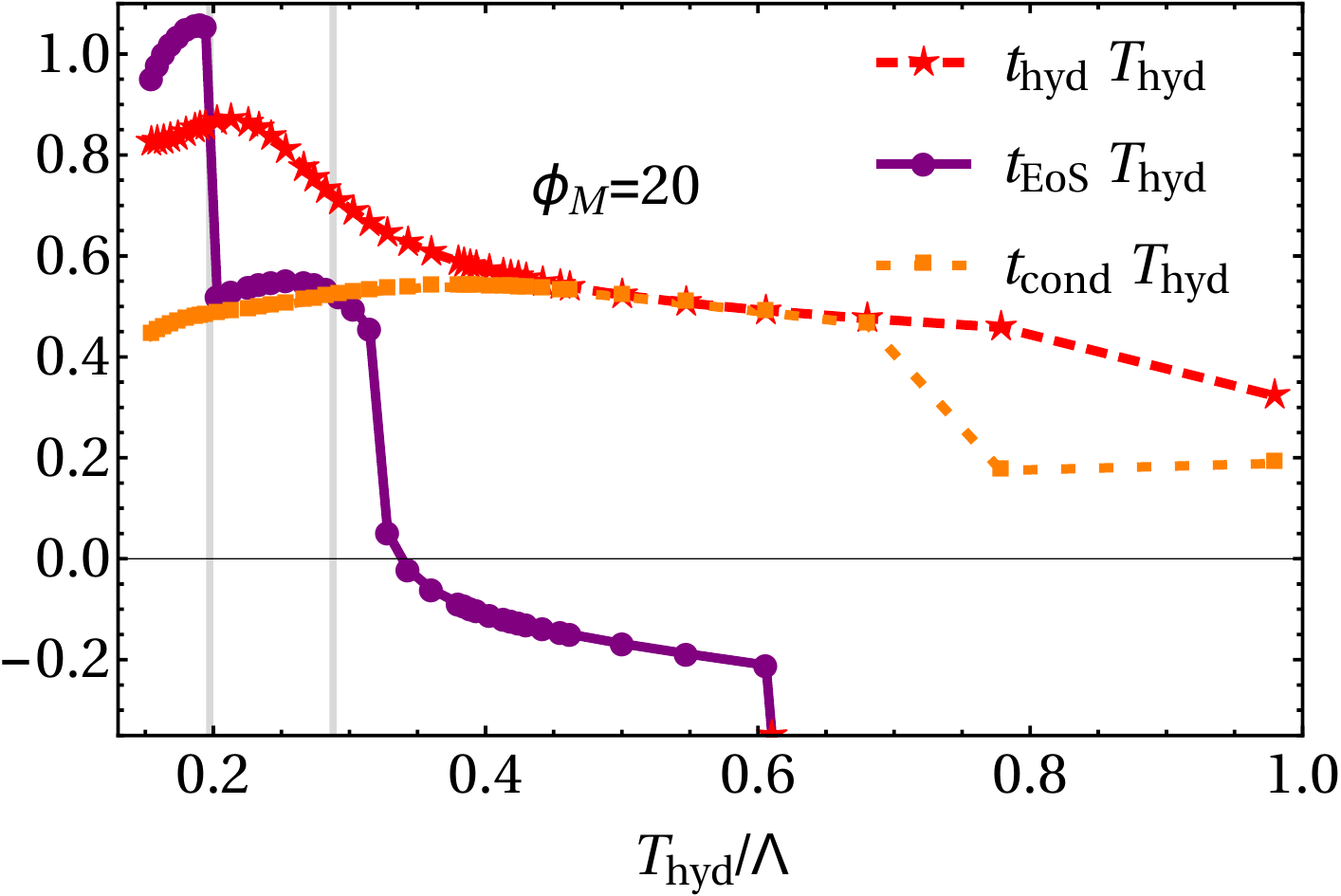}
\end{tabular}
	\caption{Hydrodynamization, EoSization and condensate relaxation times for collisions with $\mu \omega = 0.30$ for the model with $\phiM=20$ with the ``10\%'' criterion of equations \eqq{equ:criteria_hydro}, \eqq{equ:criteria_eos} and \eqq{equ:criteria_O} replaced by a 15\% criterion (left) or a 20\% criterion (right).   Regions with different orderings are separated by grey vertical lines. (These numerical simulations were performed with a 1\% regulator.)
\label{criteria} }
\end{figure}

Throughout the paper we have adopted a ``10\%'' criterion  to define the hydrodynamization, EoSization and condensate relaxation times in \eqq{equ:criteria_hydro}, \eqq{equ:criteria_eos} and \eqq{equ:criteria_O}. Since this criterion is arbitrary, it is interesting to ask what happens if the 0.1 in these equations is replaced by, say, 0.15 or 0.2. The result is summarised in figure \ref{criteria}, which shows the three equilibration times with a 15\% criterion or a 20\% criterion.
Comparison with the 10\% criterion yields the following qualitative conclusions: 
\begin{itemize}
\item
Although not shown in figure \ref{criteria}, the isotropization time is still the longest. 

\item
The conclusion that the two times $\teos$ and $\tcond$ can occur in any ordering remains true for any criterion.  

\item
The three times $\thyd$, $\teos$ and $\tcond$ can still occur in several different orderings, but which specific orderings are realized depends on the criterion. With the 15\% criterion these orderings are 1, 4, 5 and 6, whereas with the 20\% criterion we get 4, 5 and 6, and almost 1. However, it is possible that in a model with more general dynamics (larger gradients, larger bulk viscosity, etc) all possible orderings may be realized  for a given criterion.  

\item
Hydrodynamization can precede EoSization with a 15\% criterion (as with the 10\% criterion) but not with a 20\% criterion. This is not surprising since the moderate bulk viscosity of our model is only able to produce moderate deviations of the average pressure from its equilibrium value. For example, for the collisions examined in \cite{Attems:2016tby} this deviation at $\thyd$ was about 18\%. 

\item
With the two new criteria there is no collision in which hydrodynamization precedes all other equilibration processes. In other words, the orderings 2 and 3 are only realized with a 10\% criterion. Again, we expect that these orderings would be realized for  less stringent criteria  in a model with more general dynamics.

\end{itemize}

For simplicity, we have considered a model with a single scalar field, i.e.~we have focused on the dynamics in the sector in which only the conserved stress tensor and one non-conserved scalar operator are included. In a model in which several non-conserved operators are considered, the  Ward identity \eqq{eq:Ward} would relate the trace of the stress tensor to the sum of the sources times the condensates of all the non-conserved operators. Therefore any other linearly independent combination of these operators would be unconstrained by the Ward identity. It would be interesting to study  a model of this type, since presumably the dynamics would be even richer than in our one-field model.

We have also studied the post-collision deposition of energy as a function of rapidity.
As in conformal collisions, the initial longitudinal flow field is, to a surprising degree of accuracy, boost invariant. Even for the most non-conformal collisions that we have studied the size of the longitudinal gradients is insufficient to alter the longitudinal expansion of the created matter. This, together with similar results found in conformal collisions \cite{Chesler:2015fpa},  may be  viewed as dynamical evidence in support of initializing  hydrodynamic simulation of heavy ion collisions with a boost invariant flow field, even at relatively small collision energies.

Concerning the rapidity profile of the energy density, we have found that  non-conformal effects make the rapidity distribution of the collision debris  narrower than for a conformal collision with identical collision parameters. On the one hand, this is perhaps unsurprising since, at least in the hydrodynamized regime, this may be expected from the friction induced by the bulk viscosity. On the other hand, this feature highlights a main difference between the non-conformality of our model and that of QCD: In our model the theory flows at high energies to a strongly coupled fixed point, whereas QCD flows to a free fixed point. In other words, as any model that can be fully described by classical gravity \cite{Mateos:2011bs}, our model fails to reproduce asymptotic freedom. In QCD this property makes the energy rapidity profile broader and broader as the collision energy increases, since in this asymptotic regime the physics mostly responsible for setting this profile is pre-hydrodynamic weakly coupled physics. In contrast, in our model the rapidity profile saturates at high energies to that of a strongly coupled conformal theory, which is known to result in a narrower profile \cite{Chesler:2015fpa,vanderSchee:2015rta}. It would be interesting to develop hybrid approaches,  perhaps along the lines of \cite{Casalderrey-Solana:2014bpa,Iancu:2014ava}, able to address separately the strongly coupled regime at energies around $\Lambda$ via holography, and the weakly coupled regime at much higher energies via a different description.

\appendix

%%%%%%%%%%%%%%%%%%%%%%%%%%%%%%%%%%%%%%%%%%%%%%%%%%%%%%%%%%%%%%%%%%%%%%%%%%%%%%%
\section{Matching the hyperbolic equations}
\label{sec:match}
%%%%%%%%%%%%%%%%%%%%%%%%%%%%%%%%%%%%%%%%%%%%%%%%%%%%%%%%%%%%%%%%%%%%%%%%%%%%%%%

Let us consider the evolution equations for the metric variable $B$ (the corresponding ones for $\phi$ are entirely analogous). As outlined in section~\ref{sec:finite-parts}, we have two grids, grid1 and grid2, where we need to evolve $B_{g_1}$ and $B_{g_2}$ (algebraically related with the metric coefficient $B$). The two grids can overlap, but we assume for simplicity that they merely touch at point $u=u_0$, i.e., grid1 covers the region $u \in [0,u_0]$ and grid2 covers $u \in [u_0, u_{\rm max}]$, the AdS boundary being at $u=0$.

From equation~(\ref{eq:dotf}), the evolution equation for $B_{g_1}$ (the case for grid2 is entirely analogous) takes the form
\begin{align}
\label{eq:dtBg1}
\partial_t B_{g_1} & = \frac{(4 B_{g_1} + u \partial_u B_{g_1}) \left(-2 u^2 \partial_t\xi + A_{g_1} u^4 + (1 + u\xi)^2\right)+2 \dot B_{g_1}}{2 u}  \notag \\
&{} - \frac{\phi_0^2}{3} u (4B_{g_1} + u \partial_u B_{g_1}) \,,
\end{align}
which has the generic form
\begin{equation}
\label{eq:interior}
\partial_t B_{g_1} = c_{g_1}(u,z) \partial_u B_{g_1} + S_{g_1}(u,z) \,,
\end{equation}
with
\begin{equation}
\label{eq:g1-mode-speed}
c_{g_1}(u,z) = - u^2 \partial_t\xi + \frac{1}{2} A_{g_1} u^4 + \frac{1}{2} (1 + u\xi)^2
- \frac{\phi_0^2}{3} u^2 \,.
\end{equation}
$c_{g_1}(u,z)$ is locally the propagation speed and in the vicinity of $u=u_0$ we can formally write the solution of this equation (ignoring from now on the $z$ dependence) as
\begin{equation*}
B_{g_1}(t,u_0) \simeq f(u_0 + c_{g_1} t) + \int S_{g_1}
\end{equation*}
for any given function $f$.

Therefore, for $c_{g_1} > 0$ ($c_{g_1} < 0$), information is propagating from grid2 to grid1 (grid1 to grid2). In order to consistently solve this system, the procedure will then be to use equation~\eqref{eq:dtBg1} (and the corresponding one for $B_{g_2}$ on grid2) on all interior points (i.e., points where $u \neq u_0$) and for the junction point $u=u_0$ one checks the propagation speed at each $z$ point and copies the values according to the propagation direction:
\begin{itemize}
\item $c_{g_1}>0$
\begin{align}
\label{eq:plus}
& \partial_t B_{g_2}|_{u=u_0} = c_{g_2}(u_0) \partial_u B_{g_2}|_{u=u_0} + S_{g_2}(u_0) \,, \\
& \partial_t B_{g_1}|_{u=u_0} = \frac{1}{u_0^4} \partial_t B_{g_2}|_{u=u_0} \,,
\end{align}
i.e., we copy the modes leaving grid2 to grid1.

\item $c_{g_1}<0$
\begin{align}
\label{eq:minus}
& \partial_t B_{g_1}|_{u=u_0} = c_{g_1}(u_0) \partial_u B_{g_1}|_{u=u_0} + S_{g_1}(u_0) \,, \\
& \partial_t B_{g_2}|_{u=u_0} = u_0^4 \partial_t B_{g_1}|_{u=u_0} \,,
\end{align}
i.e., we copy the modes leaving grid1 to grid2.

\end{itemize}

%%%%%%%%%%%%%%%%%%%%%%%%%%%%%%%%%%%%%%%%%%%%%%%%%%%%%%%%%%%%%%%%%%%%%%%%%%%%%%%
\acknowledgments
We thank A.~Buchel, P.~Chesler, P.~Figueras, G.~Horowitz, J.~Rocha, P.~Romatschke, W.~van der Schee, B.~Schenke, U.~Sperhake, K.~Tywoniuk, R.~Venugopalan and U.~Wiedemann for discussions.
We thank the supercomputer MareNostrum at the Barcelona Supercomputing
Center for providing computational resources (project no.~ub65).
The work of MA has been supported by a Marie Sklodowska-Curie Individual Fellowship of the European Commission's Horizon 2020 Programme under contract number 658574 FastTh. JCS is a Royal Society University Research Fellow. JCS was  also supported by a Ram\'on~y~Cajal fellowship,  by  the  Marie Curie Career Integration Grant FP7-PEOPLE-2012-GIG-333786 and by the Spanish MINECO through grant FPA2013-40360-ERC.
CFS and DS acknowledge the support from contracts ESP2013-47637-P and ESP2015-67234-P (Spanish Ministry of Economy and Competitivity of Spain, MINECO).
MZ acknowledges support through the FCT (Portugal) IF programme, IF/00729/2015.
We also acknowledge funding from grants MEC FPA2013-46570-C2-1-P, MEC FPA2013-46570-C2-2-P, MDM-2014-0369 of ICCUB, 2014-SGR-104, 2014-SGR-1474, CPAN CSD2007-00042 Consolider-Ingenio 2010, and ERC Starting Grant HoloLHC-306605.

% \clearpage

\bibliographystyle{JHEP}
\bibliography{shocks_3}

\end{document}